\documentclass[sigconf]{acmart}
\usepackage{pifont}
\usepackage{array}

\usepackage{booktabs}   
\usepackage{cleveref}

\AtBeginDocument{%
  }

\copyrightyear{2025}
\acmYear{2025}
\acmConference[CHI '25]{CHI Conference on Human Factors in Computing Systems}{April 26-May 1, 2025}{Yokohama, Japan}
\acmBooktitle{CHI Conference on Human Factors in Computing Systems (CHI '25), April 26-May 1, 2025, Yokohama, Japan}\acmDOI{10.1145/3706598.3714045}
\acmISBN{979-8-4007-1394-1/2025/04}
\begin{document}

\title{\textsc{EvAlignUX}: Advancing UX Research through \\LLM-Supported Evaluation Metrics Exploration}


\author{Qingxiao Zheng}
\affiliation{%
  \institution{Computer Science and Engineering \\University at Buffalo}
  \city{Buffalo}
  \country{USA}}
\email{qingxiao@buffalo.edu}

 \author{Minrui Chen}
\affiliation{%
  \institution{Informatics \\University of Illinois Urbana-Champaign}
  \city{Champaign}
  \country{USA}}
\email{minruic2@illinois.edu}

 \author{Pranav Sharma}
\affiliation{%
  \institution{ Computer Science \\University of Illinois Urbana-Champaign}
  \city{Champaign}
  \country{USA}}
\email{pranav24@illinois.edu}

 \author{Yiliu Tang}
\affiliation{%
  \institution{Informatics \\University of Illinois Urbana-Champaign}
  \city{Champaign}
  \country{USA}}
\email{yiliut2@illinois.edu}

 \author{Mehul Oswal}
\affiliation{%
  \institution{Computer Science \\University of Illinois Urbana-Champaign}
  \city{Champaign}
  \country{USA}}
\email{mehuljo2@illinois.edu}

 \author{Yiren Liu}
\affiliation{%
  \institution{
Informatics \\University of Illinois Urbana-Champaign}
  \city{Champaign}
  \country{USA}}
\email{yirenl2@illinois.edu}

 \author{Yun Huang}
\affiliation{%
  \institution{Information Sciences \\University of Illinois at Urbana-Champaign}
  \city{Champaign}
  \country{USA}}
\email{yunhuang@illinois.edu}

\renewcommand{\shortauthors}{Zheng et al.}

\newcommand{\systemName}{\textsc{EvAlignUX~}}
\begin{abstract}
{Evaluating UX in the context of AI’s complexity, unpredictability, and generative nature presents unique challenges. \textit{How can we support HCI researchers to create comprehensive UX evaluation plans?} In this paper, we introduce \systemName, a system powered by large language models and grounded in scientific literature, designed to help HCI researchers explore evaluation metrics and their relationship to research outcomes. A user study with 19 HCI scholars showed that \systemName improved the perceived quality and confidence in UX evaluation plans while prompting deeper consideration of research impact and risks. The system enhanced participants' thought processes, leading to the creation of a \textit{“UX Question Bank”} to guide UX evaluation development. Findings also highlight how researchers’ backgrounds influence their inspiration and concerns about AI over-reliance, pointing to future research on AI’s role in fostering critical thinking. In a world where experience defines impact, we discuss the importance of shifting UX evaluation from a \textit{“method-centric”} to a \textit{“mindset-centric”} approach as the key to meaningful and lasting design evaluation.}


\end{abstract}


\begin{teaserfigure}
    \centering
    \vspace{1pt} 
    \includegraphics[width=0.9\textwidth]{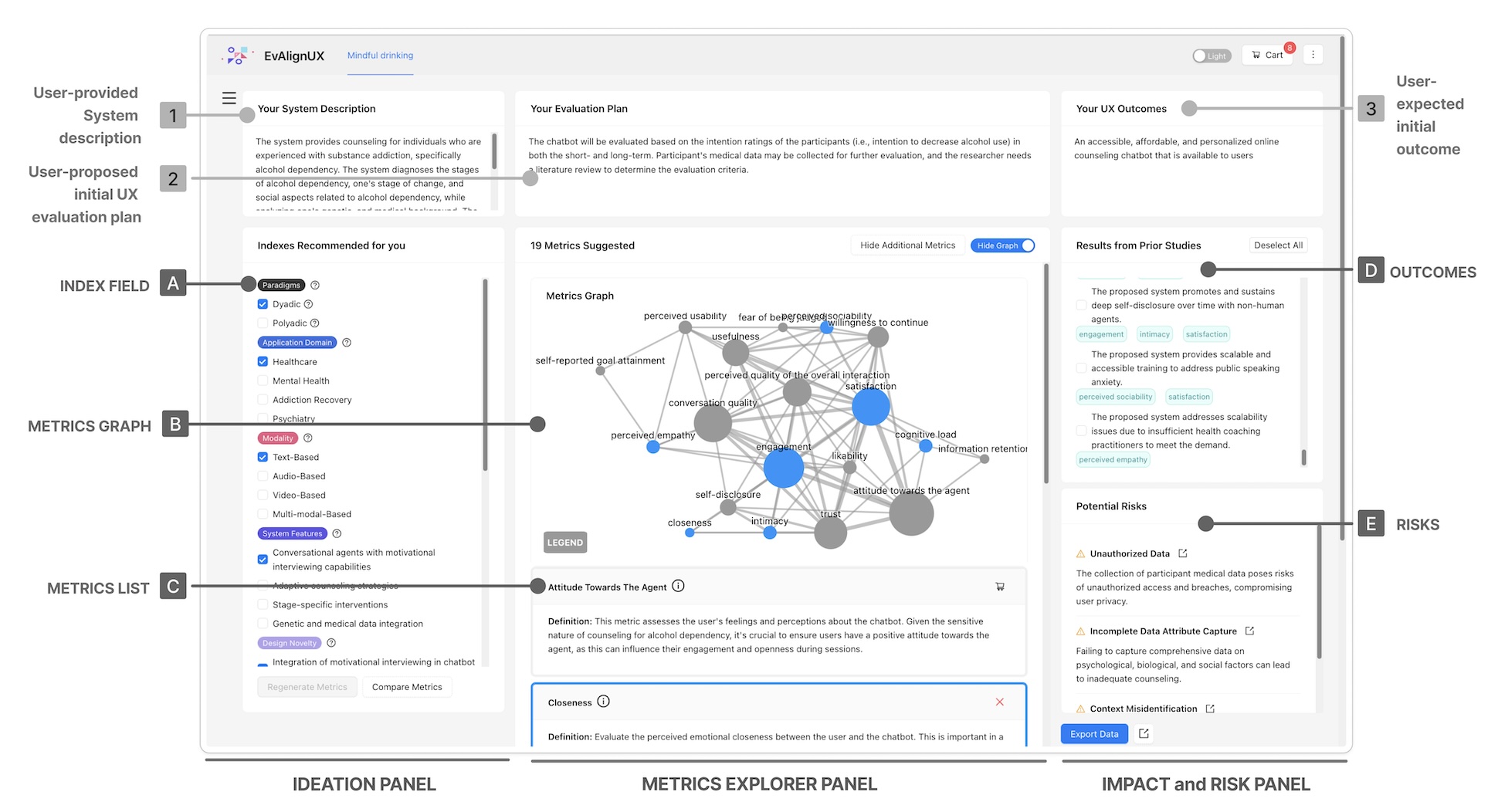}
    \caption{\systemName comprises three main panels, each powered by large language models (LLMs): the project IDEATION Panel (left), the METRICS EXPLORER Panel (center), and the IMPACT AND RISK Panel (right). After the user provides 1) a project description, 2) an initial evaluation plan, and 3) expected outcomes, they can utilize  five key features--\texttt{INDEX} [A], \texttt{METRICS GRAPH} [B], \texttt{METRICS LIST} [C], \texttt{OUTCOMES} [D], and \texttt{RISKS} [E]--to refine the UX evaluation plan and hypothesize potential results.}  
\label{fig:system-interview-overview}
\Description{The image presents the interface of EValignUX, divided into three main panels: Ideation Panel (left), Metrics Explorer Panel (center), and Impact and Risk Panel (right). The Ideation Panel contains a user-provided system description (1) and an initial UX evaluation plan (2), along with an index field (A) that suggests relevant indexes, like ‘application domain’ and ‘paradigm’ based on the project description. The Metrics Explorer Panel features a metrics graph (B) showing connections between UX metrics, which visualizes features as nodes with varying sizes and colors to indicate their importance. Below the graph is a metrics list (C), offering detailed descriptions of suggested metrics like ‘Attitude Towards the Agent’ and ‘Closeness’. The Impact and Risk Panel allows users to define expected UX evaluation outcomes (3) like ‘an accessible, affordable, and personalized online counseling chatbot’ (D), while also listing potential risks (E) such as ‘Unauthorized Data’ and ‘Incomplete Data Attribute Capture’.}
\end{teaserfigure}

\maketitle
\section{Introduction}

The rapid integration of AI into interactive systems presents a growing challenge for user experience (UX) evaluation \cite{stige2023artificial, 10.1145/3613905.3650878}. Traditional usability evaluation guided by standard frameworks such as the System Usability Scale (SUS) \cite{bangor2008empirical}, User Experience Questionnaire (UEQ) \cite{schrepp2017design}, NASA Task Load Index (NASA-TLX) \cite{hart2006nasa}, falls short in capturing the complex, dynamic interaction between humans and AI~\cite{kuang2024enhancing}. As AI systems become more context-aware and capable of generating responses in real-time, UX researchers are confronted with new questions about trust, emotional engagement, and ethical considerations \cite{10.1145/3313831.3376590, ahmad2023requirements, shneiderman2020bridging, ntoa2021user}. 

Addressing these new considerations is especially challenging for systems research, where a range of evaluation methods, often borrowing and combining techniques from software engineering, design, and usability evaluation, is triangulated to evaluate the UX of interactive computing systems \cite{marquardt2017hcitools}. Current UX evaluation methods, e.g., the Post-Study System Usability Questionnaire (PSSUQ) \cite{lewis1992psychometric}, the Subjective Mental Effort Questionnaire (SMEQ) \cite{hyland1989psychometric}, while robust for traditional systems, are inadequate for AI-powered interfaces. Researchers are striving to develop new frameworks that take into account ethical considerations, trust dynamics, intimacy, and multimodal behaviors, to adequately guide scholars and practitioners in developing their UX evaluation plans for human-AI interaction research, e.g.,~\cite{zheng2022ux, liao2024ux}. Given the plethora of well-documented evaluation methods, heuristics, and metrics (e.g. \cite{lazar2017research, 10.1145/142750.142834, vermeeren2010user}), much less is known about \textit{How researchers select appropriate metrics to conduct a comprehensive evaluation of a system}. 

Also, despite the availability of various UX evaluation methods \cite{pettersson2018bermuda, vermeeren2010user, rivero2017systematic, darin2019instrument, kieffer2019specification}, there is a lack of practical tools to guide researchers in developing comprehensive evaluation plans. Research has shown that incorporating evaluation metrics as part of the design materials in the early stages can improve the clarity in the design of human-AI interaction research~\cite{zheng2023begin}. UX is frequently regarded as a rigid set of methods, leading to calls for research and education that encourage framing UX as ``\textit{more of a mindset than a method}'' within the community \cite{gray2016s}. However, with the rapid advancement of AI models, the literature on AI-mediated multi-stakeholder interactions has expanded exponentially. This growth presents challenges for novice researchers in acquiring knowledge of relevant evaluation metrics, while seasoned researchers also face increasing pressure to stay current with the latest developments.

To bridge this gap, we designed, implemented, and evaluated \textsc{\systemName}, an interactive system that supports novice and experienced HCI researchers in developing UX evaluation plans for their research projects. \systemName is powered by the generative capabilities of Large Language Models (LLMs) to facilitate the exploration of UX evaluation ideas, as LLMs have demonstrated their potential for the ideation of scientific research by processing vast amount of information, generating context-aware insights, retrieving literature, and brainstorming \cite{doshi2024generative, glickman2024ai, liu2024ai}. Specifically, as shown in Fig.~\ref{fig:system-interview-overview}, \systemName comprises three primary panels: the \texttt{Project Ideation} Panel, the \texttt{Metric Explorer} Panel, and the \texttt{Outcomes and Risks} Panel. These panels allow users to explore UX metrics, potential outcomes, and potential risks associated with human-AI interaction research. In an evaluation with 19 HCI scholars, \systemName led to significant improvements in the clarity, concreteness, and confidence of their revised UX evaluation plans. Participants were able to better justify selected UX metrics, self-critique their plans, and raise awareness of potential risks using the system features.

This work makes several significant contributions to the HCI community. \textbf{First}, \systemName is a novel system designed to support the ideation process of developing UX evaluation plans. Powered by LLMs, it enables users to explore UX metrics and their relationships with potential research outcomes. The unique visual interface of the metrics graph supports users in examining the connections between different UX metrics. To the best of our knowledge, \systemName fills the gap in tool support available to researchers by providing insights into the research literature on human-AI evaluation metrics and facilitating better decision-making regarding UX evaluation. \textbf{Second}, our user studies demonstrate the effectiveness of \systemName in helping users achieve significantly greater clarity and feasibility in their proposed UX evaluation plans. Our findings revealed that researchers' backgrounds impacted their perceived levels of inspiration and their concerns about over-reliance on AI. They also integrated the system features into their thought processes differently, focusing on varied utilities such as the comprehensiveness of self-critique and the efficiency of implementing the ideas. \textbf{Third}, our analysis of participants' think-aloud protocols and perceived value of different features produced a comprehensive \textit{Question Bank for UX Evaluation Development}, highlighting the specific questions researchers asked at different steps of UX evaluation development and identifying how AI provides valuable support such as envisioning the research impact, assessing the risks, and justifying metric selection. \textbf{Lastly}, our findings prompt a critical discussion on shifting the focus from method to mindset in future research on UX evaluation. 
\section{Related Work}
In this section, we review the literature on (1) current UX evaluation approaches and (2) the potential of leveraging LLMs to provide tool support.

\subsection{UX Evaluation of HCI System Research}

Developing a UX evaluation plan can be challenging, as it is often unclear what constitutes "evaluation" and which metrics are appropriate \cite{ledo2018evaluation, hillman2023understanding}. Hillman et al. \cite{hillman2023understanding} elaborated on several challenges involved in developing a solid evaluation plan, such as selecting metrics that accurately reflect UX and research goals, choosing data collection methods, establishing collection protocols, and ensuring consistency across products and organizations. Additionally, long-term effects and social or cultural barriers further complicate the process \cite{hillman2023understanding, darin2019instrument, vermeeren2010user}. Several foundational conceptual frameworks serve as the backbone for UX evaluation, including Nielsen's usability engineering \cite{nielsen1994usability} and heuristic evaluation \cite{nielsen1994enhancing}, Hassenzahl’s UX framework \cite{hassenzahl2010needs}, and Norman’s emotional design theory \cite{norman2007emotional}. These frameworks emphasize multi-dimensional assessments, moving beyond performance metrics to encompass user perceptions, attitudes, and context of use.

To measure users' perceptions quantitatively,  researchers often use established surveys, such as the System Usability Scale (SUS), the User Experience Questionnaire (UEQ), the Post-Study System Usability Questionnaire (PSSUQ), the NASA Task Load Index (NASA-TLX), the Subjective Mental Effort Questionnaire (SMEQ), and the Customer Satisfaction (CSAT) \cite{bangor2008empirical, schrepp2017design, lewis1992psychometric, hart2006nasa, reichheld2003one}. PULSE metrics (e.g., page views, uptime, latency, etc.) are also commonly used to track user behavior \cite{rodden2010measuring}. However, these metrics are often “low-level or indirect measures of user experience”~\cite{rodden2010measuring}. Several evaluation frameworks relating to user-centered metrics have been introduced to the HCI and UX domains, such as the Software Development Productivity Framework \cite{sadowski2019software} and Google’s HEART framework \cite{rodden2010measuring}. For instance, the HEART framework, which includes Happiness, Engagement, Adoption, Retention, and Task Success, was introduced to measure web-based applications at scale \cite{rodden2010measuring}.

Beyond selecting appropriate metrics to evaluate systems, researchers have also developed methods and tools to enhance  survey and interview question design. For example, the Question Appraisal System (QAS-99) \cite{willis2013question} and Willis’s cognitive interviewing methodology \cite{willis2004cognitive} provide structured approaches for detecting and refining data collection items, ensuring that questions are clear, unbiased, and aligned with research goals. Similarly, QUAID \cite{graesser2006question} offers diagnostic support for identifying comprehension issues in survey questions, while Survey Sidekick \cite{hsiao2014survey} assists in creating scientifically sound surveys. Semi-automated approaches \cite{siemon2016semi} further aid in generating relevant and effective questions. Incorporating such methods into UX evaluation can yield instruments that better capture users’ experiences, perceptions, and emotional responses.

As HCI rapidly shifts toward AI-driven systems, recent studies in UX research highlight several emerging trends and challenges. A group of HCI researchers aims to identify and propose new evaluation metrics for both usability and user-centered evaluation in the age of AI. For example, Zheng et al.'s literature review of human-AI interaction research in HCI, which included user evaluations, reported a suite of metrics that were collected using surveys, system logs, and interviews to collect \cite{zheng2022ux} and argued the importance of including more user-centered metrics (e.g., sociability, socio-emotional, and AI anthropomorphism metrics) with AI's evolving role in human social interactions. Likewise, Faruk et al. proposed new subjective scales for measuring the UX of voice agents \cite{faruk2024review} and conversational agents \cite{faruk2024introducing}. In other domains, such as healthcare, Abbasian et al.  summarized some foundation metrics for evaluating the effectiveness of healthcare conversations powered by AI \cite{abbasian2024foundation}. 

However, there is a lack of tools to help researchers understand the evaluation metrics literature and design a comprehensive evaluation plan for human-AI interaction research. As Yang et al. \cite{yang2020re} systematically reported that an AI system's uncertainty and complexity can lead to unpredictable errors that damage UX and result in undesired societal impact.. Gray's finding, encapsulated in the phrase \textit{"more of a mindset than a method"} also highlighted that UX practitioners often follow a codified method rather than adopting a "mindset" that aligns with their design process, practice context, and the specific design problem \cite{gray2016s}. All these works suggest that new tools and approaches are needed to support researchers and practitioners in developing their UX evaluation plans.

\subsection{The Potential of LLM-Supported Ideation}
Large Language Models (LLMs) can process vast amounts of multi-modal information and generate context-aware, coherent text, making them ideal tools for evaluation-related tasks in various domains. A group of researchers developed novel interaction prototypes using LLMs for evaluating generative outputs \cite{desmond2024evalullm, kim2024evallm, shankar2024validates}. For example, Kim et al. developed EvalLM \cite{kim2024evallm}, an interactive system for refining prompts by evaluating multiple outputs on user-defined criteria. This helps developers compare prompt outputs to diagnose weaknesses in their prompts. Regarding LLMs' support for scientific tasks, recent studies have showcased LLM-powered methods for retrieving and recommending relevant literature \cite{huaconv}, reviewing and critiquing research drafts, writing research manuscripts, and augmenting research ideation and brainstorming for scientific writing by leveraging scientific evidence and large datasets \cite{nigam2024acceleron}. Recent advancements in HCI and AI research also have also aimed to support UX practitioners with AI-enabled tools for usability testing (e.g., LLMs are employed to help human evaluators identify additional usability issues in usability test video recordings). All these research highlights the significant potential of LLMs to support research ideation and provide valuable feedback for task evaluation. However, little is known about how researchers with varying levels of experience can benefit from using LLM-supported tools to develop their UX evaluation plans.

\section{\systemName}

To address the research gap, \systemName was designed as a tool to aid researchers in exploring evaluation metrics for developing UX plans for Human-AI research. The system was developed through an iterative process, informed by both the literature review \cite{zheng2022ux, inan2021method, vermeeren2010user, rivero2017systematic, darin2019instrument} and input from a focus group. Leveraging the authors' expertise in HCI research and practices, we began with an initial process of developing a UX evaluation plan, e.g., proposing a research topic, identifying relevant UX metrics for evaluation, and framing the expected research contribution. The design was refined over two months through multiple weekly feedback sessions with an HCI research group composed of, consisting of nine graduate students and faculty members from the HCI, Informatics, and Computer Science departments (n=9). All of the participants have worked on at least one HCI project where they designed or developed systems. Their applications covered diverse contexts: vision-based question-answering, ideation and writing, accessibility, immersive technologies, and intelligent assistants. Based on the UX evaluation challenges reported in prior literature \cite{ledo2018evaluation, hillman2023understanding}, and after piloting the initial pipeline with this group, we established three design goals (DG). We then iteratively searched for additional literature and sought feedback through weekly focus group meetings.
\begin{itemize}
    \item \textbf{DG1:} Recommend UX metrics relevant to a project; 
    \item \textbf{DG2:} Explain how the metrics are used in existing literature;  
    \item \textbf{DG3:} Inform researchers about potential risks associated with their design and evaluation plans. 
\end{itemize}

\subsection{System Design}
\begin{figure*}[h!]
    \centering
    \begin{minipage}[b]{0.65\textwidth}
        \centering
        \includegraphics[width=\textwidth]{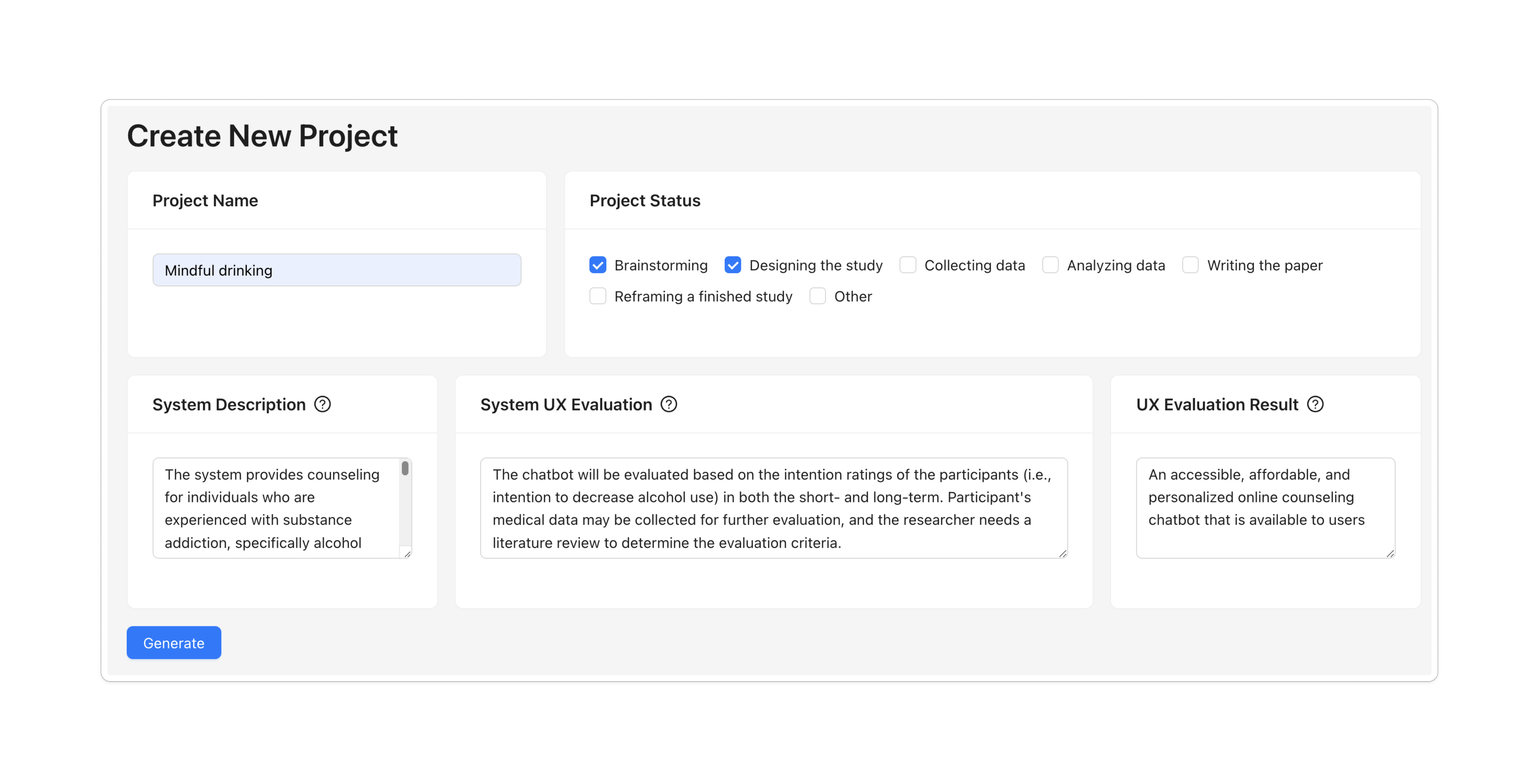} 
        \caption{\systemName Landing Page}
        \label{journey1}
    \end{minipage}%
    \hfill
    \begin{minipage}[b]{0.35\textwidth}
        {\small \textbf{User Journey 1:} \\
         Upon landing on \systemName, Sarah inputs her project details. She names the project \textit{“Mindful Drinking”}, selects “brainstorming” and “designing the study” as the current project statuses, and provides a project description, evaluation plans, and details the expected outcomes. After entering all details, she clicks “Generate” and proceeds to the main interaction page.} \\
    \end{minipage}
    \Description{The graph shows a snapshot of the EvalignUX system landing page (left) and a user journey (right). On the left, the landing page displays a form for creating a new project. This includes fields like “Project Name”, “Project Status”, “System Description”, “System UX Evaluation”, “UX Evaluation Result”. At the bottom, a “Generate” button allows the user to proceed to the next step. On the right, the accompanying text outlines the user journey. It explains that upon landing on EvalignUX, the user (Sarah) inputs project details, names the project “Mindful Drinking”, selects “brainstorming” and “designing the study” as the current project statuses, provides a system description, UX evaluation plan, and expected outcomes. After entering these details, she clicks “Generate” to continue to the main interaction page.}
\end{figure*}

Based on the DG2, \systemName{}comprises with three main panels, as shown in Figure \ref{fig:system-interview-overview}. To illustrate interactions in \systemName{}, we present a walk-through with Sarah, a UX researcher designing an AI counseling chatbot for alcohol addiction. The user journey is likened to a "shopping" process, where Sarah selects ideal evaluation plans for her proposed system. See Figure \ref{journey1} for \textbf{User Journey 1}. 

\subsubsection{Project Ideation Panel (\textbf{DG1})} This panel extracts relevant indexes to categorize the work within the existing literature. In the rapidly evolving field of UX research, evaluators often face challenges in understanding the context and relevance of newly proposed systems \cite{ledo2018evaluation}. To enhance the UX evaluator's understanding of the context in which a system operates and its significance within the broader field of UX research, we built on categorizations from prior literature reviews and key guiding literature for UX in AI design practices \cite{zheng2022ux, liao2020questioning, vermeeren2010user, kieffer2019specification}. We leverage these established frameworks to integrate critical factors into the system output. This output is designed to automatically map the proposed system based on the provided narrative, into categories including paradigms, application domains, modalities, system features, design novelty, design methods, human-AI relationship types, stakeholders, social scales, and theoretical frameworks. See Appendix \ref{index} for details. See Figure \ref{journey2} for \textbf{User Journey 2}. 

\begin{figure*}[h!]
    \centering
    \begin{minipage}[b]{0.65\textwidth}
        \centering
        \includegraphics[width=\textwidth]{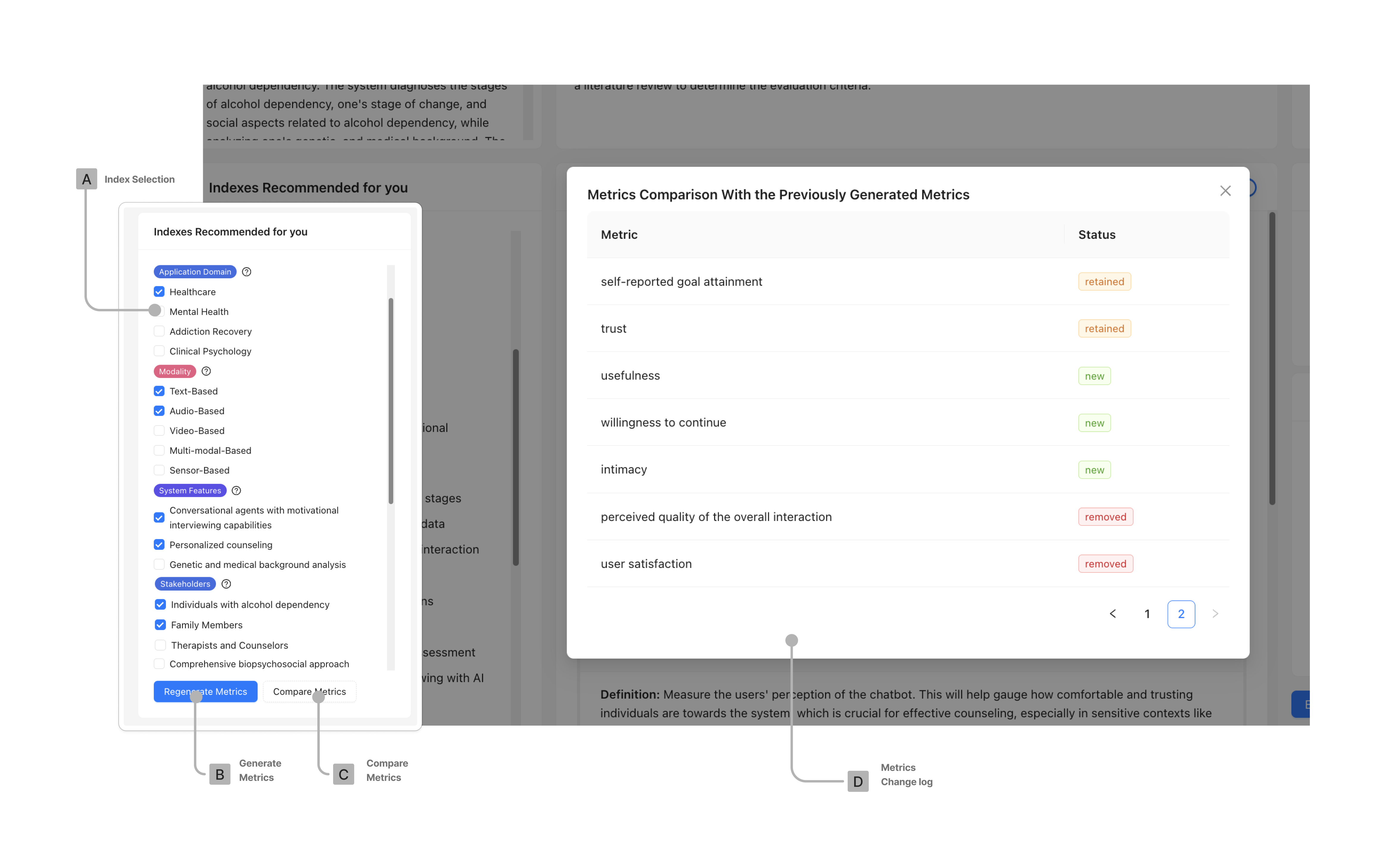} 
        \caption{Project Ideation Panel}
        \label{journey2}
    \end{minipage}%
    \hfill
    \begin{minipage}[b]{0.35\textwidth}
        \small \textbf{User Journey 2:} \\
         Sarah views a list of indexed recommendations from \systemName, noting that some are selected while others are not. She realizes that "Addiction Recovery" isn't checked under "Application Domain," so she selects it (A). She also checks "Family Members" as the evaluation target stakeholder, which she had previously overlooked. After making these changes, she clicks "Regenerate" (B) followed by "Compare Metrics" (C). This process updates her list, showing added, retained, and removed metrics (D), sparking her curiosity about the new metrics.
         \\
    \end{minipage}
    \Description{The graph shows a snapshot of the EvalignUX Project IDEATION Panel and a user journey (right). On the left, the IDEATION Panel displays indexed recommendations under indexes such as “Application Domain”, with options like “Addiction Recovery” and “Mental Health”. Users can select or deselect relevant indexes, for example, checking “Addiction Recovery” (marked as 'A') to refine the system evaluation. Below, buttons for “Regenerate Metrics”' and “Compare Metrics” allow the user to update and compare evaluation metrics. The Metrics comparison window (center) shows the changes of metrics recommended by the system before and after users change the indexes. On the right, the text outlines Sarah's journey as she reviews indexed recommendations from EvalignUX. Noticing that “Addiction Recovery” is unchecked under “Application Domain”, she selects it (A). She also selects “Family Members” as an evaluation target, which she had previously missed. After making these changes, Sarah clicks “Regenerate” (B) and “Compare Metrics” (C), which generates a list of newly added, retained, or removed metrics (D), piquing her curiosity about the updated evaluation metrics.}
\end{figure*}

\subsubsection{Metric Explorer Panel (\textbf{DG1, DG2})} 
\begin{figure*}[h!]
    \centering
    \begin{minipage}[b]{0.65\textwidth}
        \centering
    \includegraphics[width=\textwidth]{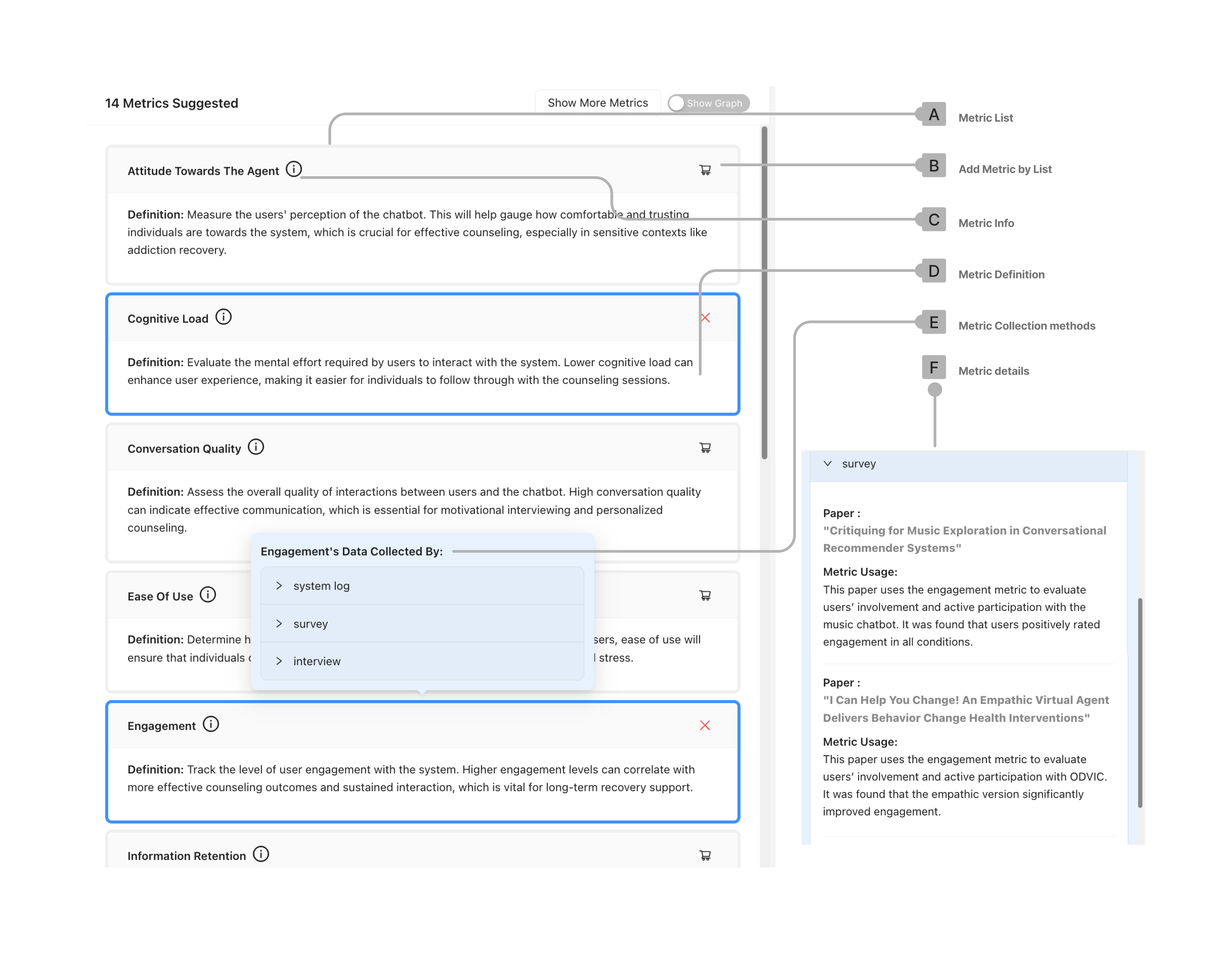} 
        \caption{Metrics Explorer Panel}
        \label{journey3}
    \end{minipage}%
    \hfill
    \begin{minipage}[b]{0.35\textwidth}
    \small \textbf{User Journey 3}: \\ Sarah starts by reviewing the suggested metrics in the list (A) and is intrigued by the "goal attainment", defined as assessing users' perceptions of achieving their recovery goals through the system (D). She clicks the metric info button (C) to explore how this metric is measured—through surveys, system logs (E), and interviews—and reviews related papers on each method's application and findings (F). Interested, she adds it to her shopping cart (B) and uses the Metric Graph feature to explore additional metrics.
    \end{minipage}
    \Description{The graph shows a snapshot of the Metrics Explorer Panel (left) and a user journey (right). The metric list (A) allows users to browse through various metrics, and by clicking the metric info button (C), users can access detailed information about each metric, including definitions (D), collection methods (E), and additional research papers or findings (F). Metrics can also be added to a “shopping cart” (B) for later use. On the right, the user journey describes Sarah's experience. She begins by reviewing the suggested metrics in the list (A) and is particularly interested in “goal attainment”, which is defined as assessing users' perceptions of achieving their recovery goals through the system (D). She clicks the metric info button (C) to explore how this metric is measured, discovering data collection methods like surveys, system logs (E), and interviews. Sarah then reviews related research papers that provide insights into each method's application and findings (F). After gaining an understanding, she adds the metric to her shopping cart (B) and uses the Metric Graph feature to explore additional metrics.}
\end{figure*}

This panel recommends relevant evaluation metrics and related literature while enabling users to explore additional metrics through both a list-view and graph-view interactions, helping them formulate a comprehensive evaluation plan. See Figure \ref{journey3} for \textbf{User Journey 3}.

Prior literature has highlighted that users are often excluded from the process of defining evaluation criteria for a system, with most participants consulted only at the end of the design process to assess the system's effects, rather than being actively involved in setting these criteria. This exclusion contradicts the goal of democratizing the design process, which seeks to empower all stakeholders to contribute meaningfully throughout the design life-cycle \cite{bossen2016evaluation, zheng2023begin}. Inspired by this challenge, we propose a metric recommendation feature that integrates UX evaluation metrics into design materials as inputs for design or evaluation. Given that graph-based interaction is a commonly used and effective approach for recommendation systems \cite{ali2020graph}, we propose incorporating a graph-view alongside the traditional list-view. This graph-view will allow users to visualize UX metrics and their related metrics, fostering inspiration and enabling more informed filtering and selection of metrics. Furthermore, prior literature outlines various methods for collecting metrics, such as system logs, surveys, and interviews \cite{zheng2022ux}. To enhance user understanding, we will provide context and examples of how these metrics have been derived and applied.

\subsubsection{Outcomes and Risks Panel (\textbf{DG2, DG3})} Assists users in framing their research contributions based on previous studies and identifying potential pitfalls by referencing databases of prior AI risks and incidents. Technologies have positively transformed society, but they have also led to undesirable consequences not anticipated at the time of design or development. Previous HCI research has explored using insights from past undesirable consequences to help researchers and practitioners gain awareness of potential adverse effects and reflect on their own experiences with technology \cite{pang2024blip}. Inspired by this, we propose a panel for generating related research outcomes and risks based on existing research and real-world cases. We sourced case studies from a database indexing real AI incidents documented by journalists — the AI, Algorithmic, and Automation Incident and Controversy (AIAAIC) repository, which previous papers have used  \cite{lee2024deepfakes}. This approach aims to strengthen UX evaluation metrics by incorporating literature-based research outcomes and relevant risks grounded in real-world incidents.

\begin{quote}
    \textbf{User Journey 4}: As Sarah adds and removes metrics from her shopping cart, she notices that the results in the Outcome feature update dynamically. She selects outcomes from prior studies that catch her interest. Then, she explores potential risks associated with her proposed UX evaluation plan by clicking on the Risk feature. \textit{“All set!”} said Sarah. She then exports the data and conducts a final review of her generated UX evaluation plan and UX outcomes report. 
\end{quote}

\subsection{System Backend and Implementation}

In this section, we describe the process of preparing the data repository, system backend, and implementation, as well as the techniques used for prompting and making recommendations. The frontend of the system is implemented using ReactJS\footnote{https://github.com/facebook/react}, TypeScript and CSS. The backend uses Python with FastAPI\footnote{https://github.com/tiangolo/fastapi/} as the RESTful API server framework. The data is represented as a Knowledge Graph using the Neo4j\footnote{https://github.com/neo4j} Graph Database to model the complex relations among research papers \cite{verma2023scholarly, kejriwal2022knowledge, wang2018acekg}. We use the GPT-4o model as our LLM engine and set the temperature to 0.7 for all components. The embeddings in the system are generated using text-embedding-3-small, with FAISS\footnote{https://github.com/facebookresearch/faiss}   as the vector database for data retrieval. Additionally, Langchain\footnote{https://github.com/langchain-ai/langchain} is used for prompting the GPT-4o model.

\subsubsection{Data Preparation}
In this section, we present how the scientific database was prepared. First, we established an \textbf{initial repository of evaluation metrics} based on the definitions specified in prior literature (e.g., \cite{zheng2022ux}). Soon, we recognized that the literature review focused on conversational AI interactions; thus two authors expanded the UX evaluation metrics to include interactions in other modalities by incorporating additional metrics that were widely applied beyond conversational interactions \cite{inan2021method, vermeeren2010user}. For example, we expanded the repository by incorporating metrics related to ethics and risks, such as "communication fairness," "user control," and "privacy concerns," which address broader dimensions of responsible interaction design. We do not claim that the repository is complete or exhaustive. It will continue to grow as new HCI literature and UX evaluation metrics emerge. Definitions for all metrics, including the newly added ones, were either directly sourced from these surveys or developed collaboratively by the authors to ensure completeness and coherence.

After building the repository of evaluation metrics, we created a \textbf{paper database} from the same survey papers and utilized GPT-4o \cite{openai2024gpt4technicalreport} to generate more fine-grained information associated with the evaluation metrics. We prompted GPT with the full text of each paper and asked it to provide a narrative of the system described in the papers, along with the challenges or the research gaps that the papers addressed. We provided index definitions and asked GPT to generate a list of ten descriptive indices based on the content of the papers. These descriptive indices include: interaction paradigms, application domain, modality, system features, design novelty, design methods, human-AI relationship types, stakeholders involved, social scale, and theoretical frameworks used.  The definitions for these indices can be found in Appendix \ref{appendix-index}. 

Between the metrics and the papers in the database, we also created usage information for each metric in the relevant papers. For example, a metric usage is formatted as follows: \textit{"This paper uses the [X] metric to evaluate users' [Y] towards [Z technology]. It was found that [S findings related to Y]."} \textit{[X]} represents the metric identified in the paper, \textit{[Y]} is the aspect of user interaction being evaluated, \textit{[Z]} is the technology or system being evaluated, and \textit{[S]} is the findings related to the user interaction being evaluated.

We then crawled metadata and appended it to each paper. The metadata includes the paper's UID, title, authors, abstract, publication date, UIDs of papers that cited it, keywords, publication venue, publisher, and authors' affiliations. Besides, we also leverage the existing AI Incident Database as a repository for risks \cite{mcgregor2020preventingrepeatedrealworld} and extracted relevant incidents and risks.
\systemName's backend recommends UX metrics based on an initial \textit{System Description}. After a series of user interactions, it generates a new Evaluation Plan and UX Outcome using an LLM and data from research papers via a Knowledge Graph, which represents complex entity relations derived from source documents.

\begin{figure*}[h!]
    \centering
    \includegraphics[width=1\linewidth]{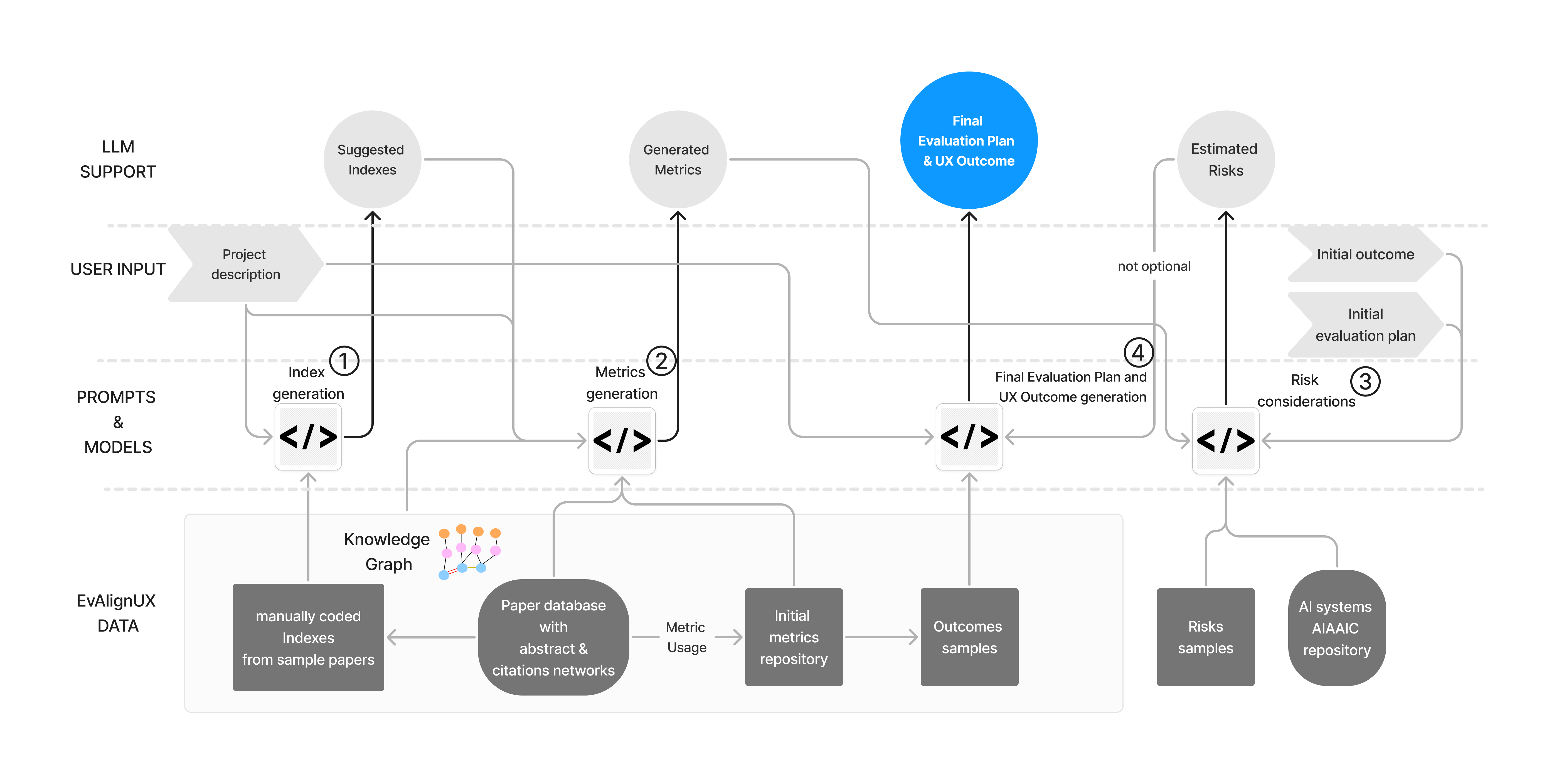}
    \caption{In \textbf{\systemName}, the user provides a \textbf{project description}, an \textbf{initial evaluation plan}, and an \textbf{initial outcome}. 
    {\large \textcircled{\small 1}} \systemName uses the \textbf{project description} to generate a list of \textbf{indexes} using the index generation prompt and a manually coded index; 
    {\large \textcircled{\small 2}} Based on either system-suggested or user-selected \textbf{indexes} and the initial \textbf{project description}, metrics are extracted from the \textbf{Knowledge Graph} and filtered through a metrics generation prompt to recommend \textbf{UX metrics}; From the user-selected \textbf{metrics}, a group of potential \textbf{outcomes} appears using information from the \textbf{Knowledge Graph}; 
    {\large \textcircled{\small 3}} Additionally, based on the user's \textbf{project description}, \textbf{initial evaluation plan}, \textbf{initial outcome}, and information from \textbf{risk samples and AI Risk repository}, several \textbf{risks} are displayed; 
   {\large \textcircled{\small 4}} A \textbf{UX evaluation plan and UX Outcome} are generated based on the selected \textbf{UX Metrics}, \textbf{project description}, \textbf{initial evaluation plan}, \textbf{initial outcome}, and \textbf{risks}.}
    \label{backend}
    \Description{This figure illustrates the workflow of EvalignUX, a system that supports HCI researchers in developing UX evaluation plans by integrating LLM support, structured data, and user inputs. Users provide a project description, initial evaluation plan, and initial outcome, which the system processes to generate relevant indexes and UX metrics using a Knowledge Graph. These metrics help define potential UX outcomes, while risk considerations are assessed using data from risk samples and an AI Risk repository. Finally, the system synthesizes the selected metrics, risks, and user inputs to generate a structured UX evaluation plan and outcome, enabling researchers to explore diverse evaluation approaches while considering impact and potential risks.}
\end{figure*} 

\subsubsection{{Index Generation and derivation of Knowledge Graph}}
\label{sec:index-generation}
Our system, \systemName, generates a list of indexes using the initial system description provided by the user. The indexes are intended to standardize the system description into fixed categories. Indexes are generated by providing the System Description, examples of other generated indexes, and the corresponding definitions of each index as input to the LLM. Subsequently, the system recommends three additional index values for each category. These indexes and the initial system description are used to retrieve a list of metrics from our Knowledge Graph. 

The Knowledge Graph, built using the Neo4j Graph Database, serves as the foundation for recommending metrics. We structured our data as a graph to capture the intricate relationships within our dataset \cite{verma2023scholarly, kejriwal2022knowledge, wang2018acekg}, enabling efficient information extraction and retrieval of global contextual information \cite{edge2024local}. The backing dataset of research papers is organized into distinct clusters of communities \cite{fortunato2010community} based on their index similarity, citations, and common metrics, using the Louvain Community detection algorithm \cite{de2011generalized}. The graph consists of three different types of nodes, each characterized by distinct properties. Table \ref{tab:nodes_attributes} in the Appendix provides a detailed overview of the distinct nodes and their associated properties. The three unique nodes are as follows:
\begin{itemize}
    \item \textbf{Paper Node:} This node contains details about each unique research paper.
    \item \textbf{Metric Node:} The node includes attributes specific to a UX metric. This node is attached to the Paper and the Outcome node via several edges. The node is shared between two paper nodes if they measure the metric.
    \item \textbf{Outcome Node:} It is attached to the Paper Node and the Outcome Node via several edges. The node is shared between two Paper Nodes if they measure the same metric. Since the metric node could be shared between multiple paper nodes, this node stores the unique information of a metric corresponding to each metric as used in a research paper.
\end{itemize}

All nodes are unique and do not duplicate. Furthermore, weighted edges exist between nodes. For instance, there exist three edges between each pair of paper nodes, namely, \textnormal{[SHARED\_METRIC]}, \textnormal{[CITES]}, and \textnormal{[CITED\_BY]}. The \textnormal{[SHARED\_METRIC]} edge exists between two paper nodes if they share at least one common metric. Similarly, if any paper A cites another paper B, there exists a \textnormal{[CITES]} edge from paper node A to paper node B, and a \textnormal{[CITED\_BY]} edge from paper node B to paper node A. Edges of type \textnormal{[CITES]} and \textnormal{[CITED\_BY]} are assigned relatively higher weight, compared to \textnormal{[SHARED\_METRIC]} edge. This is achieved by multiplying the edge weight with a constant value, due to stronger relation among the papers because of the citation relation, based on an increase in the \textit{modularity score} as discussed in \cite{blondel2008fast}. The weight assigned to these edges is influenced by the relative importance of each index category in evaluating the similarity between research papers. Index categories that indicate greater similarity receive higher weights. These weights are then multiplied by the similarity score between index values of the categories for each paper, contributing to the overall edge weight. Additionally, each Paper node is connected to a Metric node via a non-weighted edge of type \textnormal{[HAS\_METRIC]}, which is then connected to an Outcome node via another non-weighted edge of type \textnormal{[HAS\_OUTCOME]}.

\subsubsection{{Metrics Recommendation}}
The provided system description and its corresponding generated indexes are used to extract metrics from the Knowledge Graph. The system description and indexes are vectorized, and the most similar paper is determined using \textit{cosine similarity} on the existing Knowledge Graph. The community for the most similar paper is determined from the aforementioned Knowledge Graph and consequently, all metrics linked to the paper nodes belonging to that community are extracted. These metrics, along with the user system description and generated indexes, are used to further filter metrics, by prompting an LLM. Additionally, each metric is further classified into three metric collection methods, based on its usage in previous research papers, using the Knowledge Graph. These metrics are displayed in a list view and a graphical view to the user.

\subsubsection{{Potential risks and metric outcomes}}
Each metric in the Knowledge Graph is associated with outcomes corresponding to the research paper. Subsequently, these results are retrieved from the Knowledge Graph based on the final metrics list and displayed to the user so that they can filter them based on relevance to the user's system description. Each outcome is specific to a research paper and is mapped to all metrics associated with that paper. Consequently, these outcomes are further considered when generating the Potential Evaluation Plan and the Potential UX Outcome.

In addition to the recommended metric outcomes, our system generates potential risks associated with the user system using the AIAAIC repository. We extract the top three incidents that have a \textit{Euclidean distance} of less than 0.5 by performing a vector-based similarity search, using \textit{FAISS} between the user system description and the AI system description present in the AI Incident dataset. Subsequently, we extract risks associated with these AI systems, and provide them along with the user system description to a LLM to filter the most relevant risks for the system. The identified risks are contextualized based on the initial system description, and corresponding URLs for related incidents are also provided to the user.

Once the metrics have been finalized by the user, along with the corresponding metric outcomes, a Potential Evaluation Plan and Potential UX Outcome are recommended to the user. The Potential Evaluation Plan includes detailed instructions on the usage of each metric and the methods for measuring them based on the user's system description and initial evaluation plan. The potential evaluation plan is generated by providing the user's system description, a list of selected metrics, the specified metric outcomes, and the methods previously used to evaluate the UX metrics to an LLM, and then prompting the LLM to generate the plan. The Potential UX Outcomes are generated by considering the list of selected metrics, the desired results from the metric outcomes, the initial System Description, and the risks associated with the system.

\subsection{Technical Evaluation}
\begin{figure*}
    \centering
    \includegraphics[width=0.9\linewidth]{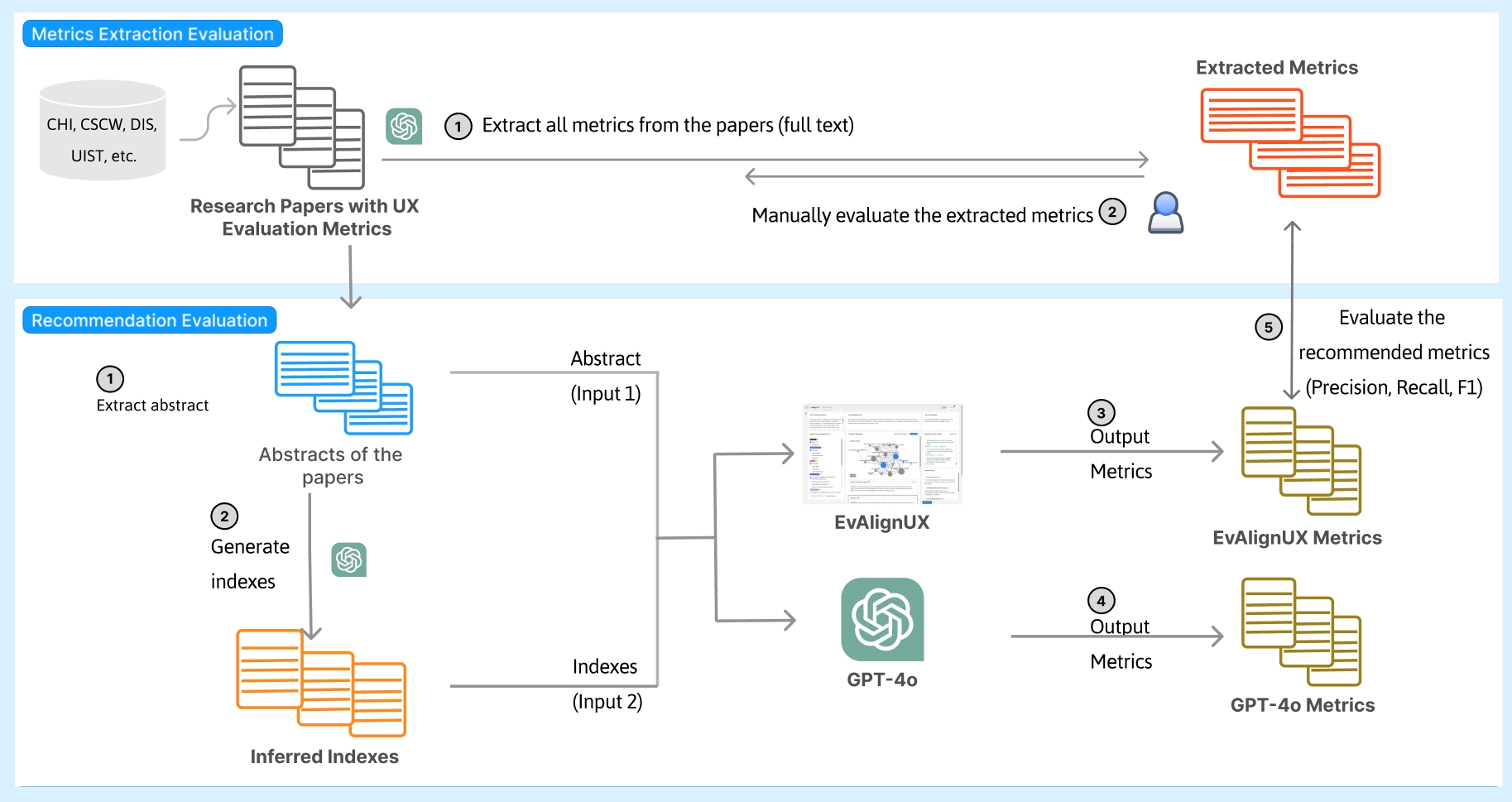}
    \caption{An overview of the workflow for evaluating \textit{Metrics Extraction} (illustrated in the upper box, details provided in Section 3.3.1) and \textit{Metrics Recommendation }(illustrated in the lower box, details provided in Section 3.3.2). For example, the metrics extracted from the full text and manually validated by researchers served as the ground truth. To evaluate the performance of the metrics \textit{Recommendation},  we compared the metrics recommended by GPT-4o (baseline) with those recommended by our  \systemName tool, using abstracts from sample papers published in HCI conferences such as CHI, CSCW, DIS, and UIST. The evaluation was based on precision, recall, and F1 scores calculated using the ground truth metrics.}
    \label{fig:evaluation-workflow}
    \Description{The graph shows a two-stage evaluation workflow: the top section illustrates the process of extracting UX evaluation metrics from full-text research papers from conferences like CHI, CSCW, DIS, and UIST are analyzed, metrics are automatically identified, and then human experts validate these extracted metrics to establish a ground truth set; the bottom section demonstrates how these ground truth metrics are used to evaluate recommendation tools, specifically comparing metrics suggested by a baseline model (GPT-4o) and our system (EvalignUX) using only paper abstracts and inferred indexes as inputs, with both recommendations assessed against the ground truth metrics using precision, recall, and F1 scores.}
\end{figure*}

We conducted a small-scale evaluation to compare \systemName's understanding of evaluation metrics with human understanding and to assess the quality of its metric suggestions based on a given narrative. \Cref{fig:evaluation-workflow} illustrates the process used to extract UX metrics applied in HCI publications and the method used to evaluate the accuracy of metric recommendations with \systemName.

\subsubsection{Evaluating Metrics Extraction}
To assess GPT's capability in extracting evaluation metrics in research papers from a predefined list of candidate metrics, we adopted a similar strategy as in \cite{pang2024blip} and randomly selected 15 papers from our initial pool of research articles. The evaluation metrics used in these papers were extracted using our GPT prompt. One author reviewed each identified metric and rated it from three aspects: whether the metric was actually measured, whether the metric was in the candidate list, and whether the reasoning or citation provided makes sense. A second author verified these assessments. A metric was considered accurate if it fulfilled the above requirements. 

We found that out of the 102 metrics identified in the 15 papers, 94\% (96/102) were indeed measured, 92\% (94/102) were in the candidate list, and 88\% were accurate (90/102), noting that two metrics overlapped between the first and second criteria. 100\% (102/102) of the reasoning and citations made sense. Overall, the results suggest that with our prompt, GPT's ability to identify metrics is largely reliable. 

A common source of hallucination is that the metric extracted was not actually measured in the evaluation section of the paper (i.e., hallucination \cite{ji2023survey}). More specifically, there were three types of hallucination: a) GPT inferred metrics based on descriptions from the paper, but these metrics were not explicitly measured in the evaluation process; b) GPT synthesized metrics from the results, but these metrics were not explicitly measured in the evaluation process; and c) GPT made recommendations for a metric that should have been used but was not actually used, based on the context of the paper. The case for (a) and (b) was often that GPT often selected a statement or conclusion from a qualitative result described in the paper. However, such qualitative results were synthesized from interviews or surveys, and therefore the identified metric was not explicitly measured in the paper. For example, GPT identified "attitude towards the agent" as an evaluation metric from \cite{wambsganss2020conversational}, quoting \textit{"students made positive comments about the interaction with CAs related to the perceived level of enjoyment and interaction."} While attitude towards the agent was not explicitly measured in the paper, it was described in the findings and was thus misidentified. For (c), an example is GPT identifying "privacy concern" as a metric from \cite{lee2021exploring}, quoting \textit{"because chatbot use can result in users' self-disclosure of sensitive topics"} from the discussion section. Although it makes sense to measure "privacy concern" based on the context of potential self-disclosure of sensitive information, this metric was not explicitly measured in the paper, so it was marked as inaccurate.

It was possible that the extracted metric was not in the list of candidate metrics. There were two reasons: (a) GPT reworded the metric in the candidate list while preserving its meaning, and (b) GPT invented a metric that was not included in the candidate list. For (a), one example is GPT identifying "utterance length" from \cite{thomas2020expressions}, while the candidate list actually uses "length of utterance". Regarding (b), GPT identified "transparency" as a metric from \cite{pecune2019model}, even though "transparency" is not part of the candidate list.

Additionally, we observed two scenarios in which certain metrics were presented in the papers but not extracted by the model. First, as noted in \cite{wang2020alexa}, GPT struggled to accurately extract metrics presented in tables or figures and omitted metrics such as "fear of being judged" and "usefulness." Second, GPT omitted metrics that were described as 'measured' in the original text but were not explicitly reported in the final results. For example, in \cite{kim2019comparing}, "usage motivation" was "measured" in the questionnaires but not reported. Similarly, in \cite{wambsganss2020conversational}, "technology acceptance" was measured but not reported; instead, "intention to use a conversational agent" was reported. 

\subsubsection{Evaluating Metrics Recommendation}  
We evaluated the performance of the metrics recommendation using a sample of 40 papers, including papers from CHI, CSCW, DIS, UIST, etc., in emerging technologies such as AI, wearable technology, VR/AR, conversational agents, and similar fields. The metrics for these papers were extracted using GPT-4o by providing the complete text of the papers to the LLM engine and they were manually verified by two authors to serve as the ground truth.

Our system is supported by GPT-4o, and we also compared our system against the same LLM engine to recommend UX metrics using precision, recall, and F1 score \cite{tamm2021quality, fayyaz2020recommendation}. The baseline GPT-4o and our system were given the same set of inputs, including the system description and the corresponding list of indexes. Unlike metrics extraction, where the entire text of the paper was provided, the evaluation against both \systemName and the baseline LLM (GPT-4o) was conducted using only the system description as the input. This approach reflects the likelihood that \systemName users would typically provide only a brief description of their system. Given the system description, a corresponding list of indexes was inferred using an LLM for each paper, as described in \cref{sec:index-generation}. We then compared \systemName and the baseline GPT-4o based on the number of metrics they recommended that were actually used in the papers, as determined by referring to the ground truth. For instance, Wan et al. \cite{wan2024building} used ``response quality'', ``conversation quality'', ``average response time'', and ``agent behavior'' to measure users' satisfaction with NPC responses, evaluate the coherence and contextual relevance of NPC dialogues, determine the response latency, and assess the emotional authenticity of NPC interactions, respectively. Given the system description and identified indexes for this work, we assessed \systemName's and the baseline's ability to recommend relevant and coherent UX metrics. 

For implementation, we sampled each input three times using our system and the baseline to generate a list of metrics. The results obtained from both systems were generalized to reflect the naming standards used in the test dataset. To perform the final comparison, we calculated the precision, recall, and F1 score for each subsequent run, and the final metric scores obtained were averaged to obtain the overall mean scores.

\begin{table}[b!]
\centering
\small
\begin{tabular}{c|c|c|c}
\hline
\textbf{Model} & \textbf{Mean Precision} & \textbf{Mean Recall} & \textbf{Mean F-1 Score} \\ \hline
GPT-4o & 0.096 & 0.203 & 0.121 \\ \hline
\systemName & 0.156 &  0.304 & 0.195\\ \hline
\end{tabular}
\caption{Comparison for metric recommendation between GPT-4o and \systemName}
\label{table:3x4_example}
\end{table}

Table 2 demonstrates that our system significantly outperforms the baseline GPT-4o, with a 6\% increase in precision, a 10.1\% increase in recall, and a 7.4\% increase in F1 score. We further conducted a pairwise t-test by processing each sample 10 times on both systems, to compare the metric mean between \systemName and the baseline ($t = 8.32, p = 0.014^{*}$). The t-test demonstrates that there is a statistically significant difference between the performance (precision, recall, and F1 score) of the two models on the same dataset. Additionally, we observed that our system recommends more relevant UX metrics for a given system description than the baseline GPT-4o. For instance, when evaluating Tan et al.'s work \cite{tan2024audioxtend}, our system recommends using \textit{Perceived Naturalness} to measure how naturally the AI-generated visuals integrate with the audiobook narrative.

However, when evaluating the baseline GPT-4o, none of the generated metrics explicitly address the naturalness of the integration between visuals and audio. Furthermore, when testing the system on Wang et al. \cite{wang2024virtuwander}, our system recommends evaluating \textit{Emotion Awareness} to monitor the system's ability to recognize and appropriately respond to users' different emotional states, which is also measured in the research. In contrast, none of the metrics recommended by the baseline model address the evaluation of the varied emotional states of the user. Additionally, when using the approach from \cite{sun2024lgtm}, the system recommends measuring \textit{Perceived Intelligence} to assess how users perceive the system's ability to generate coherent motions, which the baseline model does not explicitly recommend. The results indicate that our system outperforms the baseline GPT-4o in recommending UX metrics to users.

\section{User Study}
In this section, we outline how we evaluate \systemName's user experience and explore the following key research questions.

\textbf{RQ1}: \textit{How effectively does the use of \systemName impact researchers when developing their UX evaluation plans?} 

\textbf{RQ2}: \textit{How do users leverage the \systemName features to enhance their thought processes when developing UX evaluation plans?}

\textbf{RQ3}:  \textit{How do users' backgrounds impact their use of \systemName?}

\subsection{Participant}
Our study was approved by the university's IRB. We recruited 19 participants from social media. All participants reported that they had extensive experience with HCI/UX research: 11 participants (57.9\%) had 3–4 years of experience, 7 participants (36.8\%) had 1–2 years of experience, and only one participant had less than one year of UX experience. Regarding participants' use of AI in research studies, the majority used AI to help with simple tasks, such as coding and writing (17, 89.5\%), literature review (7, 36.8\%), and data analysis (11, 57.9\%). Notably, we found that few participants used AI for tasks that require rigor and creativity, including research ideation (7, 36.8\%), design prototyping (7, 36.8\%), research evaluation plan development (3, 15.8\%), and usability testing (1, 5.3\%). The background information of participants is shown in Appendix \ref{participants_info}. Each user study lasted 1.5 to 2 hours. Participants received 20 USD per hour as compensation for the study.

\subsection{Procedure}
\begin{figure*}
    \centering
    \includegraphics[width=1\linewidth]{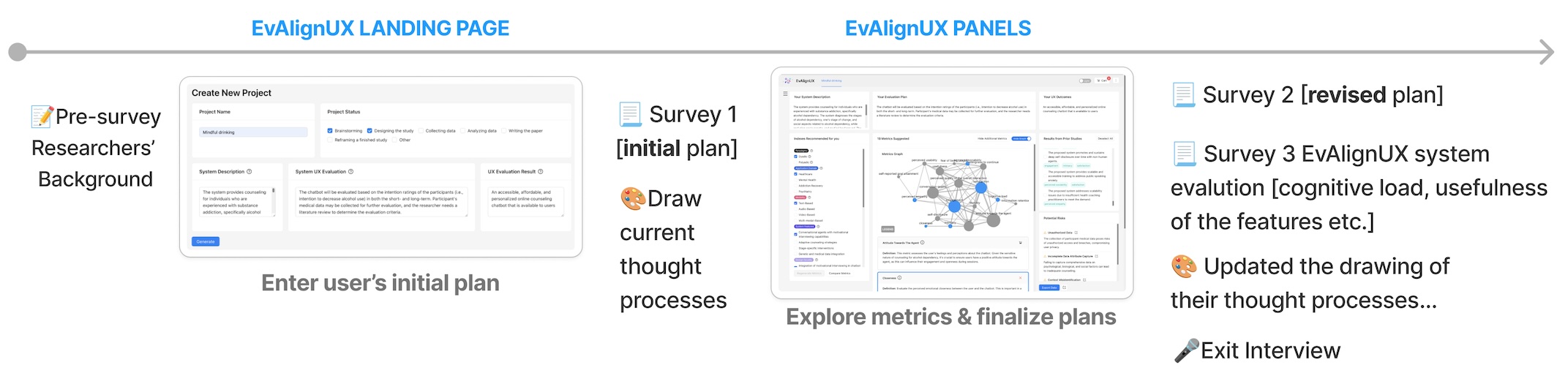}
    \caption{The Study Procedure. Participants took four surveys at different times during the study. They were asked to provide initial project ideas to \systemName and explore its three panels to select metrics and finalize their evaluation plans. They were also asked to draw their thought-processes before and after using the EvAlignUX panels.}
    \label{fig:procedure}
    \Description{This graph illustrates the study procedure for our user study. At the top, a timeline outlines the steps participants followed during the study. The process begins with participants completing a pre-survey to provide their background information. Next, they enter their initial project plan into EvalignUX. After submitting their plan, participants complete Survey 1, which captures their initial perceptions and requires them to draw their current research thought processes without the aid of AI. Following this, researchers present and explain key features of EvalignUX, allowing participants to explore metrics and finalize their plans. Participants then complete Survey 2, documenting their revised plan, followed by Survey 3, which evaluates the EvalignUX system’s features. They are also asked to update their thought process diagram using AI. The procedure concludes with an interview.}
\end{figure*}

Figure \ref{fig:procedure} demonstrates the study procedure. \ding{172} The study began with collecting background information, including participants' UX experience levels. Researchers then introduced the study context and explained the tasks. Then, they used \systemName's landing page to spend about 10 minutes entering their project descriptions, evaluation plans, and expected outcomes. After that, they completed an initial survey that captured their perceptions of the proposed project plans (survey 1, initial plan). Then, participants used the Zoom Whiteboard to draw their thought processes for developing the UX evaluation plan and engaged in a think-aloud to ensure their ideas were communicated clearly. 

In \systemName, participants were guided through the system features and asked to refine their project plans using all available features. After experimenting with the features and achieving satisfactory results, participants used the "Export Data" button to review, regenerate, and re-edit their project plans. Upon finalizing their plans, they completed exit surveys (survey 2, revised plan; survey 3, \systemName) to evaluate their perceptions of the generated plans and their experience with the system. During the \systemName process, the think-aloud method was employed to observe and record participants' interactions and thoughts. Next, participants used Zoom Whiteboard to revise their thought processes for developing the UX evaluation plan if they felt changes were needed. Finally, participants were interviewed to delve deeper into their overall experiences and gather additional insights. This step provided a comprehensive understanding of user perceptions and allowed the collection of feedback on both the proposed plans and the system features.

\subsection{Measures}
We recorded the following data to assess participants’ performance and perceptions during the study: task logs, thought process diagrams, think-aloud data, Likert-scale questionnaire responses, and interview transcripts.

\subsubsection{System Logs} For system log data, we first collected user's input to the system, including the project description (i.e., \textit{What are the main objectives of your project?}), UX evaluation plans (i.e. \textit{How do you plan to evaluate the UX of your project?}), and expected outcomes (i.e. \textit{What outcomes do you expect from this evaluation?}) as well as any updated ones. Additionally, we logged their usage behavior data, including (1) the number of times a user regenerated indexes, (2) the total number of metrics selected, (3) the number of metrics added through the graph, (4) the number of outcomes clicked including both selected and deselected outcomes, and (5) the number of risks clicked. We also logged their revised system description, UX evaluation plans, and expected outcomes. 

\subsubsection{Thought Process Diagrams} We collected users' thought process visualizations for developing a UX evaluation plan using Zoom Whiteboard before and after using the \systemName to see how AI-based assistance influenced their thinking.

\subsubsection{Questionnaire} \label{questionnaire}
The questionnaire is divided into three main parts:
\begin{itemize}
    \item Background information: Users answered questions related their past UX experiences, AI literacy, and AI usage for research purposes. 
    \item Perceptions of initial and revised UX study plan (survey 1 and 2): We collected user perceptions before and after they explored \systemName to determine if there were significant differences between their perceptions of their UX research plans. 
    The survey items are adapted from main schemes for analyzing UX evaluation methods \cite{vermeeren2010user, venable2012comprehensive, dean2006identifying}, including novelty, relevance, efficiency, completeness, clarity, concreteness, quality, feasibility, rigorous, risks considerations, and confidence. 
    \item System feature perceptions:  For each metric, we asked participants to rate their perceptions of each feature across 8 Likert-scale items (survey 3): enjoyment of use \cite{davidson2023development}, intention to use it in the future \cite{everard2005presentation}, perceived usefulness \cite{davis1989perceived},  trust in the suggestions and decisions provided \cite{everard2005presentation}, perceived cognitive load in understanding the information \cite{paas2008assessment}, inspiration for exploring and developing UX research \cite{chu2010two}, the system's contributions to the UX evaluation plan, and the system's contributions to the UX outcome.  
\end{itemize}

\subsubsection{Think Aloud}
We recorded participants' think-aloud protocols while they used the system features as well as their thought process diagrams. For example, we asked them to provide the reasoning behind their decision-making process.

\subsubsection{Interview}
Finally, we conducted a short follow-up semi-structured interview. We asked questions in three main categories, supplemented by follow-up questions based on the interviewee's responses. 
\begin{itemize}
    \item Perceptions of features and their role in decision-making for proposing a UX plan: \textit{Which feature is the most impressive? Which feature is the least impressive? How do these features support your decision-making?} 
    \item Overall user experience and potential improvements: \textit{After using the system, how useful do you feel about the system? Were there any features or information that you felt were lacking? How do you view your original and revised plans?} 
    \item Follow-up questions about their research mental model: \textit{What differences do you notice between your previous research process and the process of interacting with \systemName? How do you perceive the strengths and weaknesses of these strategies?} 
\end{itemize}

\subsection{Data Analysis}
We analyze each participant's \ding{172} thought process diagrams, \ding{173} quantitative data from surveys and logs, and \ding{174} user decision-making reasoning through think-aloud protocols, interview transcripts, and UX research plans.

\subsubsection{Thought Process Diagrams Analysis} We standardized the thought process diagrams drawn by participants by coding each illustrated step. Modifications, such as envisioning AI assistance or adding new steps, were identified. These changes were further analyzed by examining the diagrams alongside the think-aloud data. We summarized the benefits participants identified where AI could assist in proposing a UX research plan.

\subsubsection{Quantitative Data Analysis} We analyzed participants’ responses to the two post-task surveys. For all these measures, we conducted a Shapiro-Wilk test \cite{gonzalez2019shapiro} and found that the data were not normally distributed. Given the nature of our data, we applied Kendall's tau correlation to explore relationships between user experience levels, system usage, and perceptions of changes in plans and system features \cite{bolboaca2006pearson, kendall1938new}. We also reported effect sizes using standardized methodologies \cite{dixon1951introduction}. All statistical analyses were conducted using Python 3.12.

\subsubsection{Behavioral Analysis through Think-Aloud and Interviews} A thematic analysis \cite{braun2012thematic} was performed on data from think-aloud sessions and interviews. Two researchers collaboratively identified common themes through an iterative process. This analysis helped map how \systemName features address the challenges stemming from participants' thought processes while also outlining the pros and cons of these features.


\section{Findings}

This section presents findings on how participants' perceptions of their proposed UX evaluation plans changed through the use of \systemName (RQ1) in Section \ref{change}, their feature usages and perceptions (RQ2) in \ref{usage}, and how these perceptions varied based on their UX research background (RQ3) in Section \ref{bg}. Participant quotes are indicated by \textit{quotation marks}, while \texttt{system features} are formatted in code style.

\subsection{\systemName Significantly Improves the UX Evaluation Plans (RQ1)}\label{change} 

In this section, we compared participants' ratings of their original and revised UX plans across 11 measurement criteria. Of these, nine measurements showed significant improvement in the revised plans compared to the original. Five participants fully accepted the system-generated evaluation plan without modification. On average, 60.1\% (SD = 0.33) of the generated UX evaluation plan and 59.6\% (SD = 0.38) of the generated UX outcome were incorporated into their final evaluation plans. The results are shown in Figure \ref{fig:change}, with detailed information available in Appendix Table \ref{tab:combined_results}.

\begin{figure*}[h!]
    \centering
    \includegraphics[width=1\linewidth]{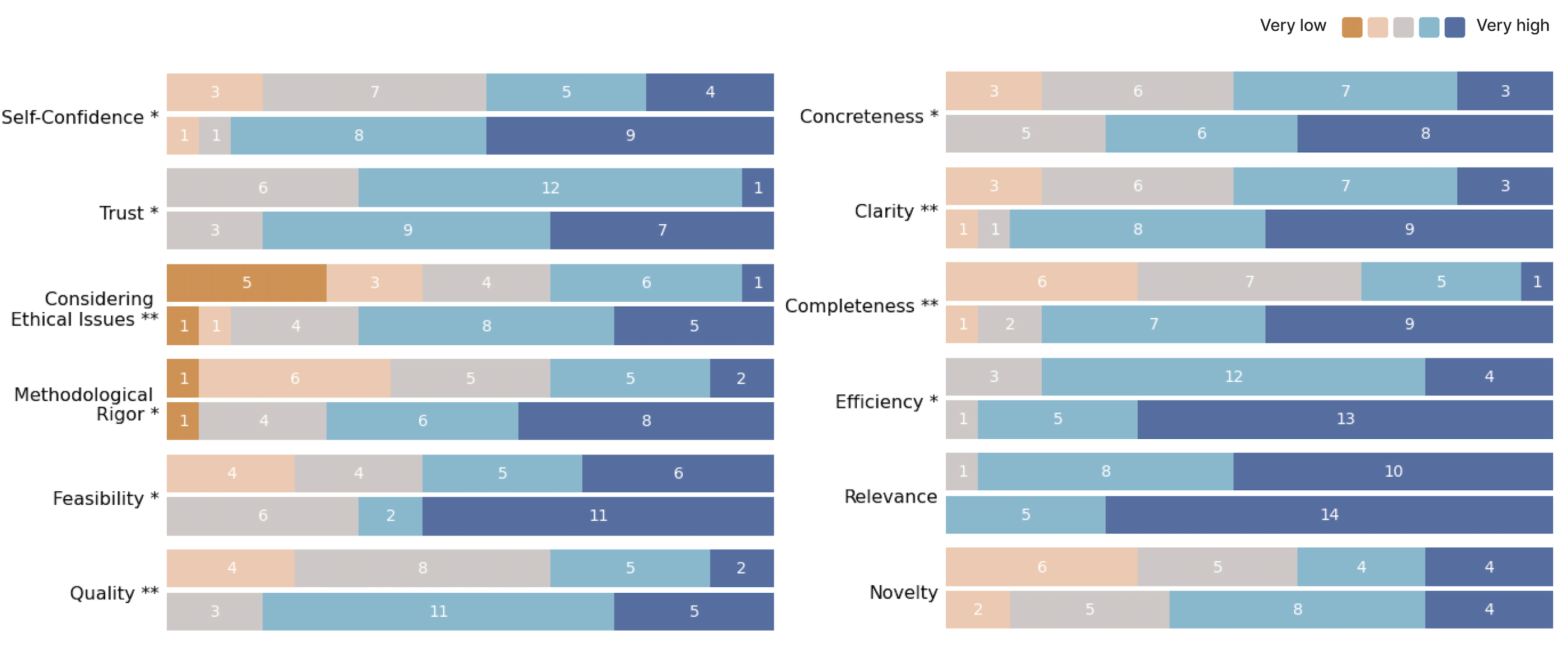}
    \caption{Distribution of participants’ ratings on their proposed UX plan before (top of each item) and after (bottom of each item) using \systemName. Participants felt that the updated plans are more efficient in achieving their research goals, more complete, clear, concrete, high-quality, feasible, rigorous, considerate ethical issues, trustworthy, and enhanced their self-confidence compared to their original proposed plans.(*:p<.05, **:p<.01). The effect size for all significant findings ranges from $r_b$= 0.51 to 0.90. }
    \label{fig:change}
    \Description{The image presents a bar chart comparing participants' ratings on their proposed UX plans before (top bar of each item) and after (bottom bar of each item) using EvalignUX. Each item is evaluated across several dimensions, including “Self-Confidence”, “Trust”, “Considering Ethical Issues”, “Methodological Rigor”, “Feasibility”, “Quality”, “Concreteness”, “Clarity”, “Completeness”, “Efficiency”, “Relevance”, and “Novelty”. The bars are color-coded to represent varying degrees of rating: from very low to very high. Each item shows a noticeable shift in participant ratings, with the ratings of after using EvalignUX generally leaning towards higher values, indicated by the darker shades of blue.}
\end{figure*}

\textbf{Overall, the revised plans were more specific}. The increased scores in Concreteness (W = 24.0, \(p\) = .032$^{*}$, $r_b = 0.74$), Completeness (W = 16.0, \(p\) = .004$^{**}$, $r_b = 0.79$), Rigor (W = 15.0, \(p\) = .009$^{**}$, $r_b = 0.75$), and Quality (W = 11.0, \(p\) = 0.004$^{**}$, $r_b = 0.82$) suggest that participants added more detailed information to their revised plan. Several participants only had a vague idea about which variables to measure (e.g., P7, P14, see Table \ref{tab:changelog}). With \systemName, they noted how it helped them improve the details in the UX plans. For example, P14 later \textit{strategically} specified six measurements by questionnaires and system logs, with each in great detail; and P7 \textit{copied} a holistic UX evaluation plan using a survey, system logs, and interview protocols, each with specific details added.

Participants not only added more details to their initial plans but also perceived them as more \textbf{concise and relevant}, as both Clarity (W = 4.0, \(p\) = 0.005$^{**}$, $r_b = 0.90$), and Efficiency in achieving research goals (W = 12.0, \(p\) = .012$^{*}$, $r_b = 0.74$) also demonstrated a considerable improvement. Participants made meaningful revisions for clarity. For instance, P5 removed text related to assessment from their initial system description, as they found it more suitable as a part of the evaluation plan instead of putting it into the system narration. Also, several participants praised the \texttt{"Metrics list"} feature for clarifying metric relevance, noting it showed \textit{"how this metric can be relevant in the context (P3)"} and \textit{"encouraged the reasoning of using this metric (P17)"}.  P8, during a think-aloud session, explained, \textit{"I just selected ‘information retention' because my project is about memory, and the definition reminds me of the importance of evaluating how well my participants retain information to measure the effectiveness of counseling sessions in my project."}

\begin{table*}
    \centering
    \begin{tabular}{|p{2cm}|p{0.5cm}|p{5cm}|p{7cm}|}
        \hline
        \textbf{Input} & \textbf{ID} & \textbf{Initial Plan} & \textbf{Revised Plan} \\
        \hline
        UX Plan & P7 & To conduct a semi-structured interview with experts to see how they collaborate with our system to analyze the video & Use a survey to measure the trust experts have in the AI’s analysis capabilities. Questions could focus on reliability, perceived competence ... Discuss their perceptions of the assistant’s expertise, any concerns about the analysis, and instances where the assistant’s suggestions were particularly effective or ineffective.\\
         \hline
        UX Plan & P14 & Ask users to experience this tool and other similar ones and then report usability, satisfaction, and others with questionnaires. & 
         Collection of short open-ended responses from participants to analyze qualitatively. Ask pre and post survey questions about their decision confidence (how accurate they feel about their prediction of what the AI will do), perception of how clear the AI visualization was, and how accurate their prediction was. Do these before and after each action.\\
         \hline
         UX Outcome &P13 & Low mental load, higher satisfaction, higher quality of subfield extraction.   & I can expect the tool to provide accurate and useful summaries for both lay people and HCI researchers, leading to high user satisfaction and effective functionality. The interactive visualizations will likely be well-received for their ability to save time and allow quick exploration.    \\
         \hline
         UX Outcome & P11 & It outperforms the baseline techniques in the education effect and subjective preferences. Achieved this results through a more engaging experience, which also lowered participants' understanding difficulty. & (...Same with initial...) There are potential risks alongside the expectations that the research outcome may not align with the intended education effect and the system may not be generalizable. However, these may could be partially reflected in the objective metrics and be acknowledged in the limitations.\\
         \hline
    \end{tabular}
    \caption{Participants' Initial and Final UX Evaluation Plans}
    \label{tab:changelog}
    \Description{Table 2 presents examples of participants’ initial and revised UX evaluation plans after using EvalignUX. Each row corresponds to either a “UX Plan” or “UX Outcome” as the input type from participants (i.e., P7, P14, P13, P11). The “Initial Plan” column summarizes the participant’s original approach to evaluating user experience. The “Revised Plan” column shows how these plans evolved, providing more specific methods with the aid of EvalignUX. Through these examples, Table 2 illustrates how participants refined their UX evaluation strategies to gather richer, more actionable insights.}
\end{table*}

Participants also felt \textbf{more confident} and believed their plans \textbf{were feasible} as Feasibility (W = 16.5, \(p\) = 0.035$^{*}$, $r_b = 0.64$) and Self-Confidence (W = 18.0, \(p\) = .014$^{*}$, $r_b = 0.70$) as indicated by  significant increases. For example, participants felt they could know how to \textit{cite (P12, P15)}, \textit{"measure (P1, P3, P14)"}, \textit{"design questions (P1, P9, P17)"}, \textit{“learn previous findings (P8, P2)”} for metrics using the metric feature, making the UX plan tangible, and even \textit{“how to collect the data specifically"} (P13, see Table \ref{tab:changelog}). They also gained confidence, especially when noticing prior work used the same metric selections. As P3 expressed, \textit{"I'm really happy someone else also uses' improvement in social skills in this context."} The validation further reinforced their self-confidence in the chosen metrics and plans. 

Participants also significantly revised their plans to include more \textbf{risk considerations} (W = 10.0, \(p\) = 0.007$^{**}$, $r_b = 0.81$). For example, building upon their original UX outcomes, P11 expanded, \textit{"There alongside the expectations the research outcome may not align with education effect the system may not be generalizable. However, these issues could be partially reflected in the objective metrics and be acknowledged in the limitations."} P11 took inspiration from concerns about generalizability from the system's output (see Table \ref{tab:changelog}, where "Generalization Failure" was mentioned, stating, \textit{"Thinking about the potential biases made me rethink how I approach user data to ensure that I don't overlook important issues."} 

However, there were no significant changes found in "Novelty" (W = 22.0, \(p\) = 0.083, $r_b = 0.51$) or UX plans' "Relevancy" to research questions. Several participants felt that the system was hard to change. P4 noted, \textit{"The core idea is already set,"} while P6 said, \textit{"It's challenging to introduce major changes."} Participants also reflected that "\textit{It is hard to justify the novelty without referring to the actual paper and verifying it on my own}" (P4).

\subsection{\systemName  Feature Use and Perceived Value (RQ2)} \label{usage}

In this section, we discuss how features of \systemName{} helped overcome various challenges in developing evaluation plans while also streamlining the thought process. In particular, three features, metrics graph, metrics list, and outcomes, were regarded as the most effective.

\begin{figure*}[ht!]
    \centering
    \setlength{\abovecaptionskip}{0pt} 
    \includegraphics[width=1.03\linewidth]{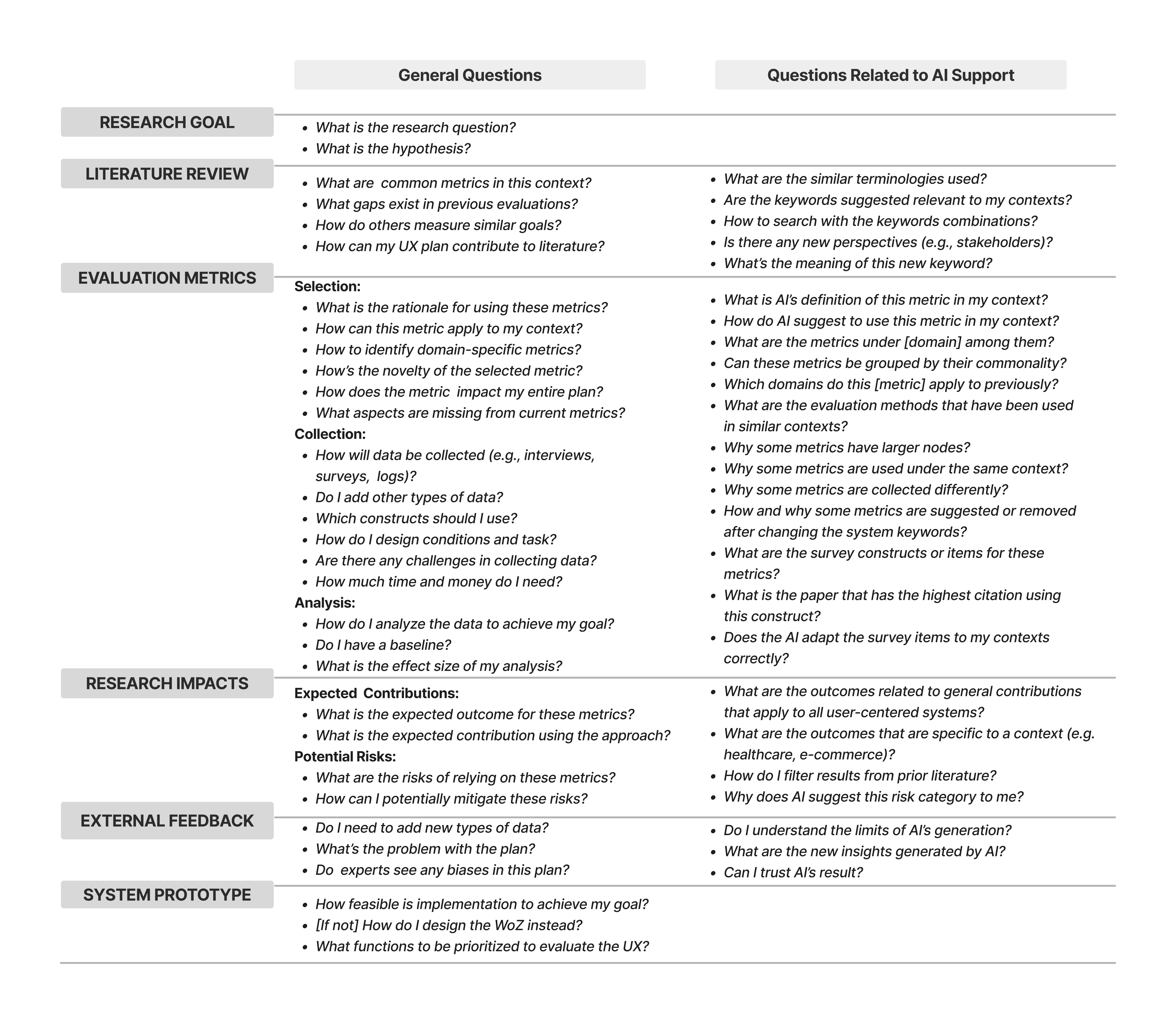}
    \caption{Question Bank for Developing UX Evaluation }
    \label{fig:mental}
    \Description{The image presents a UX Evaluation Development Questions Bank, divided into two main sections: General Questions and Questions Related to AI Support. These sections are further categorized into key areas of UX development: Research Goal, Literature Review, Evaluation Metrics, Research Impacts, External Feedback, and System Prototype. The General Questions section provides foundational queries for traditional UX evaluation, while the Questions Related to AI Support section includes prompts considering AI as the research assistant. The questions cover various stages of UX research and design, guiding users through research planning, evaluation, and the integration of AI-generated insights.}
\end{figure*}

\subsubsection{The Value of \systemName Features in Enhancing UX Evaluation}

We observed how the use of individual features and the combined use of different features in \systemName enhanced the user experience. In Figure \ref{fig:mental}, we summarized a set of prototypical questions that UX researchers asked themselves while developing their UX evaluation plans, annotated from think-aloud data collected during participants' thought processes, which highlighted five perceived benefits of AI: justify, inform, support self-critique, streamline, and vision.

Participants mentioned that AI could be used to \textbf{inform} them by providing knowledge from prior literature, e.g., to quickly \textit{“suggesting relevant literature}” (P15, 18), \textit{“summarizing existing evaluation methods based on the research context”} (P3, P11). AI also supported researchers in understanding appropriate metrics by offering \textit{“definitions”} (P8), \textit{“citations” }(P3), \textit{“how it can be used”} (P16), and information on “constructs adopted” (P17). For example, P19 combined the \textbf{\texttt{Index}} and \texttt{\textbf{Metrics List}} features. P9 initially felt that \systemName did not accurately capture their intentions. They proposed a version control system, but \systemName recommended metrics that were more likely to measure "machine learning efficiency," such as accuracy. However, after modifying a few indexes (e.g., adding two design novelties: graph-based version control, and automated design documentation), they found that \systemName suggested new metrics, such as interaction experience, overall satisfaction, perceived usability, and task success, that were much more accurate and aligned with their research.

Many participants recognized the value of using AI to visualize literature networks, which help them better \textbf{\textit{justify}} metric choices to align with research goals. This capability was particularly appreciated because it facilitates \textit{“metric selection across various venues and disciplines”} (P17), \textit{“highlight connections between metrics”} (P16) to determine whether two metrics are “interrelated or distinct” (P13) and supports decisions based on  \textit{“popularity and relevance in the field”} (P9) in a way that \textit{“aligns with their (my) research goal”} (P6). For example, P9 used \textbf{\texttt{Metrics Graph}} and \textbf{\texttt{Outcome}} together. P9 designed a conversational agent for judges. After reviewing the graph, they found that user satisfaction, accuracy, and task completion were connected, which motivated them to think about the relationship between these three metrics. However, they found that while certain metrics aligned with their research goals, not all findings were applicable to their specialized context. They remarked that, no emotions or diversity should be included as my evaluation outcomes; otherwise, the judges would be easily biased.

Participants recognized the value of using AI to help them better \textbf{\textit{self-critique}} during the development of their UX evaluation plans. This capability was particularly appreciated as it facilitates \textit{“prompt feedback”} (P1) and provides “new perspectives" (P15), acting as a \textit{“teammate"} (P6, P16) or \textit{“partner"} (P18) in scholarly settings by commenting on the \textit{“effectiveness and limitations"} of their proposed plans (P13). One participant also mentioned its potential to support \textit{"overcoming social inhibitions, such as hesitancy in seeking advice or posing questions to peers too frequently"} (P16). As P4 commented that, \textit{"It's nice also to have \systemName where you can throw ideas around, and then it will guide you directly in your research project, what to think about—not just a chat with somebody, which can sometimes be embarrassing."}

\begin{figure*}[t]
    \centering
    \includegraphics[width=0.8\linewidth]{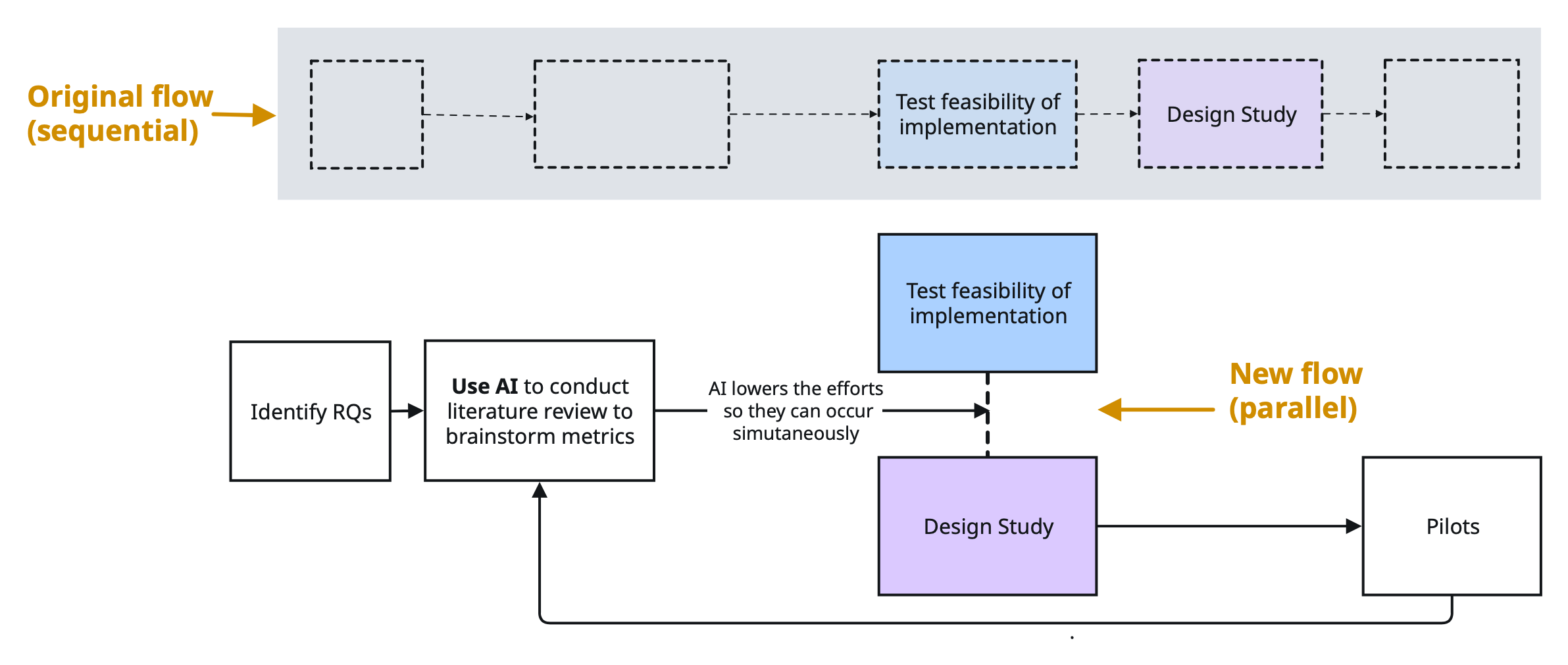}
    \caption{P2's Thought Process. P2 emphasized that AI  ``lowers the efforts so that the two steps can now occur simultaneously'' so that \textit{``rapid updates to evaluation plans as new features are added.''}}
    \label{fig:p2}
    \Description{This graph compares P2’s thought process using an original sequential workflow versus a more efficient, AI-assisted parallel workflow. In the original sequential flow, tasks are completed one after the other. It starts with identifying research questions (RQs), followed by testing the feasibility of implementation, then moving on to a design study, and continues with additional iterative steps. In the new parallel workflow, AI reduces manual effort, allowing certain tasks to occur simultaneously. Specifically, AI is used to conduct literature reviews and brainstorm metrics, which enables the design study and the feasibility test to happen in parallel. This approach, by lowering the effort required, facilitates faster updates to evaluation plans.}
\end{figure*}

Participants also recognized the value of using AI to reduce cognitive load, which helps them better \textbf{\textit{streamline}} their UX evaluation development and align it with dynamic project timelines. This capability was particularly appreciated as it \textit{"simplify decision-making processes in UX design"} (P16) to determine whether new updates are "directly contributing to the UX" (P13) of their original research goal or not. For example, P4 said, \textit{"Generally if I need to propose a UX plan, I probably need to spend like two days at least to collect all the papers. I believe your system can actually help us [HCI researchers] to shorten the time. Because every paper here comes from valid sources, so we can trust the results."} Furthermore, participants also shared the value of using the system for \textbf{\textit{re-structuring their thought process}} of the UX plan development. Figure {\ref{fig:p2}} is an example illustrating how P2 modified the literature review process by using AI to conduct the literature review and brainstorm metrics earlier in the process. 

Nearly all participants recognized the value of using AI to promote early detection of risks, which helps them better \textbf{\textit{envision}} their UX planning. This capability was particularly appreciated as it facilitates \textit{“spotting potential risks before they turn into real headaches after the study is done”} (P5), and supports decisions based on \textit{“minimizing unexpected challenges”}(P9). As P10 said mentioning the \textbf{\texttt{Risk}} feature, \textit{"It doesn’t simply show a list of risks but encourages users to click the associated link to review the actual report."} 

\begin{figure*}[h!]
    \centering
    \includegraphics[width=0.8\textwidth]{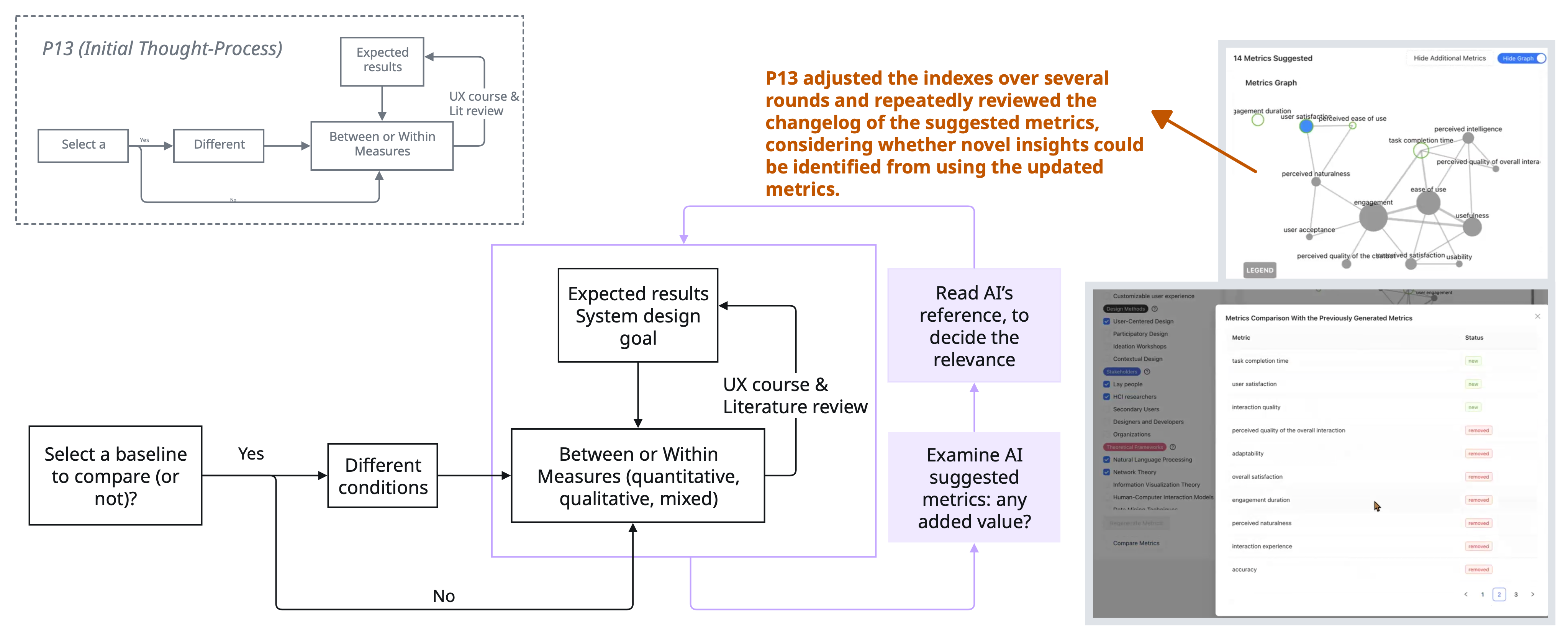}
    \caption{P13's evolving thought process when viewing AI as a teammate supporting interactive self-critique in developing UX evaluation plans for \texttt{\textbf{Index}} and \texttt{\textbf{Metric Graph}} feature. The upper left shows their initial thought process, while the purple-shaded areas indicate the perceived benefits AI introduces in the revised thought process.}
    \label{fig:p13}
    \Description{The figure shows the differences in P13’s approach to developing UX evaluation plans without and with EvalignUX, focusing on the “Index” and “Metric Graph” features. On the upper left side, a simple flowchart illustrates P13’s initial thought process, which begins with deciding whether to use a baseline for comparison and then moves through selecting different study conditions and measures—whether qualitative, quantitative, or mixed. After integrating AI as a “teammate,” the process expands to include steps like reading the AI’s references, examining the AI’s suggested metrics, and determining their relevance to the system’s design goals. On the right, a snapshot of the Metric Graph interface shows how P13 repeatedly refined the indexes by using AI’s suggestions to uncover novel insights.}
\end{figure*}

\subsubsection{Enriched Thought-Processes with the Support of \systemName} \label{thought}

\begin{figure*}[t]
    \centering
    \includegraphics[width=0.8\linewidth]{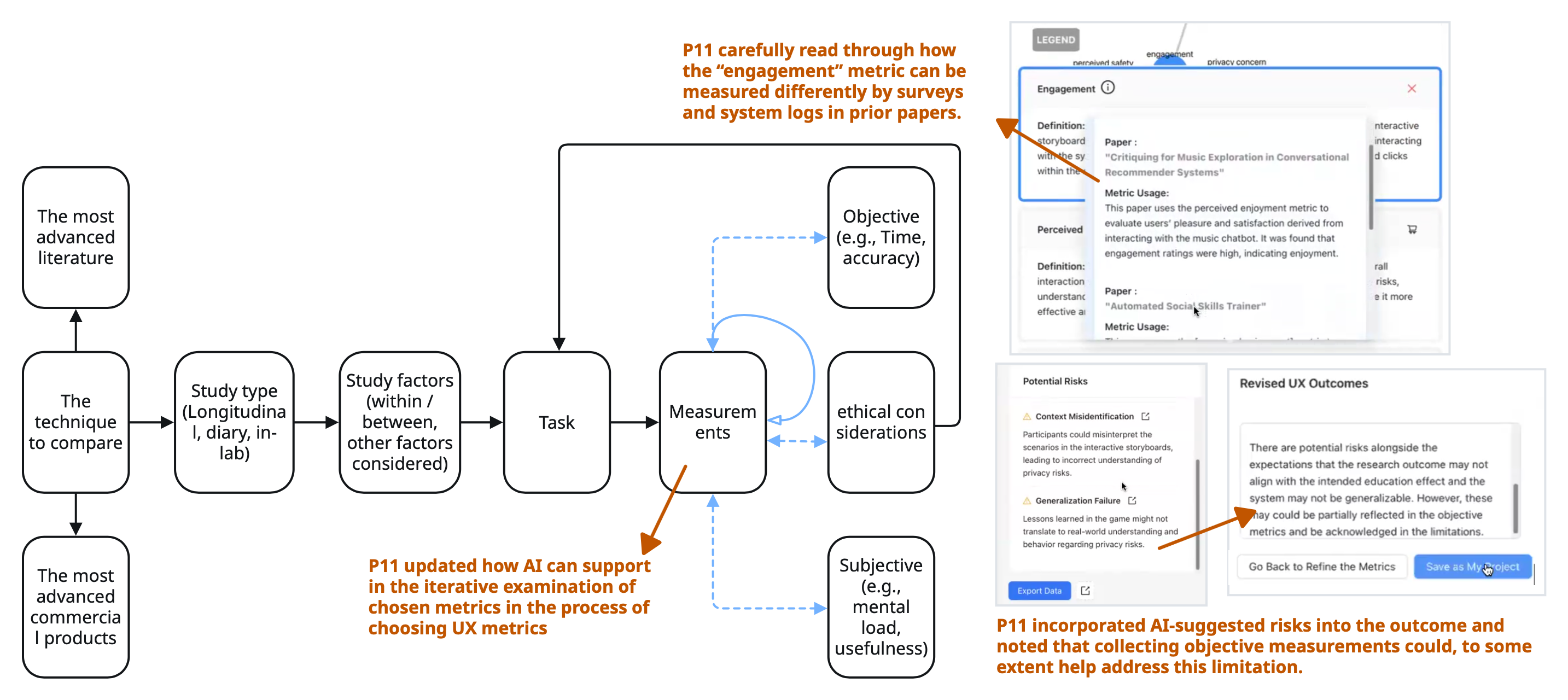}
    \caption{P11's evolving thought process when viewing AI as a teammate supporting interactive self-critique in developing UX evaluation plans for \texttt{\textbf{Metrics List}} feature. The blue lines shows their perceived AI's benefits in the iteraction on metric selection.}
    \label{fig:p11}
    \Description{The image depicts P11's evolving thought process while utilizing AI as a teammate to refine UX evaluation plans, focusing on the “Metrics List” feature. The diagram illustrates how P11 initially considered comparing techniques, study types, and study factors before performing tasks and measuring outcomes. The process then branches into different types of measurements, including objective and subjective measurements, while also considering ethical aspects. The blue lines indicate where AI's involvement is perceived to add benefits during the iterative process of metric selection. AI aids P11 by providing references on how certain metrics, like “engagement”, can be measured using surveys and system logs. Additionally, P11 incorporated AI-suggested risks, noting that AI helped refine metrics by highlighting potential risks and emphasizing the importance of collecting objective measurements to address limitations.}
\end{figure*}

Besides how participants used the \systemName features differently, they also demonstrated diverse thought processes during their development of UX evaluation plans. When updating their thought processes, they all revised their literature review process by incorporating relevant \systemName features. More than half of the participants expressed that AI facilitates \textbf{\textit{iterative loops}} during the development of an evaluation plan (8 participants). For instance, Figure {\ref{fig:p13}} illustrates P13, a UX researcher with 1-2 years experience, revised thought process after using \systemName, showing AI’s role as a supportive teammate in this process. The AI-suggested metrics, as highlighted by the purple boxes, prompted participants to critically assess the relevance and appropriateness of AI’s recommendations, ensuring alignment with system design goals and expected results. This iterative reflection, where P13 evaluates the added value of AI’s input, emphasizes AI’s role in fostering critical thinking and refinement. Similarly, Figure {\ref{fig:p11}}. demonstrates how P11, with 3-4 years of UX experience, views AI as beneficial to the iterative metric selection process.

\begin{figure*}[t]
    \centering
    \includegraphics[width=0.9\linewidth]{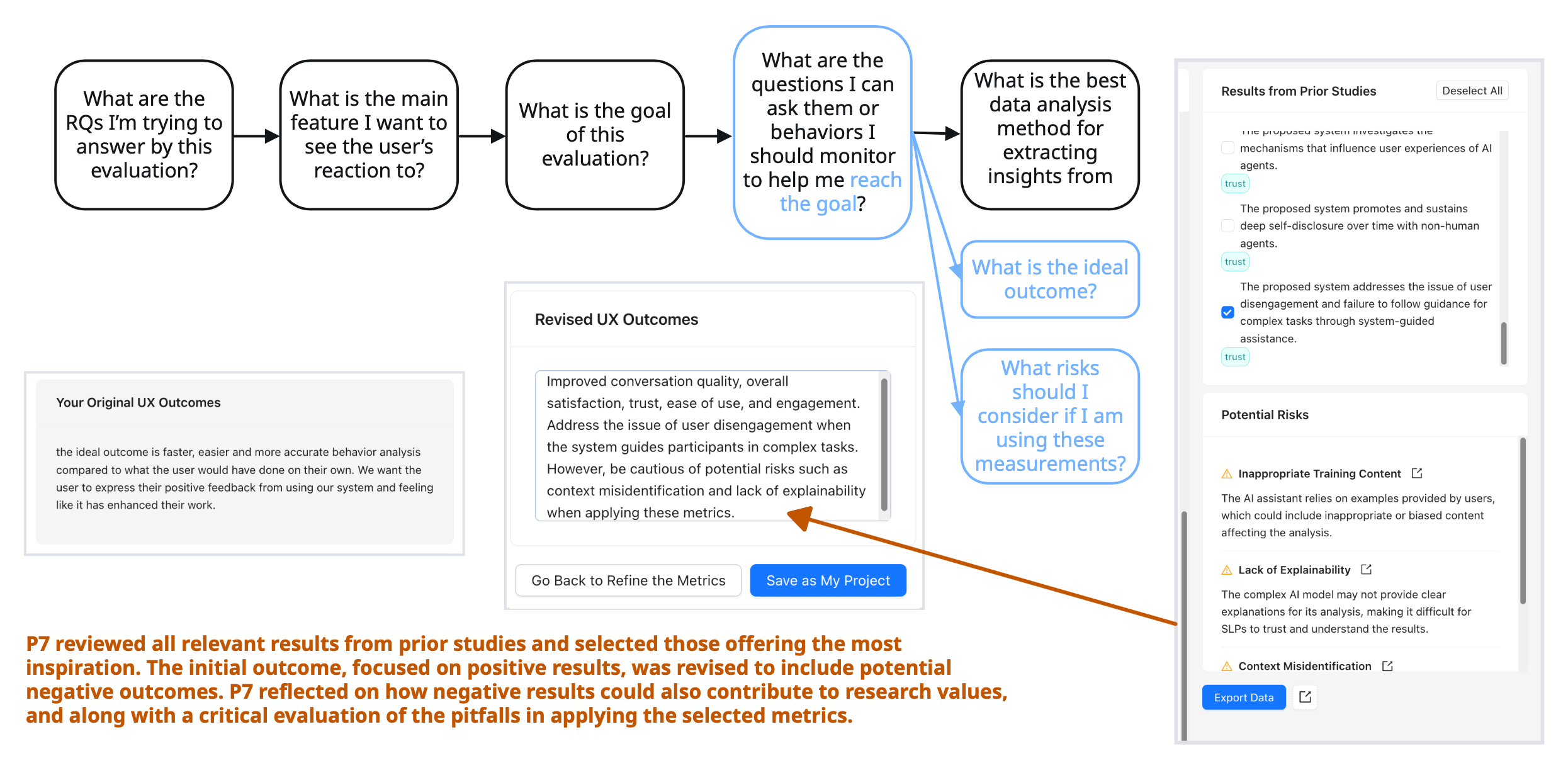}
    \caption{P7's evolving thought process by adding a new component of outcomes and risks inspired by the 
    \texttt{\textbf{Outcome}} and \texttt{\textbf{Risk}} features.}
    \label{fig:p7}
    \Description{The graph illustrates P7’s evolving thought process for developing a UX evaluation strategy with EvalignUX. Initially, P7 considers broad questions, such as the key research questions and which user interactions or features to observe—before gradually narrowing down to a specific evaluation goal. Along the way, P7 refines the plan using the Outcome and Risk features. By selecting checkboxes from “Results from Prior Studies” and “Potential Risks,” P7 reviews previous findings for inspiration, extracts meaningful insights, and acknowledges potential downsides. The revised UX outcomes now include both improvements, like enhanced conversation quality, better engagement, and ease of use, and potential pitfalls, such as user disengagement or misinterpretation of AI guidance. Arrows and speech-bubble-like shapes connect these steps, showing the progression from defining evaluation goals and desired user behaviors to selecting valuable prior results and understanding associated risks. This balanced approach demonstrates that P7 is considering both positive and negative outcomes to shape research values and identify areas for improvement.}
\end{figure*}

We also observed that UX risks and outcomes were incorporated into participants' thought processes, \textbf{regardless} of their experience level. Overall, 42\% of participants incorporated new components into their UX evaluation development thought processes after using \systemName. Notably, these included "Propose measurements" (3 junior UX-level participants, mixed research stages), "UX risks" (mixed UX levels, mixed research stages), and "Research outcomes" (mixed UX levels, mixed research stages). These additions reflect a newly gained awareness of potential challenges and impacts in their UX evaluation plan development. For example, Figure \ref{fig:p7} shows how P7, a UX researcher with 1-2 years of experience,  added two new considerations: defining the ideal outcome of the evaluation and identifying potential risks associated with the chosen measurements to their thought process after using \systemName.

\subsubsection{The Significant Use of Hypothesized \textbf{\texttt{Outcome}} and \textbf{\texttt{Metrics Exploration}} Features}

Based on these observations, we further analyzed the relationship between users' system log variables (e.g., the number of times indexes were regenerated, the number of metrics selected via the Metrics List, the number of metrics selected or deselected via Metric Graphs, the number of outcomes clicked, and the number of risks clicked) and their perceptions of five different features across 11 perception variables (e.g., enjoyment, intention, usefulness, cognitive load, inspiration), see Section \ref{questionnaire} for measurements. A total of 55 pairs were examined, with 6 showing significant results. All significant results are presented in Table \ref{tab:combined_results}.

\begin{table*}[ht!]
\centering
\begin{tabular}{p{3.5cm}p{5.5cm}p{0.75cm}p{0.75cm}p{0.75cm}p{2cm}}
\hline
\multicolumn{1}{c}{} & \multicolumn{5}{c}{\textbf{Dependent Variable}} \\
\cline{2-6}
\textbf{Feature Usage Log} & \textbf{\texttt{Feature}: Perception Scores} &\textbf{Mean} & \textbf{SD} & \textbf{$r$} & \textbf{95\% CI} \\ 
\hline
        \# of metrics added by graph & \texttt{Metrics Graph}: Cognitive Load $^*$ & 2.74 & 1.15 & -0.40 & [-0.72, -0.07] \\
        \hline
        \# of selected results & \texttt{Outcome}: Enjoyment $^{**}$ & 3.32 & 1.16 & 0.57 & [0.28, 0.82]\\
         & \texttt{Outcome}: Intention to Use $^*$ & 3.63 & 1.07 & 0.40 & [0.05, 0.66]\\
         & \texttt{Outcome}: Contribution to UX evaluation$^*$ & 3.42 & 1.39 & 0.44 & [0.08, 0.71]\\
         & \texttt{Outcome}: Contribution to UX outcomes$^{**}$ & 3.37 & 1.50 & 0.49 & [0.16, 0.75]\\
\hline
\end{tabular}
\caption{Examining the interaction between participants' feature uses (based on the system log) and their perceived value of the features (N = 19). Significant values: **$p < 0.01$, *$p < 0.05$. Note that questions about fear of over-reliance were reversed and phrased as an attention check question.} 
\label{tab:log_feature_plan}
\Description{Table 3 summarizes how participants’ feature uses based on system logs (e.g., the number of metrics they added using a metrics graph or the number of results they selected) relate to their perceptions of their perceived values of the features. The dependent variables like “Cognitive Load,” “Enjoyment,” “Intention to Use,” “Contribution to UX Evaluation,” and “Contribution to UX Outcomes,” each with a mean, standard deviation (SD), correlation coefficient (r), and 95\% confidence interval. Significant p-values are indicated, with **p < 0.01 and *p < 0.05, showing that greater engagement with the Outcome feature (i.e., selecting results) is associated with higher perceived enjoyment, greater intention to use, and a stronger sense that the features contribute positively to UX evaluation and outcomes. For the Graph feature, the correlation between the number of metrics added by graph and perceived cognitive load is negative, suggesting that the Graph feature usage may be linked to a reduction in perceived mental effort.}
\end{table*}

The results showed that selecting more potential outcome items (i.e., previous research findings) is associated with higher user ratings of the \texttt{\textbf{Outcome}} feature, both enjoyment (\textit{r} = 0.57, 95\% CI [0.28, 0.82], \textit{p} < 0.01$^{**}$) and intention to use (\textit{r} = 0.40, 95\% CI [0.05, 0.66], \textit{p} < 0.05$^*$). Additionally, users rated this feature more positively in terms of its contribution to revising their project idea (\textit{r} = 0.45, 95\% CI [0.05, 0.76], \textit{p} < 0.05$^*$), UX evaluation plan (\textit{r} = 0.44, 95\% CI [0.08, 0.71], \textit{p} < 0.05$^*$), and UX outcomes (\textit{r} = 0.49, 95\% CI [0.16, 0.75], \textit{p} < 0.05$^*$). The findings from relevant HCI domains encouraged participants to consider questions such as \textit{what demographic is left in the original plan?} (P5) \textit{What factor was not considered yet?} (P14) These thoughts prompted them to refer back to their  original project idea, UX evaluation plan, and UX outcomes. P4 selected 15 potential outcomes that were derived from prior work, they commented that for instance, \textit{"measuring trust and confidence in the chatbot's abilities and context understanding"} was \textit{"exactly what they were looking for and this point really hit the point"}. P4 further commented, they liked this feature because they felt "\textit{the generation understands my initial thoughts about the project... It helps me finalize the details of my plan."}

Additionally, the results also indicated that adding more metrics to the graph was associated with a reduction in perceived cognitive load (\textit{r} = 0.40, 95\% CI [0.07, 0.72], \textit{p} < 0.05$^*$). Participants found using the feature of \texttt{\textbf{Metrics Graph}} simple -- not only showing \textit{"an aerial view of how these metrics tie together"} (P6) but also the usage of each metric in prior research, which is \textit{"easy to understand"} (P8).

\subsection{\systemName Perception Varying by Research Background (RQ3)} \label{bg}

In this section, we examined the relationship between users' backgrounds and their perceptions of five different features across nine perception variables (e.g., enjoyment, intention, usefulness, cognitive load, inspiration, see measurements in Section \ref{questionnaire}), aiming to identify which users benefit the most from our system and which users require additional support. Seven significant results were found. All significant results are shown in Table \ref{UX}.

\begin{table*}[h]
\centering
\begin{tabular}{p{2cm}p{5.5cm}p{1cm}p{1cm}p{1cm}p{2cm}}
\hline
\multicolumn{1}{c}{} & \multicolumn{5}{c}{\textbf{Dependent Variable}} \\
\cline{2-6}
\textbf{Background} & \textbf{\texttt{Feature}: Perception Scores} & \textbf{Mean} & \textbf{SD} & \textbf{$r$} & \textbf{95\% CI} \\ 
\hline
UX experience & \texttt{{Index}:} Perceived inspiration $^*$ & 4.00 & 0.94 & -0.44 & [-0.73, -0.05] \\
&\texttt{Metrics list:} Perceived inspiration $^{**}$ & 4.11 & 0.94 & -0.67 & [-0.84, -0.43] \\
&\texttt{Metrics Graph:} Perceived inspiration $^*$ & 3.89 & 1.10& -0.49 & [-0.77, -0.13] \\
&\texttt{Outcome:} Perceived inspiration (\textit{n.s.}) & 3.68 & 1.00\\
& \texttt{Risk:} Perceived inspiration$^*$ & 3.68 & 1.33 & -0.42 & [-0.72, -0.04] \\
\hline
\end{tabular}
\caption{Examining the interaction between participants' background (based on the survey Likert 1-5) and their perceived value of the features (N = 19).Significant values: **$p < 0.01$, *$p < 0.05$. }
\label{UX}
\Description{Table 4 shows how participants’ backgrounds, measured by their self-reported UX experience, relate to their perceived inspiration drawn from system features. The table includes features, Index, Metrics list, Metrics Graph, Outcome, and Risk, along with their corresponding perceived inspiration scores (mean and standard deviation), correlation coefficients (r), and 95\% confidence intervals. Statistically significant results are denoted by * (p < 0.05) and ** (p < 0.01). For the Index, Metrics list, Metrics Graph and Risk features, there are significant negative correlations, suggesting that participants with more UX experience rated these features as less inspiring, whereas those with less experience found them more inspiring. For the Outcome feature, there is no significant difference between participants with junior UX experience and senior UX experience. In summary, this table illustrates how users’ prior UX experience levels can influence how inspirational they find certain system features.}
\end{table*}

\subsubsection{The Extent of Perceived Inspiration Varies by Users' Research Experience}
Overall, the perceived value of providing inspiration through \systemName was significantly negatively correlated with participants' background, with the exception of the \textbf{\texttt{Outcome}} feature. \textbf{Specifically, participants with less UX experience provided higher scores for perceived inspiration, except with the \textbf{\texttt{Outcome}} feature.} Despite this, all participants rated the inspirational measure highly (M = 3.58–4.00, SD = 0.94–1.33).

For \texttt{\textbf{Index}}, some participants with less experience shared that the feature helps to structure research ideas using relevant terminologies as sometimes they \textit{"just have a vague thought without knowing the exact terms"} (P9). P6, with 1 year of UX experience, shared that the feature is a \textit{"very good at pinpointing important aspects, like user-centered design and stakeholders (researchers). It did a good job of checking what the user intended. I wouldn't say anything was wrong or needs changing. Everything was spot on. There are certain things you can add, but they're already in line with what was checked before."} For experienced researchers like P2, she mentioned that the feature helps to distinguish \textit{"umbrella terms"} when they only have a vague understanding of what is being measured: \textit{"It lowered the effort to identify the literature that is actually related. For instance, I saw three terms, knowledge acquisition, knowledge retention, and knowledgability. If I just use the wrong term, I may think that no one has done it before, but it turns out people have already done it."} While experienced users acknowledged its utility for combining suggested indexes in their paper searches, they felt it lacked novelty, as many of the indexes were already familiar (P2), \textit{"I am looking for something that I have no knowledge about."} Instead, they tend to \textit{"self-validate whether they are on the right track"} (P4) rather than seek inspiration for relevant knowledge.

For ~\texttt{\textbf{Metrics List}} and ~\texttt{\textbf{Metrics Graph}}, less experienced users valued these features for revealing \textit{"metrics [they] hadn’t considered"} (P7). As P3 (with 1-2 years of UX experience) mentioned, "\textit{Because generally if I need to do that (on my own), I probably need to spend like 2 days at least to collect all the paper, which is really bad. I think \systemName can actually help us to shorten the time. Because every paper here is coming from valid sources, such as top conferences, so we can trust the result.}"  Similarly, P6 said that the graph gives her serendipity, \textit{"An aerial view of how these evaluation schemes tie together, like from usefulness to ease of use, helps when writing the paper or designing an evaluation scheme. Certain things in usefulness and trust are often correlated, as shown by the thickness of the line, which is inspiring."} Participants shared that the graph helped them explore metrics that were \textit{“novel”} (P3), as they noticed  smaller nodes, disconnected nodes, or nodes with unfamiliar names. The graph made them feel more interested in discovering \textit{“under-studied”} (P6) metrics in a domain;  and for commonly used metrics, they felt they gained insights into common practices (P7, P12). As P7 shared, \textit{"I feel like changing my selections based on the graph resulted in better and more related contributions."} 

This pattern extended to the ~\texttt{\textbf{Risk}}~feature. P6, with 1-2 years of UX research experience, found it interesting that the feature helped identify \textit{"risks that directly related to the system."} More experienced researchers, like P1, mentioned that \textit{"The risks mentioned here are not hardl ines, meaning that if users are not followed, then the system will fail. For instance, if the evaluation plan doesn’t receive IRB approval, they cannot continue to the next step; whereas the potential risk information here is more likely to be soft suggestions that may not contribute to actual revision."} 

Interestingly, inspiration from the \texttt{\textbf{Outcome}} feature did not vary with UX experience, aligning with participants \textbf{\textit{drawing}} of thought-process, where outcomes were often missing from initial UX plans regardless of participants' UX experience. The overall high mean score for rated inspiration from \texttt{{Outcome}} suggests that the feature resonated across all users, though for different reasons based on experience level. Several participants with less UX experience found the \texttt{\textbf{Outcome}}  feature useful for refining their research plans, noting how it helped them \textit{"rephrase [their] outcome to make it more detailed" }(P13). P7 said, \textit{“I haven't thought about it before, but now that I saw it here, that can be an useful contribution... It's really cool that they come from actual papers and aren't fabricated or made up."} In contrast, P4, with 3-4 years of UX experience, noted that \textit{“without going into too much like spending a day thinking about it (UX outcomes). I think that it (your suggested UX outcomes) is a really great restatement of my original one.”}

\section{Discussion}
In this section, we discuss methods for fostering a mindset-focused UX evaluation in HCI education and practice,  the potential risks of using Large Language Models to support UX evaluation, and the design implications of our findings.

\subsection{
Shifting UX Evaluation from \textit{Method-Centric} to \textit{Mindset-Centric}}

Gray’s prior work \cite{gray2016s} highlighted that many UX practitioners view UX as a \textit{"rigid set of methods"} rather than as a \textit{"mindset"}. This method-centric perspective often emphasizes the application of techniques without fully aligning with project-specific goals. Similarly, prior research \cite{goodman2013delivering, mclellan2000experience, roedl2013design, gray2014reprioritizing} has critiqued academic discussions for focusing on method creation and testing, neglecting real-world implementation contexts. This disconnect creates a \textit{"projected practice community"} \cite{gray2014reprioritizing, gray2016s}, where academic conceptions fail to align with practitioners' realities.

The interaction design and evaluation of \textsc{EvAlignUX} demonstrated that it empowered UX researchers to conduct UX evaluation more holistically. First, providing metric definitions, relevant literature, and examples that enabled for researchers to reflect on the implications of these metrics enhanced both the perceived quality of their proposed plans and their confidence in their UX evaluation plans (RQ1). Second, combining different \systemName components to refine UX plans allowed participants to self-validate, critique, and rationalize their evaluation choices (RQ2). Third, participants' thought processes and perceived inspirations from the system varied based on their UX experience and evolved from asking general questions to leveraging AI for iterative reasoning about their chosen metrics, including previously unconsidered aspects, such as outcomes and AI risks (RQ3).

\subsubsection{Mindset-Centric Evaluation: Encouraging Reflections on Insights} Unlike method-centric approaches, mindset-centric evaluation emphasizes critical reflection on why, what, and how evaluations are conducted, aligning choices with broader implications--goals, values, and the insights researchers and practitioners aim to achieve. This paradigm shift acknowledges the complexity and adaptability required in UX evaluation, bridging the gap between academic research and practical applications.
\systemName facilitates this shift by not only helping define metrics, enabling users to understand their underlying concepts, relevance, and alignment with other research areas, but also by connecting researchers with example research papers and real-world applications. This integration encouraged them to move beyond focusing solely on specific methods and to instead reflect on the insights they aim to achieve, such as fostering engagement or building trustworthiness. By shifting their question from \textit{"What method should I use?"} to \textit{"What insights am I seeking?"}, \systemName helped researchers adopt a more holistic perspective. Additionally, the inclusion of AI risk considerations encouraged users to evaluate their UX plans through new lenses, adding depth and broadening the implications of their work. This shift enriched the researchers' mental models, underscoring \systemName’s value in promoting a mindset-centric approach to UX evaluation.

\subsubsection{Implications for HCI Practice and Education} 
This mindset-centric shift holds significant implications for both HCI practice and education. 

First, it aligns with the increasing diversity and sophistication of UX evaluation approaches \cite{pettersson2018bermuda}. While numerous reviews outline UX evaluation methods \cite{pettersson2018bermuda, vermeeren2010user, rivero2017systematic, darin2019instrument, kieffer2019specification}, few interaction designs directly support the development of comprehensive evaluation plans. For example, scholars have stressed the need for adaptable, context-sensitive approaches: Vermeeren et al. \cite{vermeeren2010user} stress that no single method fits all projects; Rivero et al. \cite{rivero2017systematic} advocate for iterative frameworks that incorporate feedback; and Darin et al. \cite{darin2019instrument} call for continuous critical reflection throughout the evaluation process. Our design aligns with the intention to democratize the design process: when EvAlignUX users select metrics from the recommended options, click “more” to explore additional metrics, edit the index to update the recommendations, or adjust their selection after identifying potential risks, they provide feedback to the system. In the final UX evaluation export, users can further refine their selected metrics. Our findings demonstrate an effective interaction design that fosters a mindset-centric perspective, informing and elevating HCI systems research.

Second, our findings reveal that \systemName offers five key benefits for developing UX research plans. However, participants exhibited diverse usage patterns (RQ1, RQ2). Some enriched system-generated plans with detailed context; others prioritized applying the application of specific metrics; and some adopted the plans without critical evaluation. These patterns were, to some extent, linked to participants' UX experience (RQ3). For example, less experienced UX researchers used \systemName to explore unfamiliar metrics, domains, and terminologies, thus finding it more inspirational. In contrast, experienced researchers primarily leveraged the system as a validation tool rather than a source of inspiration. This suggests that \systemName’s primary value evolves with users' expertise—serving as an exploratory and instructive tool for novices while functioning as a confirming and refining aid for experts. Prior work emphasizes that HCI education should extend beyond teaching predefined methods to fostering adaptability, improvisation, and critical thinking skills \cite{gray2016s}. \systemName embodies these principles by enabling researchers to tailor their approaches to the unique demands of each project. Rather than enforcing rigid processes, future AI-driven UX evaluation systems and educational frameworks should encourage researchers to treat metrics as core design materials—fundamental elements that shape and guide evaluation. By adopting this mindset-centric perspective, researchers and practitioners can better navigate evolving complexities, ensuring their UX evaluations remain meaningful, impactful, and contextually relevant.

Third, the literature-based UX repository is a growing collection that could expand to include industry case studies and metrics, broadening its applicability beyond academic research. While academic UX research emphasizes systematic methodologies, theoretical frameworks, and empirical validation, industry UX case studies often prioritize practical insights, real-world constraints, and agile decision-making processes. Integrating industry contributions would make the repository more reflective of diverse UX practices, enabling designers, engineers, and customers to generate tailored plans and share them for collaborative decision-making. Future studies could explore how to structure this integration effectively, curating a broader spectrum of metrics from both peer-reviewed sources and real-world implementations, while maintaining clarity and usability. Additionally, incorporating industry case studies could introduce new ways of evaluating UX beyond traditional academic benchmarks, offering a more holistic perspective on how different sectors define and measure UX.

\subsection{Towards Responsible Use of LLMs in Interdisciplinary and Cross-Disciplinary UX Research}

Previous literature has highlighted the ethical concerns and potential harms of over-reliance on LLMs in various domains. For instance, \citet{ahmad2023impact} found that excessive dependence on AI in education led to students becoming passive, diminishing their decision-making abilities, and negatively impacting their cognitive skills.
Similarly, in human-AI co-creative writing, AI outputs are heavily influenced by the distribution of their training datasets \cite{10.1145/3308560.3317590, Argyle_Busby_Fulda_Gubler_Rytting_Wingate_2023}. Without considering users' unique needs, over-reliance on these outputs can result in research homogeneity and a lack of creativity \cite{10.1145/3630106.3658975, 10.1145/3591196.3593364}. Additionally, issues such as hallucination, lack of transparency, and algorithmic bias have been identified as major barriers to effective AI collaboration \cite{zhai2024effects}. To mitigate these risks, future iterations of \systemName and similar tools could be designed with adaptive modes or adjustable feedback mechanisms that align with a user’s experience level. Such tailoring would help novices develop core competencies over time while allowing more experienced users to refine advanced strategies, ultimately supporting meaningful skill growth rather than hindering it.
Given the limitations of LLMs, future studies need to carefully consider and address the \textbf{\textit{ethical concerns}} raised, ensuring that UX researchers remain critical, reflective, and ultimately self-reliant evaluators. Potential research directions include:

\subsubsection{Mitigating Potential Over-Reliance and Triggering Deeper Thought Processes} 
Questions about potential over-reliance have also emerged. We position the design of such a tool as an educational resource that potentially bridges formal education and real-world practice, helping novices transition into more advanced roles while also supporting experts in adopting new perspectives. It is designed to foster a mindset-centric approach, which inherently encourages reflection and critical thinking, counteracting de-skilling risks.
One potential direction is to use AI systems that provide hints, or to use a \textit{Question Bank} (see Fig. \ref{fig:mental}). This approach may aim to augment researchers' reasoning by prompting reflection and encouraging deeper engagement with their work.
Essential features include controllability and explainability, alongside clear disclaimers that warn users about the risks of over-reliance on generative AI \cite{liao2020questioning}. Previous research highlights that transparent systems with clear AI outputs foster better human-AI collaboration by encouraging users to engage more critically with the tool \cite{wang2020human, glinka2023critical}. Furthermore, studies indicate that embedding mechanisms that encourage active thinking—such as presenting alternative viewpoints or highlighting uncertainties—can help mitigate the risk of over-dependence on AI-generated solutions and foster deeper cognitive engagement \cite{bansal2021does}.

\subsubsection{Using AI to Support Self-Critique}
Our findings suggest that AI may assist in addressing challenges through self-critique mechanisms. For example, \systemName uses its literature base to provide users with relevant information that encourages self-critique. However, unlike human experts, AI systems often fail to identify nuanced flaws in research plans that stem from researchers' misunderstandings. Previous studies have shown that AI's capacity for self-critique is limited, particularly when dealing with complex, contextual errors that require domain-specific knowledge or critical thinking \cite{bansal2021does, wang2021you}. While AI's ability to critique existing research is promising, it lacks the contextual depth found in human-to-human collaboration \cite{wang2021you, xu2022demystifying}. Further research is needed to explore how AI can more closely mimic human critical evaluation in the research process.

\subsubsection{Using AI to Mitigate UX Risks}
A key design implication often overlooked is the concept of “impact visioning." While only one participant in our study considered potential risks during the initial development of their UX plans, previous studies have shown that the negative consequences of new technologies can often be predicted in early design stages \cite{lee2024deepfakes, pang2024blip}. 
AI systems, by drawing on cross-disciplinary insights, may also help researchers anticipate and mitigate potential risks in UX evaluation, raising awareness of emerging concerns in technology use and design. Prior studies in AI ethics and risk management suggest that AI can be instrumental in identifying unintended consequences and risks in various domains, including UX design, by leveraging data across disciplines to foresee issues that might not be immediately apparent \cite{vinuesa2020role, floridi2021ethical}. As the body of evaluation metrics and AI-related risks expands in the rapidly evolving landscape of human-AI interaction, practitioners can benefit from recommendations that continuously evolve.

\subsection{Limitations and Future Work} 
We acknowledge several limitations that can affect \systemName's performance and the findings. First, as an exploratory system, it incorporates a limited selection of papers, primarily focused on human-AI interaction. Also, for the \systemName backend, recommendations are confined to metrics that have been predefined in our database. Participants proposing studies in interdisciplinary fields may find the system’s suggestions less relevant or accurate; however, we do not claim that the current metrics repository is exhaustive. 
Second, although we observed differences in participants' perceptions and usage behaviors, the majority of participants were doctoral students, which may limit the generalizability of our findings. Our user evaluation was conducted by HCI researchers, but in the future, we can include industry practitioners who may lack formal UX research training but need to conduct effective system evaluations. Expanding the system’s design to better support these groups in conducting effective evaluations could be an important direction for future development.
Finally, \systemName scaffolds the conceptual phase of early-stage UX research planning. It focuses on ideation, metric selection, and critical reflection, providing a foundation for subsequent stages, such as methods, logistics, and participant considerations. For instance, selecting metrics upfront facilitates targeted discussions about participants and data collection methods. However, we acknowledge that detailed logistical support is beyond the tool’s scope and could be addressed in future enhancements.

\section{Conclusion}

User experience is vital to any Human-Computer Interaction  research, yet many UX designers and practitioners often rely on rigid evaluation methods rather than adopting a holistic approach. In this paper, we introduced \systemName, an LLM-based tool designed to support UX evaluation development. Through an evaluation with 19 HCI researchers, we found that centering UX metrics in the evaluation process fostered a mindset that balanced holistic UX impact with contextual factors, evaluation practices, and research value. This approach significantly improved participants' perceived plan quality and confidence, while also prompting their thought processes to consider risks when evaluating systems. For practical implications, we offer a \textit{Question Bank} to guide future UX evaluation efforts, and suggest that future researchers can enhance \systemName by incorporating domain-specific literature to further refine their evaluation planning. In a world where experience defines impact, a UX-driven mindset serves as the key to meaningful and lasting designs evaluation.

\begin{acks}
 
This material is based upon work supported by the National Science Foundation under Grant \#2119589. The work is also supported under the AI Research Institutes program by the National Science Foundation and the Institute of Education Sciences, U.S. Department of Education through Award \#2229873 - AI Institute for Transforming Education for Children with Speech and Language Processing Challenges. Any opinions, findings and conclusions or recommendations expressed in this material are those of the author(s) and do not necessarily reflect the views of the National Science Foundation, the Institute of Education Sciences, or the U.S. Department of Education. Additionally, this project is made possible in part by the Institute of Museum and Library Services RE-252329-OLS-22. The views, findings, conclusions or recommendations expressed in this article do not necessarily represent those of the Institute of Museum and Library Services. 
The authors acknowledge the National Artificial Intelligence Research Resource (NAIRR) Pilot (NAIRR240177) for contributing to this research.
We sincerely thank the anonymous reviewers for their constructive suggestions, which have significantly strengthened this work. 
\end{acks}

\bibliographystyle{ACM-Reference-Format}
\bibliography{sample-base.bib}


\begin{thebibliography}{110}


\ifx \showCODEN    \undefined \def \showCODEN     #1{\unskip}     \fi
\ifx \showDOI      \undefined \def \showDOI       #1{#1}\fi
\ifx \showISBNx    \undefined \def \showISBNx     #1{\unskip}     \fi
\ifx \showISBNxiii \undefined \def \showISBNxiii  #1{\unskip}     \fi
\ifx \showISSN     \undefined \def \showISSN      #1{\unskip}     \fi
\ifx \showLCCN     \undefined \def \showLCCN      #1{\unskip}     \fi
\ifx \shownote     \undefined \def \shownote      #1{#1}          \fi
\ifx \showarticletitle \undefined \def \showarticletitle #1{#1}   \fi
\ifx \showURL      \undefined \def \showURL       {\relax}        \fi
\providecommand\bibfield[2]{#2}
\providecommand\bibinfo[2]{#2}
\providecommand\natexlab[1]{#1}
\providecommand\showeprint[2][]{arXiv:#2}

\bibitem[Abbasian et~al\mbox{.}(2024)]%
        {abbasian2024foundation}
\bibfield{author}{\bibinfo{person}{Mahyar Abbasian}, \bibinfo{person}{Elahe Khatibi}, \bibinfo{person}{Iman Azimi}, \bibinfo{person}{David Oniani}, \bibinfo{person}{Zahra Shakeri Hossein~Abad}, \bibinfo{person}{Alexander Thieme}, \bibinfo{person}{Ram Sriram}, \bibinfo{person}{Zhongqi Yang}, \bibinfo{person}{Yanshan Wang}, \bibinfo{person}{Bryant Lin}, {et~al\mbox{.}}} \bibinfo{year}{2024}\natexlab{}.
\newblock \showarticletitle{Foundation metrics for evaluating effectiveness of healthcare conversations powered by generative AI}.
\newblock \bibinfo{journal}{\emph{NPJ Digital Medicine}} \bibinfo{volume}{7}, \bibinfo{number}{1} (\bibinfo{year}{2024}), \bibinfo{pages}{82}.
\newblock


\bibitem[Ahmad et~al\mbox{.}(2023a)]%
        {ahmad2023requirements}
\bibfield{author}{\bibinfo{person}{Khlood Ahmad}, \bibinfo{person}{Mohamed Abdelrazek}, \bibinfo{person}{Chetan Arora}, \bibinfo{person}{Muneera Bano}, {and} \bibinfo{person}{John Grundy}.} \bibinfo{year}{2023}\natexlab{a}.
\newblock \showarticletitle{Requirements practices and gaps when engineering human-centered Artificial Intelligence systems}.
\newblock \bibinfo{journal}{\emph{Applied Soft Computing}}  \bibinfo{volume}{143} (\bibinfo{year}{2023}), \bibinfo{pages}{110421}.
\newblock


\bibitem[Ahmad et~al\mbox{.}(2023b)]%
        {ahmad2023impact}
\bibfield{author}{\bibinfo{person}{Sayed~Fayaz Ahmad}, \bibinfo{person}{Heesup Han}, \bibinfo{person}{Muhammad~Mansoor Alam}, \bibinfo{person}{Mohd Rehmat}, \bibinfo{person}{Muhammad Irshad}, \bibinfo{person}{Marcelo Arra{\~n}o-Mu{\~n}oz}, \bibinfo{person}{Antonio Ariza-Montes}, {et~al\mbox{.}}} \bibinfo{year}{2023}\natexlab{b}.
\newblock \showarticletitle{Impact of artificial intelligence on human loss in decision making, laziness and safety in education}.
\newblock \bibinfo{journal}{\emph{Humanities and Social Sciences Communications}} \bibinfo{volume}{10}, \bibinfo{number}{1} (\bibinfo{year}{2023}), \bibinfo{pages}{1--14}.
\newblock


\bibitem[Ali et~al\mbox{.}(2020)]%
        {ali2020graph}
\bibfield{author}{\bibinfo{person}{Zafar Ali}, \bibinfo{person}{Guilin Qi}, \bibinfo{person}{Pavlos Kefalas}, \bibinfo{person}{Waheed~Ahmad Abro}, {and} \bibinfo{person}{Bahadar Ali}.} \bibinfo{year}{2020}\natexlab{}.
\newblock \showarticletitle{A graph-based taxonomy of citation recommendation models}.
\newblock \bibinfo{journal}{\emph{Artificial Intelligence Review}}  \bibinfo{volume}{53} (\bibinfo{year}{2020}), \bibinfo{pages}{5217--5260}.
\newblock


\bibitem[Argyle et~al\mbox{.}(2023)]%
        {Argyle_Busby_Fulda_Gubler_Rytting_Wingate_2023}
\bibfield{author}{\bibinfo{person}{Lisa~P. Argyle}, \bibinfo{person}{Ethan~C. Busby}, \bibinfo{person}{Nancy Fulda}, \bibinfo{person}{Joshua~R. Gubler}, \bibinfo{person}{Christopher Rytting}, {and} \bibinfo{person}{David Wingate}.} \bibinfo{year}{2023}\natexlab{}.
\newblock \showarticletitle{Out of One, Many: Using Language Models to Simulate Human Samples}.
\newblock \bibinfo{journal}{\emph{Political Analysis}} \bibinfo{volume}{31}, \bibinfo{number}{3} (\bibinfo{year}{2023}), \bibinfo{pages}{337–351}.
\newblock
\urldef\tempurl%
\url{https://doi.org/10.1017/pan.2023.2}
\showDOI{\tempurl}


\bibitem[Bangor et~al\mbox{.}(2008)]%
        {bangor2008empirical}
\bibfield{author}{\bibinfo{person}{Aaron Bangor}, \bibinfo{person}{Philip~T Kortum}, {and} \bibinfo{person}{James~T Miller}.} \bibinfo{year}{2008}\natexlab{}.
\newblock \showarticletitle{An empirical evaluation of the system usability scale}.
\newblock \bibinfo{journal}{\emph{Intl. Journal of Human--Computer Interaction}} \bibinfo{volume}{24}, \bibinfo{number}{6} (\bibinfo{year}{2008}), \bibinfo{pages}{574--594}.
\newblock


\bibitem[Bansal et~al\mbox{.}(2021)]%
        {bansal2021does}
\bibfield{author}{\bibinfo{person}{Gagan Bansal}, \bibinfo{person}{Besmira Nushi}, \bibinfo{person}{Ece Kamar}, \bibinfo{person}{Walter~S Lasecki}, \bibinfo{person}{Daniel~S Weld}, {and} \bibinfo{person}{Eric Horvitz}.} \bibinfo{year}{2021}\natexlab{}.
\newblock \showarticletitle{Does the whole exceed its parts? The effect of AI explanations on complementary team performance}. In \bibinfo{booktitle}{\emph{Proceedings of the 2021 CHI Conference on Human Factors in Computing Systems}}. \bibinfo{pages}{1--16}.
\newblock


\bibitem[Blondel et~al\mbox{.}(2008)]%
        {blondel2008fast}
\bibfield{author}{\bibinfo{person}{Vincent~D Blondel}, \bibinfo{person}{Jean-Loup Guillaume}, \bibinfo{person}{Renaud Lambiotte}, {and} \bibinfo{person}{Etienne Lefebvre}.} \bibinfo{year}{2008}\natexlab{}.
\newblock \showarticletitle{Fast unfolding of communities in large networks}.
\newblock \bibinfo{journal}{\emph{Journal of statistical mechanics: theory and experiment}} \bibinfo{volume}{2008}, \bibinfo{number}{10} (\bibinfo{year}{2008}), \bibinfo{pages}{P10008}.
\newblock


\bibitem[Bolboaca and J{\"a}ntschi(2006)]%
        {bolboaca2006pearson}
\bibfield{author}{\bibinfo{person}{Sorana-Daniela Bolboaca} {and} \bibinfo{person}{Lorentz J{\"a}ntschi}.} \bibinfo{year}{2006}\natexlab{}.
\newblock \showarticletitle{Pearson versus Spearman, Kendall’s tau correlation analysis on structure-activity relationships of biologic active compounds}.
\newblock \bibinfo{journal}{\emph{Leonardo Journal of Sciences}} \bibinfo{volume}{5}, \bibinfo{number}{9} (\bibinfo{year}{2006}), \bibinfo{pages}{179--200}.
\newblock


\bibitem[Bossen et~al\mbox{.}(2016)]%
        {bossen2016evaluation}
\bibfield{author}{\bibinfo{person}{Claus Bossen}, \bibinfo{person}{Christian Dindler}, {and} \bibinfo{person}{Ole~Sejer Iversen}.} \bibinfo{year}{2016}\natexlab{}.
\newblock \showarticletitle{Evaluation in participatory design: a literature survey}. In \bibinfo{booktitle}{\emph{Proceedings of the 14th Participatory Design Conference: Full papers-Volume 1}}. \bibinfo{pages}{151--160}.
\newblock


\bibitem[Braun and Clarke(2012)]%
        {braun2012thematic}
\bibfield{author}{\bibinfo{person}{Virginia Braun} {and} \bibinfo{person}{Victoria Clarke}.} \bibinfo{year}{2012}\natexlab{}.
\newblock \bibinfo{booktitle}{\emph{Thematic analysis.}}
\newblock \bibinfo{publisher}{American Psychological Association}.
\newblock


\bibitem[Chaudhry(2024)]%
        {10.1145/3613905.3650878}
\bibfield{author}{\bibinfo{person}{Beenish~Moalla Chaudhry}.} \bibinfo{year}{2024}\natexlab{}.
\newblock \showarticletitle{Concerns and Challenges of AI Tools in the UI/UX Design Process: A Cross-Sectional Survey}. In \bibinfo{booktitle}{\emph{Extended Abstracts of the 2024 CHI Conference on Human Factors in Computing Systems}} \emph{(\bibinfo{series}{CHI EA '24})}. \bibinfo{publisher}{Association for Computing Machinery}, \bibinfo{address}{New York, NY, USA}, Article \bibinfo{articleno}{83}, \bibinfo{numpages}{6}~pages.
\newblock
\showISBNx{9798400703317}
\urldef\tempurl%
\url{https://doi.org/10.1145/3613905.3650878}
\showDOI{\tempurl}


\bibitem[Chu et~al\mbox{.}(2010)]%
        {chu2010two}
\bibfield{author}{\bibinfo{person}{Hui-Chun Chu}, \bibinfo{person}{Gwo-Jen Hwang}, \bibinfo{person}{Chin-Chung Tsai}, {and} \bibinfo{person}{Judy~CR Tseng}.} \bibinfo{year}{2010}\natexlab{}.
\newblock \showarticletitle{A two-tier test approach to developing location-aware mobile learning systems for natural science courses}.
\newblock \bibinfo{journal}{\emph{Computers \& Education}} \bibinfo{volume}{55}, \bibinfo{number}{4} (\bibinfo{year}{2010}), \bibinfo{pages}{1618--1627}.
\newblock


\bibitem[Darin et~al\mbox{.}(2019)]%
        {darin2019instrument}
\bibfield{author}{\bibinfo{person}{Ticianne Darin}, \bibinfo{person}{Bianca Coelho}, {and} \bibinfo{person}{Bosco Borges}.} \bibinfo{year}{2019}\natexlab{}.
\newblock \showarticletitle{Which Instrument should I use? Supporting decision-making about the evaluation of user experience}. In \bibinfo{booktitle}{\emph{Design, User Experience, and Usability. Practice and Case Studies: 8th International Conference, DUXU 2019, Held as Part of the 21st HCI International Conference, HCII 2019, Orlando, FL, USA, July 26--31, 2019, Proceedings, Part IV 21}}. Springer, \bibinfo{pages}{49--67}.
\newblock


\bibitem[Davidson et~al\mbox{.}(2023)]%
        {davidson2023development}
\bibfield{author}{\bibinfo{person}{Shayn~S Davidson}, \bibinfo{person}{Joseph~R Keebler}, \bibinfo{person}{Tianxin Zhang}, \bibinfo{person}{Barbara Chaparro}, \bibinfo{person}{James Szalma}, {and} \bibinfo{person}{Christina~M Frederick}.} \bibinfo{year}{2023}\natexlab{}.
\newblock \showarticletitle{The development and validation of a universal enjoyment measure: The enjoy scale}.
\newblock \bibinfo{journal}{\emph{Current Psychology}} \bibinfo{volume}{42}, \bibinfo{number}{21} (\bibinfo{year}{2023}), \bibinfo{pages}{17733--17745}.
\newblock


\bibitem[Davis(1989)]%
        {davis1989perceived}
\bibfield{author}{\bibinfo{person}{Fred~D Davis}.} \bibinfo{year}{1989}\natexlab{}.
\newblock \showarticletitle{Perceived usefulness, perceived ease of use, and user acceptance of information technology}.
\newblock \bibinfo{journal}{\emph{MIS quarterly}} (\bibinfo{year}{1989}), \bibinfo{pages}{319--340}.
\newblock


\bibitem[De~Meo et~al\mbox{.}(2011)]%
        {de2011generalized}
\bibfield{author}{\bibinfo{person}{Pasquale De~Meo}, \bibinfo{person}{Emilio Ferrara}, \bibinfo{person}{Giacomo Fiumara}, {and} \bibinfo{person}{Alessandro Provetti}.} \bibinfo{year}{2011}\natexlab{}.
\newblock \showarticletitle{Generalized louvain method for community detection in large networks}. In \bibinfo{booktitle}{\emph{2011 11th international conference on intelligent systems design and applications}}. IEEE, \bibinfo{pages}{88--93}.
\newblock


\bibitem[Dean et~al\mbox{.}(2006)]%
        {dean2006identifying}
\bibfield{author}{\bibinfo{person}{Douglas~L Dean}, \bibinfo{person}{Jill Hender}, \bibinfo{person}{Tom Rodgers}, {and} \bibinfo{person}{Eric Santanen}.} \bibinfo{year}{2006}\natexlab{}.
\newblock \showarticletitle{Identifying good ideas: constructs and scales for idea evaluation}.
\newblock \bibinfo{journal}{\emph{Journal of Association for Information Systems}} \bibinfo{volume}{7}, \bibinfo{number}{10} (\bibinfo{year}{2006}), \bibinfo{pages}{646--699}.
\newblock


\bibitem[Desmond et~al\mbox{.}(2024)]%
        {desmond2024evalullm}
\bibfield{author}{\bibinfo{person}{Michael Desmond}, \bibinfo{person}{Zahra Ashktorab}, \bibinfo{person}{Qian Pan}, \bibinfo{person}{Casey Dugan}, {and} \bibinfo{person}{James~M Johnson}.} \bibinfo{year}{2024}\natexlab{}.
\newblock \showarticletitle{EvaluLLM: LLM assisted evaluation of generative outputs}. In \bibinfo{booktitle}{\emph{Companion Proceedings of the 29th International Conference on Intelligent User Interfaces}}. \bibinfo{pages}{30--32}.
\newblock


\bibitem[Dixon and Massey~Jr(1951)]%
        {dixon1951introduction}
\bibfield{author}{\bibinfo{person}{Wilfrid~J Dixon} {and} \bibinfo{person}{Frank~J Massey~Jr}.} \bibinfo{year}{1951}\natexlab{}.
\newblock \showarticletitle{Introduction to statistical analysis.}
\newblock  (\bibinfo{year}{1951}).
\newblock


\bibitem[Doshi and Hauser(2024)]%
        {doshi2024generative}
\bibfield{author}{\bibinfo{person}{Anil~R Doshi} {and} \bibinfo{person}{Oliver~P Hauser}.} \bibinfo{year}{2024}\natexlab{}.
\newblock \showarticletitle{Generative AI enhances individual creativity but reduces the collective diversity of novel content}.
\newblock \bibinfo{journal}{\emph{Science Advances}} \bibinfo{volume}{10}, \bibinfo{number}{28} (\bibinfo{year}{2024}), \bibinfo{pages}{eadn5290}.
\newblock


\bibitem[Edge et~al\mbox{.}(2024)]%
        {edge2024local}
\bibfield{author}{\bibinfo{person}{Darren Edge}, \bibinfo{person}{Ha Trinh}, \bibinfo{person}{Newman Cheng}, \bibinfo{person}{Joshua Bradley}, \bibinfo{person}{Alex Chao}, \bibinfo{person}{Apurva Mody}, \bibinfo{person}{Steven Truitt}, {and} \bibinfo{person}{Jonathan Larson}.} \bibinfo{year}{2024}\natexlab{}.
\newblock \showarticletitle{From local to global: A graph rag approach to query-focused summarization}.
\newblock \bibinfo{journal}{\emph{arXiv preprint arXiv:2404.16130}} (\bibinfo{year}{2024}).
\newblock


\bibitem[Everard and Galletta(2005)]%
        {everard2005presentation}
\bibfield{author}{\bibinfo{person}{Andrea Everard} {and} \bibinfo{person}{Dennis~F Galletta}.} \bibinfo{year}{2005}\natexlab{}.
\newblock \showarticletitle{How presentation flaws affect perceived site quality, trust, and intention to purchase from an online store}.
\newblock \bibinfo{journal}{\emph{Journal of management information systems}} \bibinfo{volume}{22}, \bibinfo{number}{3} (\bibinfo{year}{2005}), \bibinfo{pages}{56--95}.
\newblock


\bibitem[Faruk et~al\mbox{.}(2024a)]%
        {faruk2024review}
\bibfield{author}{\bibinfo{person}{Lawal Ibrahim~Dutsinma Faruk}, \bibinfo{person}{Mohammad~Dawood Babakerkhell}, \bibinfo{person}{Pornchai Mongkolnam}, \bibinfo{person}{Vithida Chongsuphajaisiddhi}, \bibinfo{person}{Suree Funilkul}, {and} \bibinfo{person}{Debajyoti Pal}.} \bibinfo{year}{2024}\natexlab{a}.
\newblock \showarticletitle{A review of subjective scales measuring the user experience of voice assistants}.
\newblock \bibinfo{journal}{\emph{IEEE Access}} (\bibinfo{year}{2024}).
\newblock


\bibitem[Faruk et~al\mbox{.}(2024b)]%
        {faruk2024introducing}
\bibfield{author}{\bibinfo{person}{Lawal Ibrahim~Dutsinma Faruk}, \bibinfo{person}{Debajyoti Pal}, \bibinfo{person}{Suree Funilkul}, \bibinfo{person}{Thinagaran Perumal}, {and} \bibinfo{person}{Pornchai Mongkolnam}.} \bibinfo{year}{2024}\natexlab{b}.
\newblock \showarticletitle{Introducing CASUX: A Standardized Scale for Measuring the User Experience of Artificial Intelligence Based Conversational Agents}.
\newblock \bibinfo{journal}{\emph{International Journal of Human--Computer Interaction}} (\bibinfo{year}{2024}), \bibinfo{pages}{1--25}.
\newblock


\bibitem[Fayyaz et~al\mbox{.}(2020)]%
        {fayyaz2020recommendation}
\bibfield{author}{\bibinfo{person}{Zeshan Fayyaz}, \bibinfo{person}{Mahsa Ebrahimian}, \bibinfo{person}{Dina Nawara}, \bibinfo{person}{Ahmed Ibrahim}, {and} \bibinfo{person}{Rasha Kashef}.} \bibinfo{year}{2020}\natexlab{}.
\newblock \showarticletitle{Recommendation systems: Algorithms, challenges, metrics, and business opportunities}.
\newblock \bibinfo{journal}{\emph{applied sciences}} \bibinfo{volume}{10}, \bibinfo{number}{21} (\bibinfo{year}{2020}), \bibinfo{pages}{7748}.
\newblock


\bibitem[Floridi and Chiriatti(2021)]%
        {floridi2021ethical}
\bibfield{author}{\bibinfo{person}{Luciano Floridi} {and} \bibinfo{person}{Mauro Chiriatti}.} \bibinfo{year}{2021}\natexlab{}.
\newblock \showarticletitle{Ethical challenges of AI in biomedicine}.
\newblock \bibinfo{journal}{\emph{AI in medicine}}  \bibinfo{volume}{84} (\bibinfo{year}{2021}), \bibinfo{pages}{1--5}.
\newblock


\bibitem[Fortunato(2010)]%
        {fortunato2010community}
\bibfield{author}{\bibinfo{person}{Santo Fortunato}.} \bibinfo{year}{2010}\natexlab{}.
\newblock \showarticletitle{Community detection in graphs}.
\newblock \bibinfo{journal}{\emph{Physics reports}} \bibinfo{volume}{486}, \bibinfo{number}{3-5} (\bibinfo{year}{2010}), \bibinfo{pages}{75--174}.
\newblock


\bibitem[Glickman and Zhang(2024)]%
        {glickman2024ai}
\bibfield{author}{\bibinfo{person}{Mark Glickman} {and} \bibinfo{person}{Yi Zhang}.} \bibinfo{year}{2024}\natexlab{}.
\newblock \showarticletitle{AI and generative AI for research discovery and summarization}.
\newblock \bibinfo{journal}{\emph{Harvard Data Science Review}} \bibinfo{volume}{6}, \bibinfo{number}{2} (\bibinfo{year}{2024}).
\newblock


\bibitem[Glinka and M{\"u}ller-Birn(2023)]%
        {glinka2023critical}
\bibfield{author}{\bibinfo{person}{Katrin Glinka} {and} \bibinfo{person}{Claudia M{\"u}ller-Birn}.} \bibinfo{year}{2023}\natexlab{}.
\newblock \showarticletitle{Critical-Reflective Human-AI Collaboration: Exploring Computational Tools for Art Historical Image Retrieval}.
\newblock \bibinfo{journal}{\emph{Proceedings of the ACM on Human-Computer Interaction}} \bibinfo{volume}{7}, \bibinfo{number}{CSCW2} (\bibinfo{year}{2023}), \bibinfo{pages}{1--33}.
\newblock


\bibitem[Gonz{\'a}lez-Estrada and Cosmes(2019)]%
        {gonzalez2019shapiro}
\bibfield{author}{\bibinfo{person}{Elizabeth Gonz{\'a}lez-Estrada} {and} \bibinfo{person}{Waldenia Cosmes}.} \bibinfo{year}{2019}\natexlab{}.
\newblock \showarticletitle{Shapiro--Wilk test for skew normal distributions based on data transformations}.
\newblock \bibinfo{journal}{\emph{Journal of Statistical Computation and Simulation}} \bibinfo{volume}{89}, \bibinfo{number}{17} (\bibinfo{year}{2019}), \bibinfo{pages}{3258--3272}.
\newblock


\bibitem[Goodman(2013)]%
        {goodman2013delivering}
\bibfield{author}{\bibinfo{person}{Elizabeth~Sarah Goodman}.} \bibinfo{year}{2013}\natexlab{}.
\newblock \bibinfo{booktitle}{\emph{Delivering design: Performance and materiality in professional interaction design}}.
\newblock \bibinfo{publisher}{University of California, Berkeley}.
\newblock


\bibitem[Graesser et~al\mbox{.}(2006)]%
        {graesser2006question}
\bibfield{author}{\bibinfo{person}{Arthur~C Graesser}, \bibinfo{person}{Zhiqiang Cai}, \bibinfo{person}{Max~M Louwerse}, {and} \bibinfo{person}{Frances Daniel}.} \bibinfo{year}{2006}\natexlab{}.
\newblock \showarticletitle{Question Understanding Aid (QUAID) a web facility that tests question comprehensibility}.
\newblock \bibinfo{journal}{\emph{Public Opinion Quarterly}} \bibinfo{volume}{70}, \bibinfo{number}{1} (\bibinfo{year}{2006}), \bibinfo{pages}{3--22}.
\newblock


\bibitem[Gray(2016)]%
        {gray2016s}
\bibfield{author}{\bibinfo{person}{Colin~M Gray}.} \bibinfo{year}{2016}\natexlab{}.
\newblock \showarticletitle{" It's More of a Mindset Than a Method" UX Practitioners' Conception of Design Methods}. In \bibinfo{booktitle}{\emph{Proceedings of the 2016 CHI conference on human factors in computing systems}}. \bibinfo{pages}{4044--4055}.
\newblock


\bibitem[Gray et~al\mbox{.}(2014)]%
        {gray2014reprioritizing}
\bibfield{author}{\bibinfo{person}{Colin~M Gray}, \bibinfo{person}{Erik Stolterman}, {and} \bibinfo{person}{Martin~A Siegel}.} \bibinfo{year}{2014}\natexlab{}.
\newblock \showarticletitle{Reprioritizing the relationship between HCI research and practice: bubble-up and trickle-down effects}. In \bibinfo{booktitle}{\emph{Proceedings of the 2014 conference on Designing interactive systems}}. \bibinfo{pages}{725--734}.
\newblock


\bibitem[Hart(2006)]%
        {hart2006nasa}
\bibfield{author}{\bibinfo{person}{Sandra~G Hart}.} \bibinfo{year}{2006}\natexlab{}.
\newblock \showarticletitle{NASA-task load index (NASA-TLX); 20 years later}. In \bibinfo{booktitle}{\emph{Proceedings of the human factors and ergonomics society annual meeting}}, Vol.~\bibinfo{volume}{50}. Sage publications Sage CA: Los Angeles, CA, \bibinfo{pages}{904--908}.
\newblock


\bibitem[Hassenzahl et~al\mbox{.}(2010)]%
        {hassenzahl2010needs}
\bibfield{author}{\bibinfo{person}{Marc Hassenzahl}, \bibinfo{person}{Sarah Diefenbach}, {and} \bibinfo{person}{Anja G{\"o}ritz}.} \bibinfo{year}{2010}\natexlab{}.
\newblock \showarticletitle{Needs, affect, and interactive products--Facets of user experience}.
\newblock \bibinfo{journal}{\emph{Interacting with computers}} \bibinfo{volume}{22}, \bibinfo{number}{5} (\bibinfo{year}{2010}), \bibinfo{pages}{353--362}.
\newblock


\bibitem[Hillman et~al\mbox{.}(2023)]%
        {hillman2023understanding}
\bibfield{author}{\bibinfo{person}{Serena Hillman}, \bibinfo{person}{Samira Jain}, \bibinfo{person}{Craig~M Macdonald}, \bibinfo{person}{Elizabeth~F Churchill}, \bibinfo{person}{Carolyn Pang}, \bibinfo{person}{Jofish Kaye}, {and} \bibinfo{person}{Erick Oduor}.} \bibinfo{year}{2023}\natexlab{}.
\newblock \showarticletitle{Understanding and Evaluating UX Outcomes at Scale}. In \bibinfo{booktitle}{\emph{Companion Publication of the 2023 Conference on Computer Supported Cooperative Work and Social Computing}}. \bibinfo{pages}{466--469}.
\newblock


\bibitem[Hsiao et~al\mbox{.}(2014)]%
        {hsiao2014survey}
\bibfield{author}{\bibinfo{person}{I-Han Hsiao}, \bibinfo{person}{Shuguang Han}, \bibinfo{person}{Manav Malhotra}, \bibinfo{person}{Hui~Soo Chae}, {and} \bibinfo{person}{Gary Natriello}.} \bibinfo{year}{2014}\natexlab{}.
\newblock \showarticletitle{Survey sidekick: Structuring scientifically sound surveys}. In \bibinfo{booktitle}{\emph{Intelligent Tutoring Systems: 12th International Conference, ITS 2014, Honolulu, HI, USA, June 5-9, 2014. Proceedings 12}}. Springer, \bibinfo{pages}{516--522}.
\newblock


\bibitem[Hyland et~al\mbox{.}(1989)]%
        {hyland1989psychometric}
\bibfield{author}{\bibinfo{person}{Michael~E Hyland}, \bibinfo{person}{Sidney~H Irvine}, \bibinfo{person}{Clive Thacker}, \bibinfo{person}{Peter~L Dann}, {and} \bibinfo{person}{Ian Dennis}.} \bibinfo{year}{1989}\natexlab{}.
\newblock \showarticletitle{Psychometric analysis of the Stunkard-Messick Eating Questionnaire (SMEQ) and comparison with the Dutch Eating Behavior Questionnaire (DEBQ)}.
\newblock \bibinfo{journal}{\emph{Current Psychology}} \bibinfo{volume}{8}, \bibinfo{number}{3} (\bibinfo{year}{1989}), \bibinfo{pages}{228--233}.
\newblock


\bibitem[Inan~Nur et~al\mbox{.}(2021)]%
        {inan2021method}
\bibfield{author}{\bibinfo{person}{Aulia Inan~Nur}, \bibinfo{person}{Harry B.~Santoso}, {and} \bibinfo{person}{Panca O.~Hadi~Putra}.} \bibinfo{year}{2021}\natexlab{}.
\newblock \showarticletitle{The method and metric of user experience evaluation: a systematic literature review}. In \bibinfo{booktitle}{\emph{Proceedings of the 2021 10th International Conference on Software and Computer Applications}}. \bibinfo{pages}{307--317}.
\newblock


\bibitem[Ji et~al\mbox{.}(2023)]%
        {ji2023survey}
\bibfield{author}{\bibinfo{person}{Ziwei Ji}, \bibinfo{person}{Nayeon Lee}, \bibinfo{person}{Rita Frieske}, \bibinfo{person}{Tiezheng Yu}, \bibinfo{person}{Dan Su}, \bibinfo{person}{Yan Xu}, \bibinfo{person}{Etsuko Ishii}, \bibinfo{person}{Ye~Jin Bang}, \bibinfo{person}{Andrea Madotto}, {and} \bibinfo{person}{Pascale Fung}.} \bibinfo{year}{2023}\natexlab{}.
\newblock \showarticletitle{Survey of hallucination in natural language generation}.
\newblock \bibinfo{journal}{\emph{Comput. Surveys}} \bibinfo{volume}{55}, \bibinfo{number}{12} (\bibinfo{year}{2023}), \bibinfo{pages}{1--38}.
\newblock


\bibitem[Kejriwal(2022)]%
        {kejriwal2022knowledge}
\bibfield{author}{\bibinfo{person}{Mayank Kejriwal}.} \bibinfo{year}{2022}\natexlab{}.
\newblock \showarticletitle{Knowledge graphs: A practical review of the research landscape}.
\newblock \bibinfo{journal}{\emph{Information}} \bibinfo{volume}{13}, \bibinfo{number}{4} (\bibinfo{year}{2022}), \bibinfo{pages}{161}.
\newblock


\bibitem[Kendall(1938)]%
        {kendall1938new}
\bibfield{author}{\bibinfo{person}{Maurice~G Kendall}.} \bibinfo{year}{1938}\natexlab{}.
\newblock \showarticletitle{A new measure of rank correlation}.
\newblock \bibinfo{journal}{\emph{Biometrika}} \bibinfo{volume}{30}, \bibinfo{number}{1-2} (\bibinfo{year}{1938}), \bibinfo{pages}{81--93}.
\newblock


\bibitem[Kieffer et~al\mbox{.}(2019)]%
        {kieffer2019specification}
\bibfield{author}{\bibinfo{person}{Suzanne Kieffer}, \bibinfo{person}{Luka Rukonic}, \bibinfo{person}{Vincent~Kervyn de Meerendr{\'e}}, {and} \bibinfo{person}{Jean Vanderdonckt}.} \bibinfo{year}{2019}\natexlab{}.
\newblock \showarticletitle{Specification of a UX Process Reference Model towards the Strategic Planning of UX Activities.}. In \bibinfo{booktitle}{\emph{VISIGRAPP (2: HUCAPP)}}. \bibinfo{pages}{74--85}.
\newblock


\bibitem[Kim et~al\mbox{.}(2019)]%
        {kim2019comparing}
\bibfield{author}{\bibinfo{person}{Soomin Kim}, \bibinfo{person}{Joonhwan Lee}, {and} \bibinfo{person}{Gahgene Gweon}.} \bibinfo{year}{2019}\natexlab{}.
\newblock \showarticletitle{Comparing data from chatbot and web surveys: Effects of platform and conversational style on survey response quality}. In \bibinfo{booktitle}{\emph{Proceedings of the 2019 CHI conference on human factors in computing systems}}. \bibinfo{pages}{1--12}.
\newblock


\bibitem[Kim et~al\mbox{.}(2024)]%
        {kim2024evallm}
\bibfield{author}{\bibinfo{person}{Tae~Soo Kim}, \bibinfo{person}{Yoonjoo Lee}, \bibinfo{person}{Jamin Shin}, \bibinfo{person}{Young-Ho Kim}, {and} \bibinfo{person}{Juho Kim}.} \bibinfo{year}{2024}\natexlab{}.
\newblock \showarticletitle{Evallm: Interactive evaluation of large language model prompts on user-defined criteria}. In \bibinfo{booktitle}{\emph{Proceedings of the CHI Conference on Human Factors in Computing Systems}}. \bibinfo{pages}{1--21}.
\newblock


\bibitem[Kuang et~al\mbox{.}(2024)]%
        {kuang2024enhancing}
\bibfield{author}{\bibinfo{person}{Emily Kuang}, \bibinfo{person}{Minghao Li}, \bibinfo{person}{Mingming Fan}, {and} \bibinfo{person}{Kristen Shinohara}.} \bibinfo{year}{2024}\natexlab{}.
\newblock \showarticletitle{Enhancing UX Evaluation Through Collaboration with Conversational AI Assistants: Effects of Proactive Dialogue and Timing}. In \bibinfo{booktitle}{\emph{Proceedings of the CHI Conference on Human Factors in Computing Systems}}. \bibinfo{pages}{1--16}.
\newblock


\bibitem[Lazar et~al\mbox{.}(2017)]%
        {lazar2017research}
\bibfield{author}{\bibinfo{person}{Jonathan Lazar}, \bibinfo{person}{Jinjuan~Heidi Feng}, {and} \bibinfo{person}{Harry Hochheiser}.} \bibinfo{year}{2017}\natexlab{}.
\newblock \bibinfo{booktitle}{\emph{Research methods in human-computer interaction}}.
\newblock \bibinfo{publisher}{Morgan Kaufmann}.
\newblock


\bibitem[Ledo et~al\mbox{.}(2018)]%
        {ledo2018evaluation}
\bibfield{author}{\bibinfo{person}{David Ledo}, \bibinfo{person}{Steven Houben}, \bibinfo{person}{Jo Vermeulen}, \bibinfo{person}{Nicolai Marquardt}, \bibinfo{person}{Lora Oehlberg}, {and} \bibinfo{person}{Saul Greenberg}.} \bibinfo{year}{2018}\natexlab{}.
\newblock \showarticletitle{Evaluation strategies for HCI toolkit research}. In \bibinfo{booktitle}{\emph{Proceedings of the 2018 CHI Conference on Human Factors in Computing Systems}}. \bibinfo{pages}{1--17}.
\newblock


\bibitem[Lee et~al\mbox{.}(2024b)]%
        {lee2024deepfakes}
\bibfield{author}{\bibinfo{person}{Hao-Ping Lee}, \bibinfo{person}{Yu-Ju Yang}, \bibinfo{person}{Thomas~Serban Von~Davier}, \bibinfo{person}{Jodi Forlizzi}, {and} \bibinfo{person}{Sauvik Das}.} \bibinfo{year}{2024}\natexlab{b}.
\newblock \showarticletitle{Deepfakes, Phrenology, Surveillance, and More! A Taxonomy of AI Privacy Risks}. In \bibinfo{booktitle}{\emph{Proceedings of the CHI Conference on Human Factors in Computing Systems}}. \bibinfo{pages}{1--19}.
\newblock


\bibitem[Lee et~al\mbox{.}(2024a)]%
        {10.1145/3630106.3658975}
\bibfield{author}{\bibinfo{person}{Messi~H.J. Lee}, \bibinfo{person}{Jacob~M. Montgomery}, {and} \bibinfo{person}{Calvin~K. Lai}.} \bibinfo{year}{2024}\natexlab{a}.
\newblock \showarticletitle{Large Language Models Portray Socially Subordinate Groups as More Homogeneous, Consistent with a Bias Observed in Humans}. In \bibinfo{booktitle}{\emph{Proceedings of the 2024 ACM Conference on Fairness, Accountability, and Transparency}} (Rio de Janeiro, Brazil) \emph{(\bibinfo{series}{FAccT '24})}. \bibinfo{publisher}{Association for Computing Machinery}, \bibinfo{address}{New York, NY, USA}, \bibinfo{pages}{1321–1340}.
\newblock
\showISBNx{9798400704505}
\urldef\tempurl%
\url{https://doi.org/10.1145/3630106.3658975}
\showDOI{\tempurl}


\bibitem[Lee et~al\mbox{.}(2021)]%
        {lee2021exploring}
\bibfield{author}{\bibinfo{person}{Yi-Chieh Lee}, \bibinfo{person}{Naomi Yamashita}, {and} \bibinfo{person}{Yun Huang}.} \bibinfo{year}{2021}\natexlab{}.
\newblock \showarticletitle{Exploring the effects of incorporating human experts to deliver journaling guidance through a chatbot}.
\newblock \bibinfo{journal}{\emph{Proceedings of the ACM on Human-Computer Interaction}} \bibinfo{volume}{5}, \bibinfo{number}{CSCW1} (\bibinfo{year}{2021}), \bibinfo{pages}{1--27}.
\newblock


\bibitem[Lewis(1992)]%
        {lewis1992psychometric}
\bibfield{author}{\bibinfo{person}{James~R Lewis}.} \bibinfo{year}{1992}\natexlab{}.
\newblock \showarticletitle{Psychometric evaluation of the post-study system usability questionnaire: The PSSUQ}. In \bibinfo{booktitle}{\emph{Proceedings of the human factors society annual meeting}}, Vol.~\bibinfo{volume}{36}. Sage Publications Sage CA: Los Angeles, CA, \bibinfo{pages}{1259--1260}.
\newblock


\bibitem[Liao et~al\mbox{.}(2020a)]%
        {10.1145/3313831.3376590}
\bibfield{author}{\bibinfo{person}{Q.~Vera Liao}, \bibinfo{person}{Daniel Gruen}, {and} \bibinfo{person}{Sarah Miller}.} \bibinfo{year}{2020}\natexlab{a}.
\newblock \showarticletitle{Questioning the AI: Informing Design Practices for Explainable AI User Experiences}. In \bibinfo{booktitle}{\emph{Proceedings of the 2020 CHI Conference on Human Factors in Computing Systems}} (Honolulu, HI, USA) \emph{(\bibinfo{series}{CHI '20})}. \bibinfo{publisher}{Association for Computing Machinery}, \bibinfo{address}{New York, NY, USA}, \bibinfo{pages}{1–15}.
\newblock
\showISBNx{9781450367080}
\urldef\tempurl%
\url{https://doi.org/10.1145/3313831.3376590}
\showDOI{\tempurl}


\bibitem[Liao et~al\mbox{.}(2020b)]%
        {liao2020questioning}
\bibfield{author}{\bibinfo{person}{Q~Vera Liao}, \bibinfo{person}{Daniel Gruen}, {and} \bibinfo{person}{Sarah Miller}.} \bibinfo{year}{2020}\natexlab{b}.
\newblock \showarticletitle{Questioning the AI: informing design practices for explainable AI user experiences}. In \bibinfo{booktitle}{\emph{Proceedings of the 2020 CHI conference on human factors in computing systems}}. \bibinfo{pages}{1--15}.
\newblock


\bibitem[Liao et~al\mbox{.}(2024)]%
        {liao2024ux}
\bibfield{author}{\bibinfo{person}{Q~Vera Liao}, \bibinfo{person}{Mihaela Vorvoreanu}, \bibinfo{person}{Hari Subramonyam}, {and} \bibinfo{person}{Lauren Wilcox}.} \bibinfo{year}{2024}\natexlab{}.
\newblock \showarticletitle{UX Matters: The Critical Role of UX in Responsible AI}.
\newblock \bibinfo{journal}{\emph{Interactions}} \bibinfo{volume}{31}, \bibinfo{number}{4} (\bibinfo{year}{2024}), \bibinfo{pages}{22--27}.
\newblock


\bibitem[Liu et~al\mbox{.}(2024)]%
        {liu2024ai}
\bibfield{author}{\bibinfo{person}{Yiren Liu}, \bibinfo{person}{Si Chen}, \bibinfo{person}{Haocong Cheng}, \bibinfo{person}{Mengxia Yu}, \bibinfo{person}{Xiao Ran}, \bibinfo{person}{Andrew Mo}, \bibinfo{person}{Yiliu Tang}, {and} \bibinfo{person}{Yun Huang}.} \bibinfo{year}{2024}\natexlab{}.
\newblock \showarticletitle{How ai processing delays foster creativity: Exploring research question co-creation with an llm-based agent}. In \bibinfo{booktitle}{\emph{Proceedings of the CHI Conference on Human Factors in Computing Systems}}. \bibinfo{pages}{1--25}.
\newblock


\bibitem[Marquardt et~al\mbox{.}(2017)]%
        {marquardt2017hcitools}
\bibfield{author}{\bibinfo{person}{Nicolai Marquardt}, \bibinfo{person}{Steven Houben}, \bibinfo{person}{Michel Beaudouin-Lafon}, {and} \bibinfo{person}{Andrew~D Wilson}.} \bibinfo{year}{2017}\natexlab{}.
\newblock \showarticletitle{HCITools: Strategies and best practices for designing, evaluating and sharing technical HCI toolkits}. In \bibinfo{booktitle}{\emph{Proceedings of the 2017 CHI Conference Extended Abstracts on Human Factors in Computing Systems}}. \bibinfo{pages}{624--627}.
\newblock


\bibitem[McGregor(2020)]%
        {mcgregor2020preventingrepeatedrealworld}
\bibfield{author}{\bibinfo{person}{Sean McGregor}.} \bibinfo{year}{2020}\natexlab{}.
\newblock \bibinfo{title}{Preventing Repeated Real World AI Failures by Cataloging Incidents: The AI Incident Database}.
\newblock
\newblock
\showeprint[arxiv]{2011.08512}~[cs.CY]
\urldef\tempurl%
\url{https://arxiv.org/abs/2011.08512}
\showURL{%
\tempurl}


\bibitem[McLellan(2000)]%
        {mclellan2000experience}
\bibfield{author}{\bibinfo{person}{Hilary McLellan}.} \bibinfo{year}{2000}\natexlab{}.
\newblock \showarticletitle{Experience design}.
\newblock \bibinfo{journal}{\emph{Cyberpsychology and behavior}} \bibinfo{volume}{3}, \bibinfo{number}{1} (\bibinfo{year}{2000}), \bibinfo{pages}{59--69}.
\newblock


\bibitem[Ng et~al\mbox{.}(2022)]%
        {ng2022examination}
\bibfield{author}{\bibinfo{person}{DTK Ng}, \bibinfo{person}{WY Luo}, \bibinfo{person}{HMY Chan}, {and} \bibinfo{person}{SKW Chu}.} \bibinfo{year}{2022}\natexlab{}.
\newblock \showarticletitle{An examination on primary students’ development in AI Literacy through digital story writing}.
\newblock \bibinfo{journal}{\emph{Computers \& Education: Artificial Intelligence}}  \bibinfo{volume}{100054} (\bibinfo{year}{2022}).
\newblock


\bibitem[Nielsen(1992)]%
        {10.1145/142750.142834}
\bibfield{author}{\bibinfo{person}{Jakob Nielsen}.} \bibinfo{year}{1992}\natexlab{}.
\newblock \showarticletitle{Finding usability problems through heuristic evaluation}. In \bibinfo{booktitle}{\emph{Proceedings of the SIGCHI Conference on Human Factors in Computing Systems}} (Monterey, California, USA) \emph{(\bibinfo{series}{CHI '92})}. \bibinfo{publisher}{Association for Computing Machinery}, \bibinfo{address}{New York, NY, USA}, \bibinfo{pages}{373–380}.
\newblock
\showISBNx{0897915135}
\urldef\tempurl%
\url{https://doi.org/10.1145/142750.142834}
\showDOI{\tempurl}


\bibitem[Nielsen(1994a)]%
        {nielsen1994enhancing}
\bibfield{author}{\bibinfo{person}{Jakob Nielsen}.} \bibinfo{year}{1994}\natexlab{a}.
\newblock \showarticletitle{Enhancing the explanatory power of usability heuristics}. In \bibinfo{booktitle}{\emph{Proceedings of the SIGCHI conference on Human Factors in Computing Systems}}. \bibinfo{pages}{152--158}.
\newblock


\bibitem[Nielsen(1994b)]%
        {nielsen1994usability}
\bibfield{author}{\bibinfo{person}{Jakob Nielsen}.} \bibinfo{year}{1994}\natexlab{b}.
\newblock \bibinfo{booktitle}{\emph{Usability engineering}}.
\newblock \bibinfo{publisher}{Morgan Kaufmann}.
\newblock


\bibitem[Nigam et~al\mbox{.}(2024)]%
        {nigam2024acceleron}
\bibfield{author}{\bibinfo{person}{Hitesh Nigam}, \bibinfo{person}{Mayur Patwardhan}, \bibinfo{person}{Lovekesh Vig}, {and} \bibinfo{person}{Gautam Shroff}.} \bibinfo{year}{2024}\natexlab{}.
\newblock \showarticletitle{Acceleron: A Tool to Accelerate Research Ideation}.
\newblock \bibinfo{journal}{\emph{arXiv preprint arXiv:2403.04382}} (\bibinfo{year}{2024}).
\newblock


\bibitem[Norman(2007)]%
        {norman2007emotional}
\bibfield{author}{\bibinfo{person}{Don Norman}.} \bibinfo{year}{2007}\natexlab{}.
\newblock \bibinfo{booktitle}{\emph{Emotional design: Why we love (or hate) everyday things}}.
\newblock \bibinfo{publisher}{Basic books}.
\newblock


\bibitem[Ntoa et~al\mbox{.}(2021)]%
        {ntoa2021user}
\bibfield{author}{\bibinfo{person}{Stavroula Ntoa}, \bibinfo{person}{George Margetis}, \bibinfo{person}{Margherita Antona}, {and} \bibinfo{person}{Constantine Stephanidis}.} \bibinfo{year}{2021}\natexlab{}.
\newblock \showarticletitle{User experience evaluation in intelligent environments: A comprehensive framework}.
\newblock \bibinfo{journal}{\emph{Technologies}} \bibinfo{volume}{9}, \bibinfo{number}{2} (\bibinfo{year}{2021}), \bibinfo{pages}{41}.
\newblock


\bibitem[OpenAI et~al\mbox{.}(2024)]%
        {openai2024gpt4technicalreport}
\bibfield{author}{\bibinfo{person}{OpenAI}, \bibinfo{person}{Josh Achiam}, \bibinfo{person}{Steven Adler}, {and} \bibinfo{person}{et al.}} \bibinfo{year}{2024}\natexlab{}.
\newblock \bibinfo{title}{GPT-4 Technical Report}.
\newblock
\newblock
\showeprint[arxiv]{2303.08774}~[cs.CL]
\urldef\tempurl%
\url{https://arxiv.org/abs/2303.08774}
\showURL{%
\tempurl}


\bibitem[Paas et~al\mbox{.}(2008)]%
        {paas2008assessment}
\bibfield{author}{\bibinfo{person}{FGWC Paas}, \bibinfo{person}{Paul Ayres}, {and} \bibinfo{person}{Mariya Pachman}.} \bibinfo{year}{2008}\natexlab{}.
\newblock \showarticletitle{Assessment of cognitive load in multimedia learning}.
\newblock \bibinfo{journal}{\emph{Recent Innovations in Educational Technology That Facilitate Student Learning, Information Age Publishing Inc., Charlotte, NC}} (\bibinfo{year}{2008}), \bibinfo{pages}{11--35}.
\newblock


\bibitem[Paas(1992)]%
        {paas1992training}
\bibfield{author}{\bibinfo{person}{Fred~GWC Paas}.} \bibinfo{year}{1992}\natexlab{}.
\newblock \showarticletitle{Training strategies for attaining transfer of problem-solving skill in statistics: a cognitive-load approach.}
\newblock \bibinfo{journal}{\emph{Journal of educational psychology}} \bibinfo{volume}{84}, \bibinfo{number}{4} (\bibinfo{year}{1992}), \bibinfo{pages}{429}.
\newblock


\bibitem[Pang et~al\mbox{.}(2024)]%
        {pang2024blip}
\bibfield{author}{\bibinfo{person}{Rock~Yuren Pang}, \bibinfo{person}{Sebastin Santy}, \bibinfo{person}{Ren{\'e} Just}, {and} \bibinfo{person}{Katharina Reinecke}.} \bibinfo{year}{2024}\natexlab{}.
\newblock \showarticletitle{BLIP: Facilitating the Exploration of Undesirable Consequences of Digital Technologies}. In \bibinfo{booktitle}{\emph{Proceedings of the CHI Conference on Human Factors in Computing Systems}}. \bibinfo{pages}{1--18}.
\newblock


\bibitem[Pecune et~al\mbox{.}(2019)]%
        {pecune2019model}
\bibfield{author}{\bibinfo{person}{Florian Pecune}, \bibinfo{person}{Shruti Murali}, \bibinfo{person}{Vivian Tsai}, \bibinfo{person}{Yoichi Matsuyama}, {and} \bibinfo{person}{Justine Cassell}.} \bibinfo{year}{2019}\natexlab{}.
\newblock \showarticletitle{A model of social explanations for a conversational movie recommendation system}. In \bibinfo{booktitle}{\emph{Proceedings of the 7th International Conference on Human-Agent Interaction}}. \bibinfo{pages}{135--143}.
\newblock


\bibitem[Pettersson et~al\mbox{.}(2018)]%
        {pettersson2018bermuda}
\bibfield{author}{\bibinfo{person}{Ingrid Pettersson}, \bibinfo{person}{Florian Lachner}, \bibinfo{person}{Anna-Katharina Frison}, \bibinfo{person}{Andreas Riener}, {and} \bibinfo{person}{Andreas Butz}.} \bibinfo{year}{2018}\natexlab{}.
\newblock \showarticletitle{A Bermuda triangle? A Review of method application and triangulation in user experience evaluation}. In \bibinfo{booktitle}{\emph{Proceedings of the 2018 CHI conference on human factors in computing systems}}. \bibinfo{pages}{1--16}.
\newblock


\bibitem[Reichheld(2003)]%
        {reichheld2003one}
\bibfield{author}{\bibinfo{person}{Frederick~F Reichheld}.} \bibinfo{year}{2003}\natexlab{}.
\newblock \showarticletitle{The one number you need to grow}.
\newblock \bibinfo{journal}{\emph{Harvard business review}} \bibinfo{volume}{81}, \bibinfo{number}{12} (\bibinfo{year}{2003}), \bibinfo{pages}{46--55}.
\newblock


\bibitem[Rezwana and Maher(2023)]%
        {10.1145/3591196.3593364}
\bibfield{author}{\bibinfo{person}{Jeba Rezwana} {and} \bibinfo{person}{Mary~Lou Maher}.} \bibinfo{year}{2023}\natexlab{}.
\newblock \showarticletitle{User Perspectives on Ethical Challenges in Human-AI Co-Creativity: A Design Fiction Study}. In \bibinfo{booktitle}{\emph{Proceedings of the 15th Conference on Creativity and Cognition}} (Virtual Event, USA) \emph{(\bibinfo{series}{Creativity and Cognition '23})}. \bibinfo{publisher}{Association for Computing Machinery}, \bibinfo{address}{New York, NY, USA}, \bibinfo{pages}{62–74}.
\newblock
\showISBNx{9798400701801}
\urldef\tempurl%
\url{https://doi.org/10.1145/3591196.3593364}
\showDOI{\tempurl}


\bibitem[Rivero and Conte(2017)]%
        {rivero2017systematic}
\bibfield{author}{\bibinfo{person}{Luis Rivero} {and} \bibinfo{person}{Tayana Conte}.} \bibinfo{year}{2017}\natexlab{}.
\newblock \showarticletitle{A systematic mapping study on research contributions on UX evaluation technologies}. In \bibinfo{booktitle}{\emph{Proceedings of the XVI Brazilian symposium on human factors in computing systems}}. \bibinfo{pages}{1--10}.
\newblock


\bibitem[Rodden et~al\mbox{.}(2010)]%
        {rodden2010measuring}
\bibfield{author}{\bibinfo{person}{Kerry Rodden}, \bibinfo{person}{Hilary Hutchinson}, {and} \bibinfo{person}{Xin Fu}.} \bibinfo{year}{2010}\natexlab{}.
\newblock \showarticletitle{Measuring the user experience on a large scale: user-centered metrics for web applications}. In \bibinfo{booktitle}{\emph{Proceedings of the SIGCHI conference on human factors in computing systems}}. \bibinfo{pages}{2395--2398}.
\newblock


\bibitem[Roedl and Stolterman(2013)]%
        {roedl2013design}
\bibfield{author}{\bibinfo{person}{David~J Roedl} {and} \bibinfo{person}{Erik Stolterman}.} \bibinfo{year}{2013}\natexlab{}.
\newblock \showarticletitle{Design research at CHI and its applicability to design practice}. In \bibinfo{booktitle}{\emph{Proceedings of the SIGCHI Conference on Human Factors in Computing Systems}}. \bibinfo{pages}{1951--1954}.
\newblock


\bibitem[Roselli et~al\mbox{.}(2019)]%
        {10.1145/3308560.3317590}
\bibfield{author}{\bibinfo{person}{Drew Roselli}, \bibinfo{person}{Jeanna Matthews}, {and} \bibinfo{person}{Nisha Talagala}.} \bibinfo{year}{2019}\natexlab{}.
\newblock \showarticletitle{Managing Bias in AI}. In \bibinfo{booktitle}{\emph{Companion Proceedings of The 2019 World Wide Web Conference}} (San Francisco, USA) \emph{(\bibinfo{series}{WWW '19})}. \bibinfo{publisher}{Association for Computing Machinery}, \bibinfo{address}{New York, NY, USA}, \bibinfo{pages}{539–544}.
\newblock
\showISBNx{9781450366755}
\urldef\tempurl%
\url{https://doi.org/10.1145/3308560.3317590}
\showDOI{\tempurl}


\bibitem[Sadowski et~al\mbox{.}(2019)]%
        {sadowski2019software}
\bibfield{author}{\bibinfo{person}{Caitlin Sadowski}, \bibinfo{person}{Margaret-Anne Storey}, {and} \bibinfo{person}{Robert Feldt}.} \bibinfo{year}{2019}\natexlab{}.
\newblock \showarticletitle{A software development productivity framework}.
\newblock \bibinfo{journal}{\emph{Rethinking Productivity in Software Engineering}} (\bibinfo{year}{2019}), \bibinfo{pages}{39--47}.
\newblock


\bibitem[Schrepp et~al\mbox{.}(2017)]%
        {schrepp2017design}
\bibfield{author}{\bibinfo{person}{Martin Schrepp}, \bibinfo{person}{Andreas Hinderks}, {and} \bibinfo{person}{J{\"o}rg Thomaschewski}.} \bibinfo{year}{2017}\natexlab{}.
\newblock \showarticletitle{Design and evaluation of a short version of the user experience questionnaire (UEQ-S)}.
\newblock \bibinfo{journal}{\emph{International Journal of Interactive Multimedia and Artificial Intelligence, 4 (6), 103-108.}} (\bibinfo{year}{2017}).
\newblock


\bibitem[Shankar et~al\mbox{.}(2024)]%
        {shankar2024validates}
\bibfield{author}{\bibinfo{person}{Shreya Shankar}, \bibinfo{person}{JD Zamfirescu-Pereira}, \bibinfo{person}{Bj{\"o}rn Hartmann}, \bibinfo{person}{Aditya~G Parameswaran}, {and} \bibinfo{person}{Ian Arawjo}.} \bibinfo{year}{2024}\natexlab{}.
\newblock \showarticletitle{Who Validates the Validators? Aligning LLM-Assisted Evaluation of LLM Outputs with Human Preferences}.
\newblock \bibinfo{journal}{\emph{arXiv preprint arXiv:2404.12272}} (\bibinfo{year}{2024}).
\newblock


\bibitem[Shen et~al\mbox{.}(2023)]%
        {huaconv}
\bibfield{author}{\bibinfo{person}{Hua Shen}, \bibinfo{person}{Chieh-Yang Huang}, \bibinfo{person}{Tongshuang Wu}, {and} \bibinfo{person}{Ting-Hao~Kenneth Huang}.} \bibinfo{year}{2023}\natexlab{}.
\newblock \showarticletitle{ConvXAI: Delivering Heterogeneous AI Explanations via Conversations to Support Human-AI Scientific Writing}. In \bibinfo{booktitle}{\emph{Companion Publication of the 2023 Conference on Computer Supported Cooperative Work and Social Computing}} (Minneapolis, MN, USA) \emph{(\bibinfo{series}{CSCW '23 Companion})}. \bibinfo{publisher}{Association for Computing Machinery}, \bibinfo{address}{New York, NY, USA}, \bibinfo{pages}{384–387}.
\newblock
\showISBNx{9798400701290}
\urldef\tempurl%
\url{https://doi.org/10.1145/3584931.3607492}
\showDOI{\tempurl}


\bibitem[Shneiderman(2020)]%
        {shneiderman2020bridging}
\bibfield{author}{\bibinfo{person}{Ben Shneiderman}.} \bibinfo{year}{2020}\natexlab{}.
\newblock \showarticletitle{Bridging the gap between ethics and practice: guidelines for reliable, safe, and trustworthy human-centered AI systems}.
\newblock \bibinfo{journal}{\emph{ACM Transactions on Interactive Intelligent Systems (TiiS)}} \bibinfo{volume}{10}, \bibinfo{number}{4} (\bibinfo{year}{2020}), \bibinfo{pages}{1--31}.
\newblock


\bibitem[Siemon et~al\mbox{.}(2016)]%
        {siemon2016semi}
\bibfield{author}{\bibinfo{person}{Dominik Siemon}, \bibinfo{person}{Taras Rarog}, {and} \bibinfo{person}{Susanne Robra-Bissantz}.} \bibinfo{year}{2016}\natexlab{}.
\newblock \showarticletitle{Semi-automated questions as a cognitive stimulus in idea generation}. In \bibinfo{booktitle}{\emph{2016 49th Hawaii International Conference on System Sciences (HICSS)}}. IEEE, \bibinfo{pages}{257--266}.
\newblock


\bibitem[Stankov et~al\mbox{.}(2012)]%
        {stankov2012confidence}
\bibfield{author}{\bibinfo{person}{Lazar Stankov}, \bibinfo{person}{Jihyun Lee}, \bibinfo{person}{Wenshu Luo}, {and} \bibinfo{person}{David~J Hogan}.} \bibinfo{year}{2012}\natexlab{}.
\newblock \showarticletitle{Confidence: A better predictor of academic achievement than self-efficacy, self-concept and anxiety?}
\newblock \bibinfo{journal}{\emph{Learning and individual differences}} \bibinfo{volume}{22}, \bibinfo{number}{6} (\bibinfo{year}{2012}), \bibinfo{pages}{747--758}.
\newblock


\bibitem[Stige et~al\mbox{.}(2023)]%
        {stige2023artificial}
\bibfield{author}{\bibinfo{person}{{\AA}sne Stige}, \bibinfo{person}{Efpraxia~D Zamani}, \bibinfo{person}{Patrick Mikalef}, {and} \bibinfo{person}{Yuzhen Zhu}.} \bibinfo{year}{2023}\natexlab{}.
\newblock \showarticletitle{Artificial intelligence (AI) for user experience (UX) design: a systematic literature review and future research agenda}.
\newblock \bibinfo{journal}{\emph{Information Technology \& People}} (\bibinfo{year}{2023}).
\newblock


\bibitem[Sun et~al\mbox{.}(2024)]%
        {sun2024lgtm}
\bibfield{author}{\bibinfo{person}{Haowen Sun}, \bibinfo{person}{Ruikun Zheng}, \bibinfo{person}{Haibin Huang}, \bibinfo{person}{Chongyang Ma}, \bibinfo{person}{Hui Huang}, {and} \bibinfo{person}{Ruizhen Hu}.} \bibinfo{year}{2024}\natexlab{}.
\newblock \showarticletitle{LGTM: Local-to-Global Text-Driven Human Motion Diffusion Model}. In \bibinfo{booktitle}{\emph{ACM SIGGRAPH 2024 Conference Papers}}. \bibinfo{pages}{1--9}.
\newblock


\bibitem[Tamm et~al\mbox{.}(2021)]%
        {tamm2021quality}
\bibfield{author}{\bibinfo{person}{Yan-Martin Tamm}, \bibinfo{person}{Rinchin Damdinov}, {and} \bibinfo{person}{Alexey Vasilev}.} \bibinfo{year}{2021}\natexlab{}.
\newblock \showarticletitle{Quality metrics in recommender systems: Do we calculate metrics consistently?}. In \bibinfo{booktitle}{\emph{Proceedings of the 15th ACM Conference on Recommender Systems}}. \bibinfo{pages}{708--713}.
\newblock


\bibitem[Tan et~al\mbox{.}(2024)]%
        {tan2024audioxtend}
\bibfield{author}{\bibinfo{person}{Felicia Fang-Yi Tan}, \bibinfo{person}{Peisen Xu}, \bibinfo{person}{Ashwin Ram}, \bibinfo{person}{Wei~Zhen Suen}, \bibinfo{person}{Shengdong Zhao}, \bibinfo{person}{Yun Huang}, {and} \bibinfo{person}{Christophe Hurter}.} \bibinfo{year}{2024}\natexlab{}.
\newblock \showarticletitle{AudioXtend: Assisted Reality Visual Accompaniments for Audiobook Storytelling During Everyday Routine Tasks}. In \bibinfo{booktitle}{\emph{Proceedings of the CHI Conference on Human Factors in Computing Systems}}. \bibinfo{pages}{1--22}.
\newblock


\bibitem[Thomas et~al\mbox{.}(2020)]%
        {thomas2020expressions}
\bibfield{author}{\bibinfo{person}{Paul Thomas}, \bibinfo{person}{Daniel McDuff}, \bibinfo{person}{Mary Czerwinski}, {and} \bibinfo{person}{Nick Craswell}.} \bibinfo{year}{2020}\natexlab{}.
\newblock \showarticletitle{Expressions of style in information seeking conversation with an agent}. In \bibinfo{booktitle}{\emph{Proceedings of the 43rd International ACM SIGIR Conference on Research and Development in Information Retrieval}}. \bibinfo{pages}{1171--1180}.
\newblock


\bibitem[Venable et~al\mbox{.}(2012)]%
        {venable2012comprehensive}
\bibfield{author}{\bibinfo{person}{John Venable}, \bibinfo{person}{Jan Pries-Heje}, {and} \bibinfo{person}{Richard Baskerville}.} \bibinfo{year}{2012}\natexlab{}.
\newblock \showarticletitle{A comprehensive framework for evaluation in design science research}. In \bibinfo{booktitle}{\emph{Design Science Research in Information Systems. Advances in Theory and Practice: 7th International Conference, DESRIST 2012, Las Vegas, NV, USA, May 14-15, 2012. Proceedings 7}}. Springer, \bibinfo{pages}{423--438}.
\newblock


\bibitem[Verma et~al\mbox{.}(2023)]%
        {verma2023scholarly}
\bibfield{author}{\bibinfo{person}{Shilpa Verma}, \bibinfo{person}{Rajesh Bhatia}, \bibinfo{person}{Sandeep Harit}, {and} \bibinfo{person}{Sanjay Batish}.} \bibinfo{year}{2023}\natexlab{}.
\newblock \showarticletitle{Scholarly knowledge graphs through structuring scholarly communication: a review}.
\newblock \bibinfo{journal}{\emph{Complex \& intelligent systems}} \bibinfo{volume}{9}, \bibinfo{number}{1} (\bibinfo{year}{2023}), \bibinfo{pages}{1059--1095}.
\newblock


\bibitem[Vermeeren et~al\mbox{.}(2010)]%
        {vermeeren2010user}
\bibfield{author}{\bibinfo{person}{Arnold~POS Vermeeren}, \bibinfo{person}{Effie Lai-Chong Law}, \bibinfo{person}{Virpi Roto}, \bibinfo{person}{Marianna Obrist}, \bibinfo{person}{Jettie Hoonhout}, {and} \bibinfo{person}{Kaisa V{\"a}{\"a}n{\"a}nen-Vainio-Mattila}.} \bibinfo{year}{2010}\natexlab{}.
\newblock \showarticletitle{User experience evaluation methods: current state and development needs}. In \bibinfo{booktitle}{\emph{Proceedings of the 6th Nordic conference on human-computer interaction: Extending boundaries}}. \bibinfo{pages}{521--530}.
\newblock


\bibitem[Vinuesa et~al\mbox{.}(2020)]%
        {vinuesa2020role}
\bibfield{author}{\bibinfo{person}{Ricardo Vinuesa}, \bibinfo{person}{Hossein Azizpour}, \bibinfo{person}{Iolanda Leite}, \bibinfo{person}{Madeline Balaam}, \bibinfo{person}{Virginia Dignum}, \bibinfo{person}{Sami Domisch}, \bibinfo{person}{Anna Fell{\"a}nder}, \bibinfo{person}{Simone~D Langhans}, \bibinfo{person}{Max Tegmark}, {and} \bibinfo{person}{Francesco Fuso~Nerini}.} \bibinfo{year}{2020}\natexlab{}.
\newblock \showarticletitle{The role of artificial intelligence in achieving the Sustainable Development Goals}.
\newblock \bibinfo{journal}{\emph{Nature communications}} \bibinfo{volume}{11}, \bibinfo{number}{1} (\bibinfo{year}{2020}), \bibinfo{pages}{1--10}.
\newblock


\bibitem[Wambsganss et~al\mbox{.}(2020)]%
        {wambsganss2020conversational}
\bibfield{author}{\bibinfo{person}{Thiemo Wambsganss}, \bibinfo{person}{Rainer Winkler}, \bibinfo{person}{Matthias S{\"o}llner}, {and} \bibinfo{person}{Jan~Marco Leimeister}.} \bibinfo{year}{2020}\natexlab{}.
\newblock \showarticletitle{A conversational agent to improve response quality in course evaluations}. In \bibinfo{booktitle}{\emph{Extended Abstracts of the 2020 CHI conference on human factors in computing systems}}. \bibinfo{pages}{1--9}.
\newblock


\bibitem[Wan et~al\mbox{.}(2024)]%
        {wan2024building}
\bibfield{author}{\bibinfo{person}{Hongyu Wan}, \bibinfo{person}{Jinda Zhang}, \bibinfo{person}{Abdulaziz~Arif Suria}, \bibinfo{person}{Bingsheng Yao}, \bibinfo{person}{Dakuo Wang}, \bibinfo{person}{Yvonne Coady}, {and} \bibinfo{person}{Mirjana Prpa}.} \bibinfo{year}{2024}\natexlab{}.
\newblock \showarticletitle{Building LLM-based AI Agents in Social Virtual Reality}. In \bibinfo{booktitle}{\emph{Extended Abstracts of the CHI Conference on Human Factors in Computing Systems}}. \bibinfo{pages}{1--7}.
\newblock


\bibitem[Wang et~al\mbox{.}(2020a)]%
        {wang2020human}
\bibfield{author}{\bibinfo{person}{Dakuo Wang}, \bibinfo{person}{Elizabeth Churchill}, \bibinfo{person}{Pattie Maes}, \bibinfo{person}{Xiangmin Fan}, \bibinfo{person}{Ben Shneiderman}, \bibinfo{person}{Yuanchun Shi}, {and} \bibinfo{person}{Qianying Wang}.} \bibinfo{year}{2020}\natexlab{a}.
\newblock \showarticletitle{From human-human collaboration to Human-AI collaboration: Designing AI systems that can work together with people}. In \bibinfo{booktitle}{\emph{Extended abstracts of the 2020 CHI conference on human factors in computing systems}}. \bibinfo{pages}{1--6}.
\newblock


\bibitem[Wang et~al\mbox{.}(2021)]%
        {wang2021you}
\bibfield{author}{\bibinfo{person}{Di Wang}, \bibinfo{person}{Ming Yin}, {and} \bibinfo{person}{Gloria Mark}.} \bibinfo{year}{2021}\natexlab{}.
\newblock \showarticletitle{Do you trust me to trust you? Mental model transparency and personalized explainability in human-AI collaboration}. In \bibinfo{booktitle}{\emph{Proceedings of the 2021 CHI Conference on Human Factors in Computing Systems}}. \bibinfo{pages}{1--13}.
\newblock


\bibitem[Wang et~al\mbox{.}(2020b)]%
        {wang2020alexa}
\bibfield{author}{\bibinfo{person}{Jinping Wang}, \bibinfo{person}{Hyun Yang}, \bibinfo{person}{Ruosi Shao}, \bibinfo{person}{Saeed Abdullah}, {and} \bibinfo{person}{S~Shyam Sundar}.} \bibinfo{year}{2020}\natexlab{b}.
\newblock \showarticletitle{Alexa as coach: Leveraging smart speakers to build social agents that reduce public speaking anxiety}. In \bibinfo{booktitle}{\emph{Proceedings of the 2020 CHI conference on human factors in computing systems}}. \bibinfo{pages}{1--13}.
\newblock


\bibitem[Wang et~al\mbox{.}(2018)]%
        {wang2018acekg}
\bibfield{author}{\bibinfo{person}{Ruijie Wang}, \bibinfo{person}{Yuchen Yan}, \bibinfo{person}{Jialu Wang}, \bibinfo{person}{Yuting Jia}, \bibinfo{person}{Ye Zhang}, \bibinfo{person}{Weinan Zhang}, {and} \bibinfo{person}{Xinbing Wang}.} \bibinfo{year}{2018}\natexlab{}.
\newblock \showarticletitle{Acekg: A large-scale knowledge graph for academic data mining}. In \bibinfo{booktitle}{\emph{Proceedings of the 27th ACM international conference on information and knowledge management}}. \bibinfo{pages}{1487--1490}.
\newblock


\bibitem[Wang et~al\mbox{.}(2024)]%
        {wang2024virtuwander}
\bibfield{author}{\bibinfo{person}{Zhan Wang}, \bibinfo{person}{Lin-Ping Yuan}, \bibinfo{person}{Liangwei Wang}, \bibinfo{person}{Bingchuan Jiang}, {and} \bibinfo{person}{Wei Zeng}.} \bibinfo{year}{2024}\natexlab{}.
\newblock \showarticletitle{Virtuwander: Enhancing multi-modal interaction for virtual tour guidance through large language models}. In \bibinfo{booktitle}{\emph{Proceedings of the CHI conference on human factors in computing systems}}. \bibinfo{pages}{1--20}.
\newblock


\bibitem[Willis and Lessler(2013)]%
        {willis2013question}
\bibfield{author}{\bibinfo{person}{GB Willis} {and} \bibinfo{person}{JT Lessler}.} \bibinfo{year}{2013}\natexlab{}.
\newblock \bibinfo{title}{Question appraisal system: QAS 99. National Cancer Institute}.
\newblock
\newblock


\bibitem[Willis(2004)]%
        {willis2004cognitive}
\bibfield{author}{\bibinfo{person}{Gordon~B Willis}.} \bibinfo{year}{2004}\natexlab{}.
\newblock \bibinfo{booktitle}{\emph{Cognitive interviewing: A tool for improving questionnaire design}}.
\newblock \bibinfo{publisher}{sage publications}.
\newblock


\bibitem[Xu and Mulligan(2022)]%
        {xu2022demystifying}
\bibfield{author}{\bibinfo{person}{Horvitz Eric Zeng~Eric Xu, Wesley} {and} \bibinfo{person}{Deirdre~K Mulligan}.} \bibinfo{year}{2022}\natexlab{}.
\newblock \showarticletitle{Demystifying AI among end users: A taxonomy of AI literacy}. In \bibinfo{booktitle}{\emph{Proceedings of the 2022 CHI Conference on Human Factors in Computing Systems}}. \bibinfo{pages}{1--19}.
\newblock


\bibitem[Yang et~al\mbox{.}(2020)]%
        {yang2020re}
\bibfield{author}{\bibinfo{person}{Qian Yang}, \bibinfo{person}{Aaron Steinfeld}, \bibinfo{person}{Carolyn Ros{\'e}}, {and} \bibinfo{person}{John Zimmerman}.} \bibinfo{year}{2020}\natexlab{}.
\newblock \showarticletitle{Re-examining whether, why, and how human-AI interaction is uniquely difficult to design}. In \bibinfo{booktitle}{\emph{Proceedings of the 2020 chi conference on human factors in computing systems}}. \bibinfo{pages}{1--13}.
\newblock


\bibitem[Zhai et~al\mbox{.}(2024)]%
        {zhai2024effects}
\bibfield{author}{\bibinfo{person}{Chunpeng Zhai}, \bibinfo{person}{Santoso Wibowo}, {and} \bibinfo{person}{Lily~D Li}.} \bibinfo{year}{2024}\natexlab{}.
\newblock \showarticletitle{The effects of over-reliance on AI dialogue systems on students' cognitive abilities: a systematic review}.
\newblock \bibinfo{journal}{\emph{Smart Learning Environments}} \bibinfo{volume}{11}, \bibinfo{number}{1} (\bibinfo{year}{2024}), \bibinfo{pages}{28}.
\newblock


\bibitem[Zheng and Huang(2023)]%
        {zheng2023begin}
\bibfield{author}{\bibinfo{person}{Qingxiao Zheng} {and} \bibinfo{person}{Yun Huang}.} \bibinfo{year}{2023}\natexlab{}.
\newblock \showarticletitle{" Begin with the End in Mind": Incorporating UX Evaluation Metrics into Design Materials of Participatory Design}. In \bibinfo{booktitle}{\emph{Extended Abstracts of the 2023 CHI Conference on Human Factors in Computing Systems}}. \bibinfo{pages}{1--7}.
\newblock


\bibitem[Zheng et~al\mbox{.}(2022)]%
        {zheng2022ux}
\bibfield{author}{\bibinfo{person}{Qingxiao Zheng}, \bibinfo{person}{Yiliu Tang}, \bibinfo{person}{Yiren Liu}, \bibinfo{person}{Weizi Liu}, {and} \bibinfo{person}{Yun Huang}.} \bibinfo{year}{2022}\natexlab{}.
\newblock \showarticletitle{UX research on conversational human-AI interaction: A literature review of the ACM digital library}. In \bibinfo{booktitle}{\emph{Proceedings of the 2022 CHI Conference on Human Factors in Computing Systems}}. \bibinfo{pages}{1--24}.
\newblock


\end{thebibliography}

\appendix{}

\section{Appendix}

\subsection{Definitions for Index}\label{appendix-index}
\begin{enumerate}
    \item \textbf{Paradigms:}
        \begin{itemize}
            \item \textit{Dyadic:} One-on-one interaction between human and AI. Dyadic human-AI interaction, where end users interact with AI systems, and end users are not AI creators.
            \item \textit{Polyadic:} Multiple social interactions mediated by AI. Where AI sits between two or many types of end users, and these end users are not AI creators either.
        \end{itemize}
    \item \textbf{Application domain:} The specific field or sector in which the system is applied, such as education, healthcare, finance, urban planning.
    \item \textbf{Modality:} The form or channel through which users interact with the system, including Text-Based, Voice-Based, Multi-Modal (Text and Voice Based), Visual Based, Sensor Based, Haptic (touch) Based.
    \item \textbf{System Features:} The characteristics and attributes of the system, including embodiment, level of intelligence, autonomy, adaptability, and learning capabilities.
    \item \textbf{Design Novelty:} The uniqueness and innovativeness of the system's design, indicating how novel the system is compared to existing technologies.
    \item \textbf{Design methods:} Refers to the systematic approaches, strategies, and techniques employed in the creation and development of designs. These methods guide designers in the process of generating ideas, solving problems, and producing effective, functional, and aesthetically pleasing solutions. Example design methods can be:
        \begin{itemize}
            \item Framework-based design (proof-of-concept designs guided by prior frameworks, models, or theories)
            \item Participatory design
            \item Need-finding interviews
            \item Ideation workshops
            \item User-Centered Design
            \item Research through design
            \item Design fiction
            \item Human-Centered Design
            \item Contextual design
            \item Service design
            \item Inclusive design
            \item Ethnographic design
            \item Critical design
            \item Reflective design
            \item Sustainable design
            \item Wizard of Oz
        \end{itemize}
    \item \textbf{Human-AI relationship types:} The nature of the interaction and relationship between humans and the system, ranging from assistant, collaborator, tool, to advisor.
    \item \textbf{Stakeholders:} Individuals or groups who have an interest in, are affected by, or can influence the design, development, deployment, and use of the system. This includes:
        \begin{itemize}
            \item Primary Users: Main direct users.
            \item Secondary Users: Occasional or indirect users.
            \item Tertiary Users: Affected non-users.
            \item Designers and Developers: System creators and maintainers.
            \item Organizations: Entities managing or funding the system.
            \item Policymakers: Regulators setting guidelines.
            \item Support Teams: Maintenance personnel.
            \item Researchers: Those studying and evaluating the system.
        \end{itemize}
    \item \textbf{Social scale:} The scale at which the system operates, such as individual, group, organizational, societal.
    \item \textbf{Theoretical Frameworks:} The underlying theories and concepts that support the development and application of the system, which may include theories from cognitive science, social science, computer science.
\label{index}
\end{enumerate}

\subsection{Graph}\label{tab:nodes_attributes}
Table \ref{TAB:graphtable} is the table for the graph database nodes and their attributes.

\begin{table*}
    \centering
    \begin{tabular}
    {|p{2.5cm}|p{5cm}|p{8cm}|}
    \hline
        \textbf{Node Type} & \textbf{Attributes} & \textbf{Explanation} \\
        \hline
        Paper & Title \newline Narrative \newline Indexes \newline CommunityID \newline \newline Narrative Embedding \newline Index Embedding 
        & The research paper title \newline The narrative/abstract of the research paper \newline The results corresponding to each index category  \newline The communityID to which the research paper is classified into 
        \newline The embedding of the research paper narrative
        \newline The embedding of each index value \\
        \hline
        Metric & Metric Name \newline Metric Type & The metric name category to which the metric is assigned \\
        \hline
        Outcome & Outcome Achieved \newline Paper Title \newline Paper Citation \newline Metric Method \newline Metric Usage \newline Metric Challenges & Metric outcome/result \newline The research paper title \newline The reason for using the metric \newline Method used to evaluate the metric \newline How the metric is used in the research paper \newline The challenges or research gap identified in the research paper \\
        \hline
    \end{tabular}
    \caption{Graph Database Nodes and Their Attributes}
    \label{TAB:graphtable}
\end{table*}

\subsection{User Study Survey}

\begin{enumerate}
    \item Background collection
    \begin{itemize}
        \item What's your name?
        \item What's your email address?
        \item What's your age? (optional)
        \item How do you identify your gender? Male, Female, Prefer not to say, Other, please specify
        \item What degree program are you currently enrolled in? Bachelor's Degree, Master's Degree, Doctorate (PhD or equivalent), Other, please specify
        \item UX research experience: How many years of UX research experience do you have?
        None, less than 1 year, 2-3 years, 3-4 years, 5+ years
        \item Coding experience: How many years of coding experience do you have?
        None, less than 1 year, 2-3 years, 3-4 years, 5+ years
        \item AI literacy \cite{ng2022examination}: How strongly do you agree or disagree with the following statements?  
        \begin{itemize}
            \item I can operate AI applications in everyday life.
            \item I can use AI meaningfully to achieve my everyday goals.
            \item I can assess what the limitations and opportunities of using an AI are.    
        \end{itemize}
        \item AI Assistance in UX Research: Currently, what types of research tasks do you use AI (e.g., ChatGPT) to assist with?
        Research Ideation, Literature review, Evaluation plan development, Data analysis, Prototyping and design, Usability testing, Coding, Writing, Other please specify
    \end{itemize}

    \item Survey 1 and 2 Initial/Final UX Study Plan: We are going to ask about your perceptions of the proposed UX plans for your project.
    \begin{itemize}
    \item How do you feel about the following statements about your proposed UX plans? Strongly disagree, Disagree, Neither agree nor disagree, Agree, Strongly agree
        \begin{itemize} 
            \item Novelty \cite{dean2006identifying}: My proposed UX plans are novel. 
            \item Relevance \cite{dean2006identifying}: My proposed UX plans are relevant to my research goal. 
            \item Efficiency \cite{venable2012comprehensive}: My proposed UX plans are efficient in supporting my research goals.
            \item Completeness \cite{dean2006identifying}: My proposed UX plans are complete.
            \item Clarity \cite{dean2006identifying}: My proposed UX plans are clear.
            \item Concreteness \cite{dean2006identifying}: My proposed UX plans are concrete.
            \item Quality \cite{dean2006identifying}: My proposed UX plans are of high quality.
            \item Feasibility \cite{dean2006identifying}: My proposed UX plans is workable and can be easily implemented. 
            \item Rigorous \cite{venable2012comprehensive}: My proposed UX plans are methodologically rigorous.
            \item Risk considerations \cite{venable2012comprehensive}: My proposed UX plans are in consider of potential issues.
            \item Confidence \cite{stankov2012confidence}: My proposed UX plans are making me feel confident about them.
        \end{itemize}
    \end{itemize}
    
    \item Survey 3 System Features: We are going to ask about your perceptions of each feature in our system. How do you feel about the following statements regarding the Index/ Metric Graph/ Metric List/ Outcome/ Risk feature? (We repeated the same set of questions for each feature.) Strongly disagree, Disagree, Neither agree nor disagree, Agree, Strongly agree
    \begin{itemize}
        \item I enjoy using this feature. \cite{davidson2023development}
        \item I will use this feature in the future if it is publicly available \cite{everard2005presentation}.
        \item I find this feature useful \cite{davis1989perceived}.
        \item I trusted the suggestions and decisions provided by it \cite{everard2005presentation}.
        \item It took significant mental effort to understand this feature \cite{paas1992training}.
        \item I was more interested in exploring and developing my research projects/UX evaluation plan/UX outcomes after using it \cite{chu2010two}.
        \item This feature contributes to revising my UX evaluation plan.
        \item This feature contributes to revising my UX outcomes.
    \end{itemize}
\end{enumerate}

\subsection{Demographic}
Table \ref{participants_info} is the demographic table.

\begin{table*}[h]
\small
\caption{Participants Demographic. LR = Literature Review,
RI = Research Ideation,
EPD = Evaluation Plan Development,
PD = Prototyping \& Design,
DA = Data Analysis,
C = Coding,
W = Writing,
UT = Usability Testing.}
\centering
\resizebox{\textwidth}{!}{%
\scriptsize
\begin{tabular}{cp{1cm}p{1cm}p{1cm}p{1.5cm}p{3cm}}
\toprule
\textbf{ID} & \textbf{Education Level} & \textbf{UX Experience} & \textbf{Coding Experience} & \textbf{AI Literacy \cite{ng2022examination}} & \textbf{AI Usage in Research} \\
\midrule
P1  & Master's                     & 3-4 years & 1-2 years & 4.33 & RI, EPD, C, W. \\
P2  & Master's                     & 3-4 years & 5+ years & 5.00 & RI, EPD, PD, DA, C, W. \\
P3  & Doctorate  & 1-2 years & 3-4 years & 4.67 & LR, PD, DA, C, W. \\
P4  & Master's                     & 3-4 years & 5+ years & 4.67 & RI, LR, DA, C, W. \\
P5  & Postdoc                      & 1-2 years &  3-4 years & 4.00 & LR, UT, C, W. \\
P6  & Doctorate & 1-2 years & 3-4 years & 5.00 & PD, DA, C, W. \\
P7  & Doctorate & 1-2 years & 5+ years & 4.00 & DA, C, W. \\
P8  & Doctorate  & 3-4 years & 5+ years & 5.00 & LR, DA, C, W. \\
P9  & Doctorate  & 1-2 years & 3-4 years & 4.00 & RI, C, W. \\
P10 & Doctorate & 3-4 years & 3-4 years & 4.00 & DA, C, W. \\
P11 & Doctorate  & 3-4 years & 5+ years & 3.67 & RI, DA, C, W, Other \\
P12 & Master's                     & 1-2 years & 3-4 years & 5.00 & RI, LR, EPD, PD, DA, C, W. \\
P13 & Doctorate  & 1-2 years & 3-4 years & 4.00 & C, W. \\
P14 & Bachelor's                   & 3-4 years & 3-4 years & 5.00 & LR, PD, DA, C, W. \\
P15 & Doctorate & 3-4 years & 5+ years & 4.00 & C, W. \\
P16 & Doctorate & < 1 year & 5+ years & 4.33 & Other \\
P17 & Doctorate  & 3-4 years & 5+ years & 4.67 & Other \\
P18 & Doctorate  & 3-4 years & 5+ years & 5.00 & RI, LR, PD, DA, C, W. \\
P19 & Master's & 3-4 years & 5+ years & 4.00 & PD, C, W. \\
\bottomrule
\end{tabular}}
\label{participants_info}
\end{table*}

\subsection{Data Analysis}
Table \ref{tab:combined_results} is the data analysis for changes in user perceptions towards the proposed UX plan result.

\begin{table*}[h]
\caption{Descriptive Analysis and Wilcoxon Signed Rank Test Results for Changes in User Perceptions Towards a Proposed UX Plan (N = 19). Significant values: **$p < 0.01$, *$p < 0.05$.}
\centering
\begin{tabular}{p{3cm}p{1.5cm}p{1.5cm}p{1.5cm}p{1.5cm}p{1.5cm}p{1.5cm}p{1.5cm}}
\hline
\textbf{Constructs} & \textbf{Before Median} & \textbf{Before IQR} & \textbf{After Median} & \textbf{After IQR} & \textbf{Test Statistic} & \textbf{Effect Size $r_b$} \\ 
\hline
Novelty & 3 & 2 & 4 & 1 & 22 & 0.51 \\
Relevance & 5 & 1 & 5 & 0.5 & 8 & 0.56 \\
Efficiency$^*$ & 4 & 0 & 5 & 1  & 12 & 0.74 \\
Completeness$^{**}$ & 3 & 2 & 4 & 1 & 16 & 0.79 \\
Clarity$^{**}$ & 4 & 1 & 4 & 1 & 4 & 0.90 \\
Concreteness$^*$ & 4 & 1 & 4 & 1.5 & 24 & 0.74 \\
Quality$^{**}$ & 3 & 1 & 4 & 0.5 & 11 & 0.82 \\
Workability$^*$ & 4 & 2 & 5 & 2 & 16.5 & 0.64 \\
Rigor$^*$ & 3 & 2 & 4 & 1.5 & 15 & 0.75 \\
Ethics$^{**}$ & 3 & 2.5 & 4 & 1.5 & 10 & 0.81 \\
Trust$^*$ & 4 & 1 & 4 & 1 & 5 & 0.82 \\
Confidence$^*$ & 3 & 1 & 4 & 1 & 18 & 0.70 \\

\hline
\end{tabular}
\label{tab:combined_results}
\end{table*}

\end{document}